\documentclass[doi=false]{aa}  

\usepackage{graphicx}
\usepackage{txfonts}
\usepackage{float}
\usepackage{pdflscape}
\usepackage{threeparttable}
\usepackage{bbding}
\usepackage{tikz}
\usepackage{multirow}
\usepackage{orcidlink}
\usepackage{natbib}

\def\asec{$^{\prime\prime}$}

\def\farcs{\hbox{$.\!\!^{\prime\prime}$}}

\def\h{$^{\mathrm{h}}$}

\def\tlvco{$^{12}$CO}
\def\thrco{$^{13}$CO}
\def\ceteno{C$^{18}$O}
\def\h2co{H$_{2}$CO}
\def\C3h2{$c-$C$_3$H$_2$}
\def\ch3oh{CH$_{3}$OH}
\def\htwocolow{3$_{0,3}$--2$_{0,2}$}
\def\htwocomid{3$_{2,1}$--2$_{2,0}$}
\def\htwocohigh{3$_{2,2}$--2$_{2,1}$}

\newcommand*\circled[1]{\tikz[baseline=(char.base)]{
            \node[shape=circle,draw,inner sep=0.5pt] (char) {#1};}}

\begin{document} 

   \title{Early Planet Formation in Embedded Disks (eDisk). XX. Constraining the chemical tracers of young protostellar sources}

   \author{Rajeeb Sharma
          \inst{1}
          \orcidlink{0000-0002-0549-544X},
          Jes K. J\o rgensen
          \inst{1}
          \orcidlink{0000-0001-9133-8047},
          Merel L. R. van 't Hoff
          \inst{2,3}
          \orcidlink{0000-0002-2555-9869},
          Jeong-Eun Lee
          \inst{4}
          \orcidlink{0000-0003-3119-2087},
          Yuri Aikawa
          \inst{5}
          \orcidlink{0000-0003-3283-6884},
          Sacha Gavino
          \inst{1,6}
          \orcidlink{0000-0001-5782-915X},
          Yao-Lun Yang
          \inst{7}
          \orcidlink{0000-0001-8227-2816},
          Nagayoshi Ohashi
          \inst{8}
          \orcidlink{0000-0003-0998-5064},
          John J. Tobin
          \inst{9}
          \orcidlink{0000-0002-6195-0152},
          Patrick M. Koch
          \inst{8}
          \orcidlink{0000-0003-2777-5861},
          Zhi-Yun Li
          \inst{10}
          \orcidlink{0000-0002-7402-6487},
          Leslie W. Looney
          \inst{11,9}
          \orcidlink{0000-0002-4540-6587},
          Mayank Narang
          \inst{8}
          \orcidlink{0000-0002-0554-1151},
          Suchitra Narayanan
          \inst{12}
          \orcidlink{0000-0002-0244-6650},
          \and
          Travis J.\ Thieme
          \inst{8}
          \orcidlink{0000-0003-0334-1583}
          }

   \institute{Niels Bohr Institute, University of Copenhagen, Jagtvej 155A, 2200 Copenhagen N., Denmark\\
              \email{rajeeb.sharma@nbi.ku.dk}
         \and
             Department of Astronomy, University of Michigan, 1085 S. University Ave., Ann Arbor, MI 48109-1107, USA
         \and
             Department of Physics and Astronomy, Purdue University, 525 Northwestern Avenue, West Lafayette, IN 47907, USA
         \and
             Department of Physics and Astronomy, SNU Astronomy Research Center, Seoul National University, 1 Gwanak-ro, Gwanak-gu, Seoul 08826, Korea
         \and
            Department of Astronomy, Graduate School of Science, The University of Tokyo, 113-0033 Tokyo, Japan
         \and
            Dipartimento di Fisica e Astronomia, Università di Bologna, Via Gobetti 93/2, 40122 Bologna, Italy
         \and
            Star and Planet Formation Laboratory, RIKEN Cluster for Pioneering Research, Wako-shi, Saitama, 351-0106, Japan
         \and
            Institute of Astronomy and Astrophysics, Academia Sinica. 11F of Astronomy-Mathematics Building, AS/NTU No.1, Sec. 4, Roosevelt Rd, Taipei 106216, Taiwan, R.O.C.
         \and
            National Radio Astronomy Observatory, 520 Edgemont Rd., Charlottesville, VA 22903 USA
         \and
            University of Virginia, 530 McCormick Rd., Charlottesville, Virginia 22904, USA
         \and
            Department of Astronomy, University of Illinois, 1002 West Green St, Urbana, IL 61801, USA
         \and
            Institute for Astronomy, University of Hawai\`{i} at Mānoa, 2680 Woodlawn Dr., Honolulu, HI 96822, USA           
             }


\authorrunning{R. Sharma et al.}
\titlerunning{eDisk: Constraining the chemical tracers of young protostellar sources}

 
  \abstract
   {Recent studies indicate that the formation of planets in protoplanetary disks begins early in the embedded Class 0/I phases of protostellar evolution. The physical and chemical makeup of the embedded phase can provide valuable insights into the process of star and planet formation.} 
   {This study aims to provide a thorough overview of the various morphologies for molecular emissions observed on disk scales ($\lesssim$100 au) toward nearby embedded sources.}
   {We present high angular resolution (0\farcs1, $\sim$15 au) molecular line emissions for \tlvco, \thrco, \ceteno, SO, SiO, DCN, \ch3oh, \h2co, and \C3h2 toward 19 nearby protostellar sources in the context of the Atacama Large Millimeter/submillimeter Array (ALMA) Large Program ``Early Planet Formation in Embedded Disks (eDisk).''} 
   {Emissions in \tlvco~are seen toward all sources and primarily trace outflowing materials. A few sources also show high-velocity jets in SiO emission and high-velocity channel maps of \tlvco. The \thrco~and \ceteno~emissions are well-known tracers of high-density regions and trace the inner envelope and disk regions with clear signs of rotation seen at continuum scales. The large-scale emissions of \thrco~also delineate the outflow cavity walls where the outflowing and infalling materials interact with each other, and exposure to UV radiation leads to the formation of hydrocarbons such as \C3h2. Both DCN and \ch3oh, when detected, show compact emissions from the inner envelope and disk regions that peak at the position of the protostar. The \ch3oh~emissions are contained within the region of DCN emissions, which suggests that \ch3oh~traces the hot core regions. Likewise, a few sources, also display emissions in \ch3oh~toward the outflow. Both SO and \h2co~show complex morphology among the sources, suggesting that they are formed through multiple processes in protostellar systems.}
  {}


   \maketitle

%

\section{Introduction}

The formation of low-mass stars such as our Sun begins with the gravitational collapse of dense prestellar cores in cold molecular clouds. As the temperature and the pressure increase, the collapsing core eventually forms a protostar, signaling a shift from the prestellar to the protostellar phase. In the earliest phases of protostellar evolution, the Class 0 and Class I stages, the protostar is embedded in an infalling envelope that accounts for a substantial fraction of the mass of the system \citep{Lada_1984, Andre_1993, Andre_2000}. The embedded nature of these young sources has long proved to be an obstacle for directly observing protostellar disks. Consequently, most studies in the past have focused predominantly on disks in more evolved Class II sources, where most of the envelope has been accreted or dissipated \citep[e.g.,][]{Alma_2015,Andrews_2018}.

\begin{figure}[ht!]
\centering
    \includegraphics[width=1.0\linewidth]{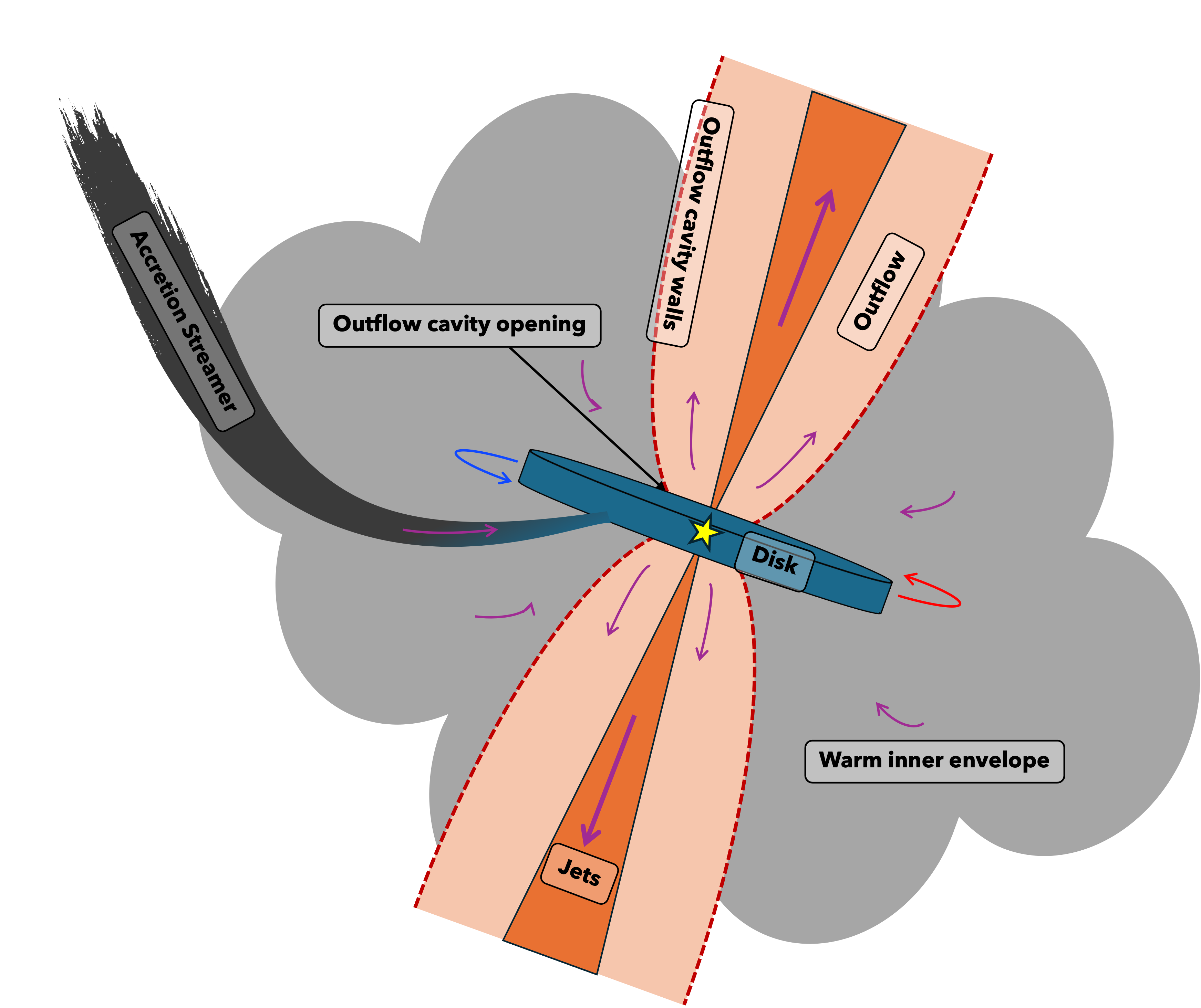}
\caption{Illustration of the different components present in an embedded protostellar system.  \label{fig:embedded_source}}
\end{figure}

Over its first decade of operations, the Atacama Large Millimeter/submillimeter Array (ALMA) has revolutionized studies of the inner regions of envelopes and structures of protoplanetary disks surrounding young stars in their earliest stages. Figure~\ref{fig:embedded_source} presents an overview of our current understanding of the different components of an embedded protostar, such as its disk, envelope, and outflows. These components play an important role during the formation of the protostar and the planets. For instance, protostellar disks regulate the mass accreted by the protostar from the envelope and provide the necessary conditions for the dust grain growth that seeds planet formation \citep{Testi_2014, Maury_2019}. Likewise, protostellar outflows and winds eject mass back into the molecular cloud, decreasing the protostar's efficiency of mass accretion and replenishing the turbulent motions via feedback mechanisms \citep{Arce_2007,Nakamura_2014}. 

Understanding and characterizing the components of young protostellar systems in detail is therefore crucial in forming a comprehensive picture of star and planet formation. One way to achieve this is to study the molecular line emissions that trace the various components within a protostellar system. These molecules serve as a powerful diagnostic tool for these components and provide valuable insight into the various physical and chemical processes that occur during the formation of stars and planets. 

The chemical makeup of a protostellar system is both complex and dynamic. Many molecules ranging from simple diatomic molecules to complex organic molecules (COMs; C-bearing molecules with six or more atoms) can already form during the prestellar phase \citep[e.g., see reviews by][]{Jorgensen_2020,Ceccarelli_2023}. Initially, much of the chemistry of a system is at least partially inherited from the prestellar core and infalling envelope, where most molecules are primarily frozen onto the dust grains \citep{Pontoppidan_2014,Boogert_2015}. As the protostar evolves, the wide range of temperatures, densities, and physical processes occurring within its structures facilitate multiple new reaction pathways. Consequently, the inherited chemical composition undergoes significant changes, giving rise to a variety of new species \citep[e.g., see reviews by][]{Herbst_2009,Jorgensen_2020,Oberg_2021}.

Astrochemical studies have long aimed to enhance our understanding of protostellar systems through molecular observations. This goal involves two primary objectives: first, identifying unique chemical tracers of specific components of protostellar systems on different spatial scales and in various evolutionary stages of young stars, and second, thoroughly investigating the chemistry of these various components. An attempt at doing this systematically using ALMA was presented by \citet{Tychoniec_2021}, who examined the spectral line emissions of various molecules toward 16 Class 0/I protostars in nearby star-forming regions. By combining high-resolution spectral line data from several ALMA observations at different wavelengths (1.3 mm, 2 mm, and 3 mm observations), they analyzed emissions from several molecules and developed a comprehensive reference of molecules that traces the various morphologies in Class 0/I sources.

In this work, we present high-resolution ALMA spectral line observations of 19 nearby Class 0 and I protostars from the Early Planet Formation in Embedded Disks (eDisk) survey \citep{Ohashi_2023}. This work extends on the findings from \citet{Tychoniec_2021}, offering two key advantages over the previous study. First, unlike the sources in \citet{Tychoniec_2021}, which were drawn from various studies, the eDisk program utilizes a uniform and well-characterized sample of sources, all observed in the same manner. This provides a more consistent and reliable baseline for comparison between the different sources. Second, the eDisk observations achieve higher spatial resolutions compared to those in the previous study ($\sim$0\farcs1 than $\sim$0\farcs5 -- 3\arcsec). This enables us to investigate the physical and chemical processes in the innermost regions of protostellar disks.

Initial results from continuum observations and kinematic analyses of individual eDisk sources have already been presented in a series of first-look papers. These papers show that in contrast to Class II sources, where rings and gaps are ubiquitous, clear substructures are only seen in 2 out of the 19 sources, L1489IRS \citep{Yamato_2023} and OphIRS63 \citep{Flores_2023}. Additionally, position velocity (PV) analysis of spectral line emissions has shown that Keplerian-rotating disks are common even in the youngest Class 0 protostars \citep{Aso_2023,Hoff_2023,Kido_2023,Sai_2023,Sharma_2023,Thieme_2023}. However, while these first-look papers have discussed cases of specific sources, a comprehensive synthesis of the data that identifies and contrasts the different chemical tracers across the entire sample is still lacking. This paper aims to address this gap by systematically analyzing the molecular tracers across all eDisk sources, providing a holistic view of the molecular tracers in early protostellar systems.

The paper is structured as follows. Sect.~\ref{sec:data} briefly describes the observations and data reduction processes. The empirical results from the observations of the molecular line emissions are presented in Sect.~\ref{sec:results}. The implications of the results are discussed in Sect.~\ref{sec:discussion}, and the conclusions are presented in Sect.~\ref{sec:conclusion}.

\section{Observations}\label{sec:data}

Observations of the sources were conducted in multiple sessions that spanned April 2021 to July 2022 as part of the eDisk ALMA Large Program (2019.1.00261.L, PI: N. Ohashi) in Band 6 at a wavelength of 1.3 mm, with supplemental data obtained from the ALMA DDT observations (2019.A.00034.S, PI: J. Tobin), also observed in Band 6 at a wavelength of 1.3 mm. The projected baselines of the observations ranged from 15 m to 12594.5 m, allowing for spatial resolutions of $\sim$0\farcs04 for the continuum and $\sim$0\farcs1 for the spectral lines, with slight variations among the sources due to differences in observing conditions and exact array configurations. The spectral setup was set up to probe the dust continuum and a suite of molecular species, including \tlvco, \thrco, \ceteno, SO, SiO, DCN, \C3h2, \h2co, and \ch3oh. Table~\ref{tab:lines} provides an overview of these targeted molecules, together with the velocity resolutions at which they were observed. 

Both short and long baseline data were obtained for all sources in our sample, with the exception of OphIRS63, for which only short baseline data could be obtained due to scheduling constraints. The short baseline observations increase the maximum recoverable scale, ($\theta_{MRS}$), of our observations to $\approx 3$\arcsec, allowing us to investigate the relatively extended emissions around the sources. A detailed description of the observational parameters, including the frequency ranges of the basebands, specific observation dates, number of antennas used, precipitable water vapor levels, and the complete spectral and correlator setups, can be found in \citet{Ohashi_2023}.

The data were initially calibrated using the standard ALMA pipeline calibrations, which involved bandpass, flux, and phase calibrations. Subsequent reduction and imaging were performed using the Common Astronomy Software Application \citep[CASA;][]{Mcmullin_2007} version 6.2.1, following a script developed for the eDisk Large Program.\footnote{\url{https://doi.org/10.5281/zenodo.7986682}} To enhance the data quality and signal-to-noise ratio, both phase-only and phase-and-amplitude self-calibration were applied to the continuum data. The number of self-calibrations applied varied by source. The spectral line cubes were then generated by subtracting the continuum emission from the visibility data using the CASA task {\it uvcontsub}. The derived solutions from the continuum self-calibrations were also applied to the line data. For each source, the final continuum images were created with a range of robust parameters from -2.0 to 2.0, and the line images were created with robust parameters of 0.5 and 2.0. We adopted the robust value of 2.0 for both the continuum contours and the spectral line images used in this work to increase the signal-to-noise ratio, particularly of the weaker spectral line emissions. For OphIRS63, which only has short baseline observations, we adopt the robust value of 0.5. This choice allowed us to keep the beam sizes of its line observations within a factor of $\lesssim$2-3 of the other sources, without significantly compromising the signal-to-noise ratio of the images.

In addition, comparable data for the protostars TMC1A  \citep[2015.1.01415.S;][]{Bjerkeli_2016,Harsono_2018} and B335 \citep[2013.1.00879.S, 2017.1.00288.S;][]{Yen_2015,Bjerkeli_2019} were taken from the ALMA archive. Alongside the continuum emission, the archival data for TMC1A and B335 included molecular emissions from \tlvco, \thrco, and \ceteno. These raw datasets for these sources were reduced and imaged using the same script developed for the eDisk Large program to ensure consistency.

\begin{table*}[]
\centering
\begin{threeparttable}
\caption{Summary of spectral line transitions covered.}
\label{tab:lines}
\begin{tabular}{lccccccc}
\hline\hline
Molecule & Transition & Rest Frequency & Velocity Resolution & $A_{ij}$$^{a}$ & $E_{up}$$^{b}$ & RMS range$^{c}$ & \\
 & & (GHz) & (km s$^{-1}$) & (s$^{-1}$) & (K) &  (mJy beam$^{-1}$) & \\
\hline
$^{12}$CO    & $2-1$                 & 230.538000 & 0.635 & $6.910 \times 10^{-7}$ & 16.6 & $0.83 - 2.14$ & \\
$^{13}$CO    & $2-1$                 & 220.398684 & 0.167 & $5.066 \times 10^{-7}$ & 15.9 & $1.58 - 4.61$ & \\
C$^{18}$O    & $2-1$                 & 219.560354 & 0.167 & $6.011 \times 10^{-7}$ & 15.8 & $1.36 - 3.49$ & \\
SiO          & $5-4$                 & 217.104980  & 1.340 & $5.196 \times 10^{-4}$ & $31.3$ &  $0.52 - 1.80$ &\\
SO           & $6_5-5_4$         & 219.949442 & 0.167 & $1.335 \times 10^{-4}$ & 35.0 & $1.75 - 4.33$ & \\
CH$_3$OH     & $4_2-3_1$, E      & 218.440063 & 1.340 & $4.686 \times 10^{-5}$ & 45.6 & $0.46 - 1.24$ & \\
DCN          & $3-2$                & 217.238538 & 1.340 & $4.575 \times 10^{-4}$ & 20.9 & $0.51 - 1.68$ & \\
H$_2$CO      & $3_{0,3}-2_{0,2}$ & 218.222192 & 0.167 & $2.818 \times 10^{-4}$ & 21.0 & $0.45 - 1.14$ & \\
H$_2$CO      & $3_{2,1}-2_{2,0}$ & 218.760066 & 1.340 & $1.577 \times 10^{-4}$ & 68.1 & $1.30 - 3.26$ & \\
H$_2$CO      & $3_{2,2}-2_{2,1}$ & 218.475632 & 1.340 & $1.571 \times 10^{-4}$ & 68.1 & $0.45 - 1.24$ & \\
c-C$_3$H$_2$$^{\mathrm{\dagger}}$ & $6_{0,6}-5_{1,5}$ & 217.822148 & 1.340 & $5.396 \times 10^{-4}$ & 38.6 & $0.47 - 1.36$ & \\
c-C$_3$H$_2$$^{\mathrm{\dagger}}$ & $6_{1,6}-5_{0,5}$ & 217.822148 & 1.340 & $5.396 \times 10^{-4}$ & 38.6 & $0.47 - 1.36$ & \\
c-C$_3$H$_2$ & $5_{1,4}-4_{2,3}$ & 217.940046 & 1.340 & $4.026 \times 10^{-4}$ & 35.4 & $0.53 - 1.16$ & \\
c-C$_3$H$_2$ & $5_{2,4}-4_{1,3}$ & 218.160456 & 1.340 & $4.041 \times 10^{-4}$ & 35.4 & $0.45 - 1.19$ & \\
\hline
                             
\end{tabular}
\begin{tablenotes}
    
    \item[\textdagger]These two lines are blended in eDisk observations.
    \item[$a$]Einstein A-coefficients.
    \item[$b$]Upper-state energy of the transition.
    \item[$c$]This range represents the range of the RMS values obtained toward different sources.

\end{tablenotes}
\end{threeparttable}
\end{table*}

\section{Results}\label{sec:results}

\begin{table*}[]
\centering
\begin{threeparttable}
\caption{Summary of the eDisk sources.} \label{tab:continuum}
\begin{tabular}{lcccccccc}
\hline\hline
Source Name  & Hereafter & Class & Distance & T$_{\mathrm{bol}}$ & L$_{\mathrm{bol}}$ & Inclination & $v_{\mathrm{sys}}$ & References \\
 & & & (pc) & (K) & ($L_{\odot}$) & ($^{\circ}$) & (km s$^{-1}$) & \\
 \hline
BHR 71 IRS1      & BHR71 IRS1 & 0 & 176 & 66  & 10  & 39 & --4.45 & (1)    \\
BHR 71 IRS2      & BHR71 IRS2 & 0 & 176 & 39  & 1.1 & 31 & --4.45 & (1)    \\
Ced110 IRS4$^{\mathrm{\dagger}}$      & Ced110IRS4 & 0 & 189 & 68  & 1.0 & 75 & 4.67 & (2)    \\
GSS30 IRS3       & GSS30IRS3  & 0 & 138 & 50  & 1.7 & 72 & 2.84 & (3)    \\
IRAS 04166+2706  & IRAS04166  & 0 & 156 & 61  & 0.4 & 47 & 6.80 & (4)    \\
IRAS 15398--3359 & IRAS15398  & 0 & 155 & 50  & 1.4 & 51 & 5.40 & (5)    \\
IRAS 16253--2429 & IRAS16253  & 0 & 139 & 42  & 0.16 & 68 & 4.00 & (6)   \\
IRAS 16544--1604 & IRAS16544  & 0 & 151 & 50  & 0.89 & 73 & 4.96 & (7)   \\
L1527 IRS        & L1527      & 0 & 140 & 41  & 1.3 & 75 & 5.90 & (8)    \\
R CrA IRAS 32$^{\mathrm{\dagger}}$   & IRAS32     & 0 & 150 & 64  & 1.6 & 69 & 5.86 & (9)    \\
R CrA IRS5N      & IRS5N      & 0 & 147 & 59  & 1.4 & 65 & 6.65 & (10)    \\
IRAS 04169+2702  & IRAS04169  & I & 156 & 163 & 1.5 & 44 & 6.90 & (11)    \\
IRAS 04302+2247  & IRAS04302  & I & 160 & 88  & 0.43 & 84 & 5.70 & (12)   \\
L1489 IRS        & L1489      & I & 146 & 213 & 3.4 & 71 & 7.38 & (13)     \\
Oph IRS43$^{\mathrm{\dagger}}$        & OphIRS43   & I & 137 & 193 & 4.1 & 78 & 3.90 & (14)    \\
Oph IRS63        & OphIRS63   & I & 132 & 348 & 1.3 & 47 & 2.80 & (15)    \\
R CrA IRS7B$^{\mathrm{\dagger}}$      & IRS7B      & I & 152 & 88  & 5.1 & 68 & 5.90 & (16, 17)    \\
\hline
B335             & B335       & 0 & 165 & 41  & 1.4 & 37 & 8.30 & (9) \\
TMC-1A           & TMC1A      & I & 137 & 183 & 2.3 & 52 & 6.80 & (9)  \\
\hline
\end{tabular}
\begin{tablenotes}
    \item[\textdagger] These sources have close binaries with projected seperations of $\sim$1\farcs30 ($\sim$250 au) for Ced110IRS4, $\sim$1\farcs38 ($\sim$207 au) for IRAS32, $\sim$0\farcs55 ($\sim$74 au) for OphIRS43, and $\sim$0\farcs70 ($\sim$106 au) for IRS7B. The inclination angle for these binaries are based on their primary companion.
    \item \textbf{References.} (1)~\citet{Gavino_2024}; (2) \citet{Sai_2023}; (3) \citet{Santamaria_2024}; (4) \citet{Phuong_2025}; (5) \citet{Thieme_2023}; (6) \citet{Aso_2023}; (7) \citet{Kido_2023}; (8) \citet{Hoff_2023}; (9) \citet{Ohashi_2023}; (10) \citet{Sharma_2023}; (11) \citet{Han_2025}; (12) \citet{Lin_2023}; (13) \citet{Yamato_2023}; (14) \citet{Narayanan_2023}; (15) \citet{Flores_2023}; (15) \citet{Takakuwa_2024}; (16) \citet{Ohashi_2023}.
\end{tablenotes}
\end{threeparttable}
\end{table*}

\begin{figure*}[ht!]
    \includegraphics[width=1.0\textwidth]{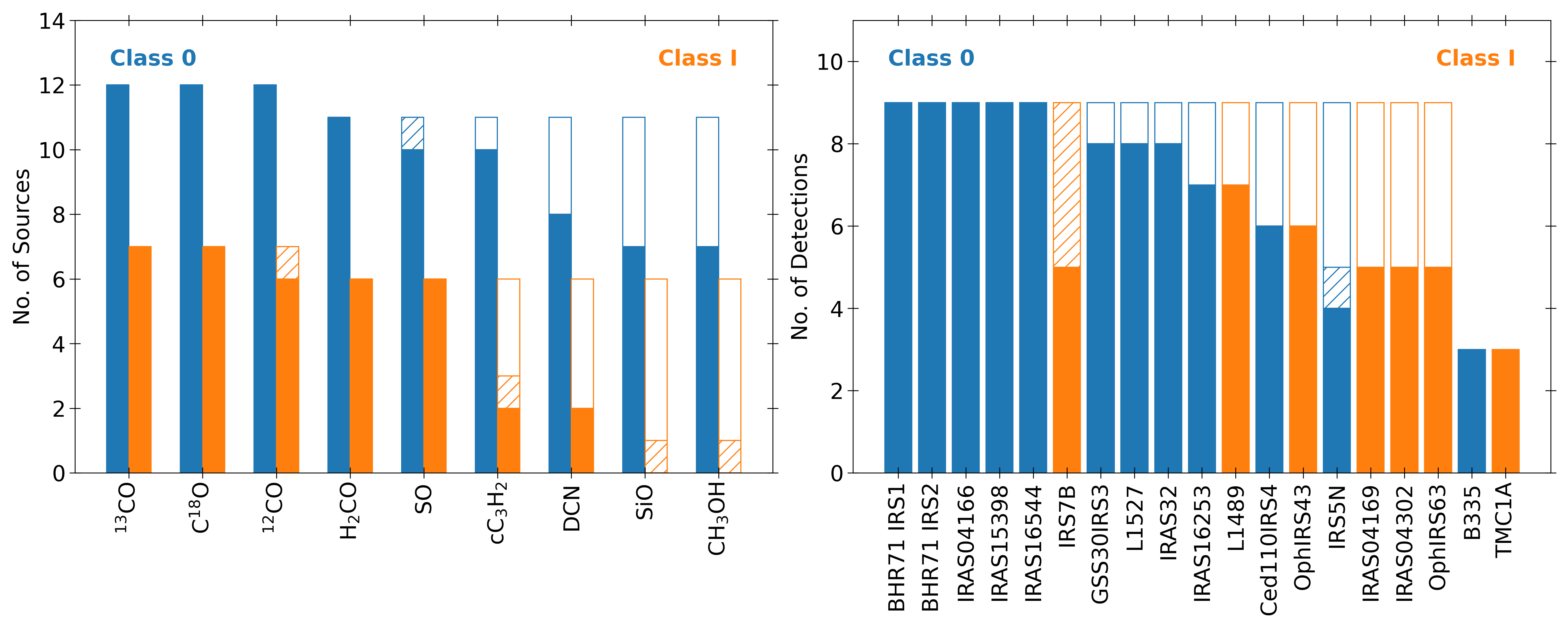}
\caption{Overview of the detection statistics showcasing plots illustrating the number of sources where emissions of each molecule are observed (\textit{left}) and the number of molecules observed toward each source (\textit{right}). The dashed regions represent cases where emissions are observed toward a source but are unlikely to be associated with that source. \label{fig:detection_stats}}
\end{figure*}

Table~\ref{tab:continuum} provides a summary of the continuum observations for the eDisk sample. All sources are located within 200 pc and have inclinations greater than 30 degrees, offering favorable viewing angles to study the disks. The bolometric luminosities (L$_{\mathrm{bol}}$) of these sources range from 0.16 L$_{\odot}$ to 10 L$_{\odot}$, with a median value of $\sim$1 L$_{\odot}$, reflecting a diverse set of protostellar objects.
In order to characterize the different morphologies and the kinematics observed toward the eDisk sources, we create peak intensity (moment 8) and peak velocity (moment 9) maps for each of the targeted molecules. These maps are made only using pixels where emissions are detected at a level of $\geq3\sigma$. Here, a ``clear'' detection in the molecular emission is defined to be a spatial structure in the moment maps that extends beyond the area of a synthesized beam. This ensures that any marginal emissions or isolated peaks that can arise from noise fluctuations and exceed the 3$\sigma$ threshold are not identified as false positives. Figure~\ref{fig:detection_stats} presents an overview of the overall detection statistics, detailing the number of sources that display emissions for each molecule on the left and the number of molecules detected toward each source on the right. Certain molecules such as CO isotopologues, SO, and \h2co are found throughout the sample, while molecules such as SiO, DCN, and \ch3oh~are predominantly detected toward the Class 0 sources. Table~\ref{tab:molecules_sources} summarizes the molecular lines detected toward each source in this study. 

In this section, we explore the origins of each of the molecules observed toward the eDisk sources. For the purpose of this study, we rely exclusively on the results seen in the moment 8 and moment 9 plots to identify and define the different components. For each of the molecular species observed, we display only a representative subset of sources that provide key insights into the emission morphology described in the main text. These representative sources are selected on the basis of the clarity and distinctiveness of the various emission morphologies observed in the moment maps. 
The complete maps of all sources for each molecule are shown in Appendices~\ref{appendix:large_scale_emissions} and \ref{appendix:small_scale_emissions}, with references to these figures made whenever the corresponding sources are discussed throughout the sections. Furthermore, for ease of access and exploration, an interactive version of all the molecular maps for each source is available online.\footnote{\url{https://group.asiaa.sinica.edu.tw/almaLP_edisk/data.php}}

\begin{table*}[]
\centering
\begin{threeparttable}
\caption{Summary of molecules observed toward eDisk sources.} \label{tab:molecules_sources}
\begin{tabular}{lccccccccc}
\hline\hline
Source Name  & \tlvco & \thrco & \ceteno & SiO & SO & DCN & \ch3oh & \h2co & \C3h2 \\
 \hline
BHR71 IRS1 & \checkmark & \checkmark & \checkmark & \checkmark & \checkmark & \checkmark & \checkmark & \checkmark & \checkmark \\
BHR71 IRS2 & \checkmark & \checkmark & \checkmark & \checkmark & \checkmark & \checkmark & \checkmark & \checkmark & \checkmark \\
Ced110IRS4 & \checkmark & \checkmark & \checkmark & \XSolid & \checkmark & \XSolid & \XSolid & \checkmark & \checkmark \\
GSS30IRS3  & \checkmark & \checkmark & \checkmark & \checkmark & \checkmark & \XSolid & \checkmark & \checkmark & \checkmark \\
IRAS04166  & \checkmark & \checkmark & \checkmark & \checkmark & \checkmark & \checkmark & \checkmark & \checkmark & \checkmark \\
IRAS15398  & \checkmark & \checkmark & \checkmark & \checkmark & \checkmark & \checkmark & \checkmark & \checkmark & \checkmark \\
IRAS16253  & \checkmark & \checkmark & \checkmark & \XSolid & \checkmark & \checkmark & \XSolid & \checkmark & \checkmark \\
IRAS16544  & \checkmark & \checkmark & \checkmark & \checkmark & \checkmark & \checkmark & \checkmark & \checkmark & \checkmark \\
L1527      & \checkmark & \checkmark & \checkmark & \checkmark & \checkmark & \checkmark & \XSolid & \checkmark & \checkmark \\
IRAS32     & \checkmark & \checkmark & \checkmark & \XSolid & \checkmark & \checkmark & \checkmark & \checkmark & \checkmark \\
IRS5N      & \checkmark & \checkmark & \checkmark & \XSolid & \circled{\checkmark} & \XSolid & \XSolid & \checkmark & \XSolid \\
IRAS04169  & \checkmark & \checkmark & \checkmark & \XSolid & \checkmark & \XSolid & \XSolid & \checkmark & \XSolid \\
IRAS04302  & \checkmark & \checkmark & \checkmark & \XSolid & \checkmark & \XSolid & \XSolid & \checkmark & \XSolid \\
L1489      & \checkmark & \checkmark & \checkmark & \XSolid & \checkmark & \checkmark & \XSolid & \checkmark & \checkmark \\
OphIRS43   & \checkmark & \checkmark & \checkmark & \XSolid & \checkmark & \XSolid & \XSolid & \checkmark & \checkmark \\
OphIRS63   & \checkmark & \checkmark & \checkmark & \XSolid & \checkmark & \XSolid & \XSolid & \checkmark & \XSolid \\
IRS7B      & \circled{\checkmark} & \checkmark & \checkmark & \circled{\checkmark} & \checkmark & \checkmark & \circled{\checkmark} &  \checkmark & \circled{\checkmark} \\
B335       & \checkmark & \checkmark & \checkmark & -- & -- & -- & -- & -- & -- \\
TMC1A      & \checkmark & \checkmark & \checkmark & -- & -- & -- & -- & -- & -- \\
 \hline
\end{tabular}
\begin{tablenotes}
\item[\checkmark] Emission is detected at $\geq 3 \sigma$ level toward the source.
\item[\XSolid] No emission is detected at $3 \sigma$ level.
\item[\circled{\checkmark}] Emission is detected at $\geq 3 \sigma$ level, but is unlikely to be directly associated with the protostellar system.
\end{tablenotes}
\end{threeparttable}
\end{table*}

\subsection{\tlvco}

\begin{figure*}[ht!]
    \includegraphics[width=1.0\textwidth]{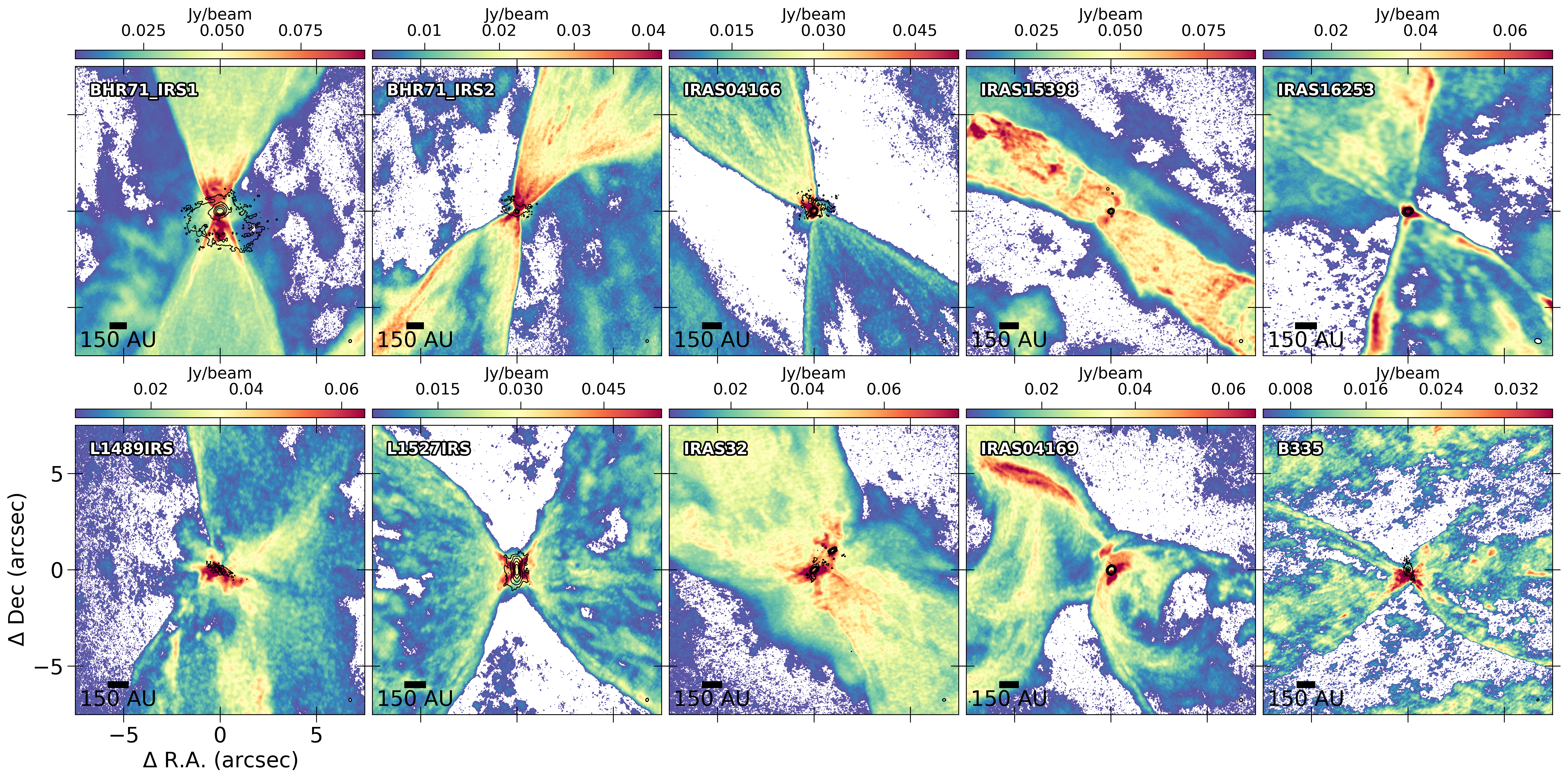}
\caption{Moment 8 maps of \tlvco~($J$=2--1) created using $\geq3\sigma$ emissions toward ten representative eDisk sources. The contour lines display the continuum emission at thresholds of 5$\sigma$, 15$\sigma$, 45$\sigma$, 135$\sigma$, and 320$\sigma$ for each source. The scale bar is located at the bottom left, and the synthesized beam is indicated in white at the bottom-right corner of each image. \label{fig:12co_mom8}}
\end{figure*}

\tlvco~(2--1) emissions are detected toward all eDisk sources. Apart from IRS5N, Ced110IRS4, IRS7B, and OphIRS43, these emissions are primarily associated with molecular outflows and also delineate the walls of the outflow cavities. Figure~\ref{fig:12co_mom8} presents moment 8 maps that display the large-scale emission of \tlvco~(2--1) toward ten eDisk sources. These maps reveal the different varieties of outflows observed in our sample. Of the 15 sources that clearly display molecular outflow in \tlvco, 13 exhibit wide-angle outflows that generally display curvature or have shoulder-like structures and deviate from being strictly conical or parabolic (see Figure~\ref{appendix:12co_mom8_mom9}). In comparison, the outflow toward IRAS15398 is much more collimated and displays a U-shaped morphology, with the outflow cavity maintaining a relatively constant cross-sectional area. The outflow toward IRAS04302 is faint and lacks a clearly identifiable morphology (see Figure~\ref{appendix:12co_mom8_mom9}). However, the extended blue-shifted emission seen toward the east shifts away from the protostar at higher velocities in the channel map (\citealt{Feeney-Johansson_prep}; see also \citealt{Lin_2023}). 

The origin of the outflows appears to be deep within the inner disk region, very close to the corresponding positions of the protostars. The emissions are generally perpendicular to the major axis of the elongated continuum emission and typically exhibit distinct velocity gradients with red- and blue-shifted emissions tracing individual outflow lobes (see Figure~\ref{appendix:12co_mom8_mom9}). The outflow originating from the main source of the close binary system IRAS32 appears to interact and blend with the material surrounding the secondary source. In addition to the outflow, the warm inner envelope and disk regions surrounding the continuum can also be traced by the \tlvco~emission, especially in the velocity channels close to the systemic velocity ($v_{sys}$) of the source. These emissions likely originate from CO molecules sublimating off dust grains in the inner envelope and disk regions. This can occur due to the presence of viscous accretion heating that raises the dust temperatures above 20 K, the sublimation point of CO, even toward the midplane of the disk~\citep{Takakuwa_2024}.

The \tlvco~emissions toward the sources Ced110IRS4, OphIRS43, IRS7B, and IRS5N do not appear to trace any apparent outflow or jets associated with the protostar (see Figure~\ref{appendix:12co_mom8_mom9}). Three of these sources are close binaries where the emissions mainly seem to be associated with the main source. These emissions then appear to interact and blend with the material surrounding the secondary source, as with IRAS32. In Ced110IRS4 and OphIRS43, the emission appears along the direction of the major axis of the dust disks and appears to trace the rotation of the disk instead of the outflow, similar to \ceteno~emission (see Figures~\ref{appendix:12co_mom8_mom9} and~\ref{appendix:c18o_mom8_mom9} for the corresponding \tlvco~and \ceteno~images, respectively). However, it also appears to trace the base of the outflow cavities along the minor axis of the continuum for both sources. In IRS7B, although extended emission is seen surrounding the source, it does not seem to be directly associated with the protostar. As for IRS5N, although extended emissions from the large-scale surroundings are clearly present, there are some emissions that are most likely associated with the protostar. The most notable of these is the spiral feature seen toward the west of the source that is likely tracing infalling material \citep{Sharma_2023}. Both IRS7B and IRS5N are part of the complex Coronet region, which harbors multiple highly energetic molecular hydrogen emission-line objects (MHOs) and dozens of Herbig–Haro (HH) objects that likely interact and conceal any emission from these sources \citep[see][and references therein]{Wang_2004}. Furthermore, any low-velocity emissions close to the $v_{sys}$ of these sources are difficult to distinguish due to line opacity effects and spatial filtering.

\subsection{\thrco}

\begin{figure*}[ht!]
    \includegraphics[width=1.0\textwidth]{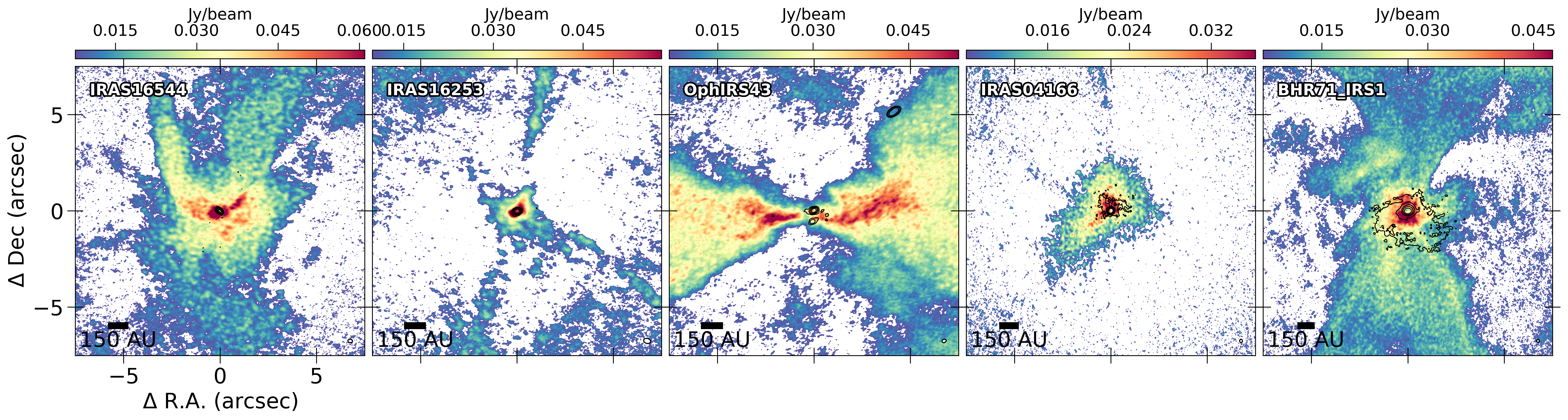}
\caption{Moment 8 maps of \thrco~($J$=2--1) created using $\geq3\sigma$ emissions toward five representative eDisk sources. The contours display the continuum emission at the same levels as Figure~\ref{fig:12co_mom8}. The scale bar is located at the bottom left, and the synthesized beam is indicated in white at the bottom-right corner of each image.\label{fig:13co_mom8}}
\end{figure*}

Emissions in \thrco~(2--1) are also observed toward all eDisk sources. Figure~\ref{fig:13co_mom8} displays the \thrco~moment 8 maps toward five eDisk sources. The emissions primarily trace the inner $\sim$3-4\arcsec area and peak in the disk and the warm inner envelope region near the protostar. Moment 9 maps show clear signs of rotation with the separation of blue- and red-shifted emissions along the major axis of the continuum (see Figure~\ref{appendix:13co_mom8_mom9_zoomed}). The extended large-scale \thrco~emissions trace the outflows in a few sources (e.g., IRAS16544 and BHR71 IRS1; see Figure~\ref{fig:13co_mom8}), and the sections of the outflow cavity walls in others (e.g., IRAS04166 and IRAS16253; see Figure~\ref{fig:13co_mom8}). \thrco~is a well-known tracer of dense regions, as it is sensitive to higher column densities than \tlvco. The extended emissions are likely resulting from the outflows and the cavity walls. The cavity walls have higher densities than those of the surrounding envelope, and their temperatures are expected to be above 20 K, the sublimation temperature of CO \citep{Collings_2004}. In addition, the materials in the cavity walls are exposed to passive heating from the UV radiation of the protostar, which can further elevate temperatures in this region, sublimating more CO from ices and facilitating the production of hydrocarbons (see Sect.~\ref{subsubsection:cavity_walls} and also \citealt{Lee_2014,Lee_2015}).

The \thrco~emission toward IRS5N and IRS7B once again appears to be affected by the surrounding environment, although to a lesser extent than the \tlvco~emission. The zoomed-in moment 9 map of IRS7B shows a distinct region coinciding with the continuum that exhibits a much different velocity profile compared to the surrounding emission (see Figure~\ref{appendix:13co_mom8_mom9_zoomed}). This emission likely originates from the disk and inner envelope of IRS7B. Likewise, the \thrco~emission observed along the north-south direction of IRS5N is closer to the systemic velocity and slightly red-shifted, which is in contrast with the predominantly blue-shifted large-scale emission surrounding the source.

\subsection{\ceteno}

\begin{figure*}[ht!]
    \includegraphics[width=1.0\textwidth]{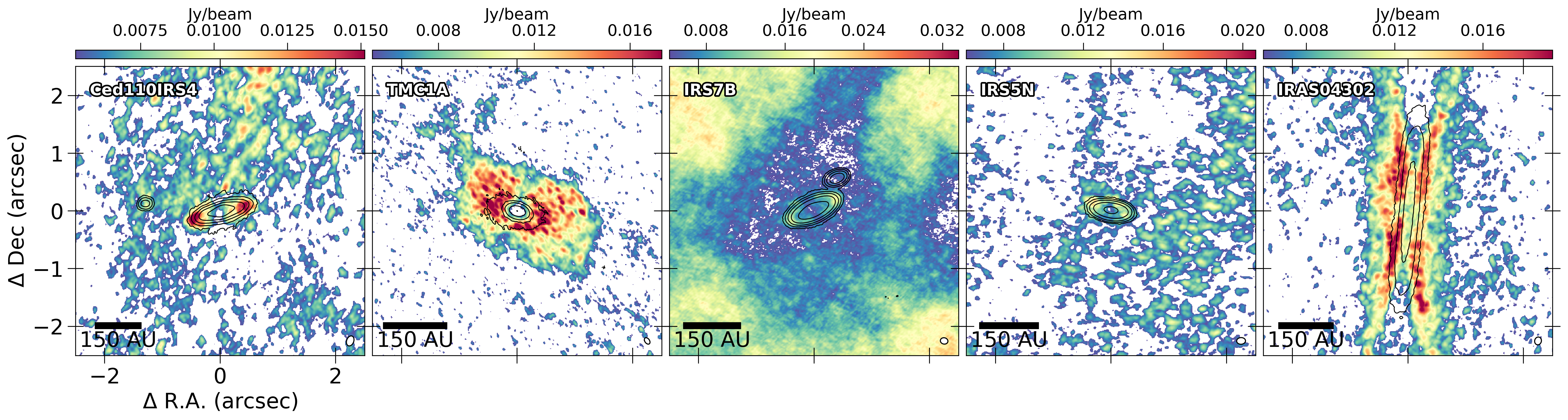}
    \includegraphics[width=1.0\textwidth]{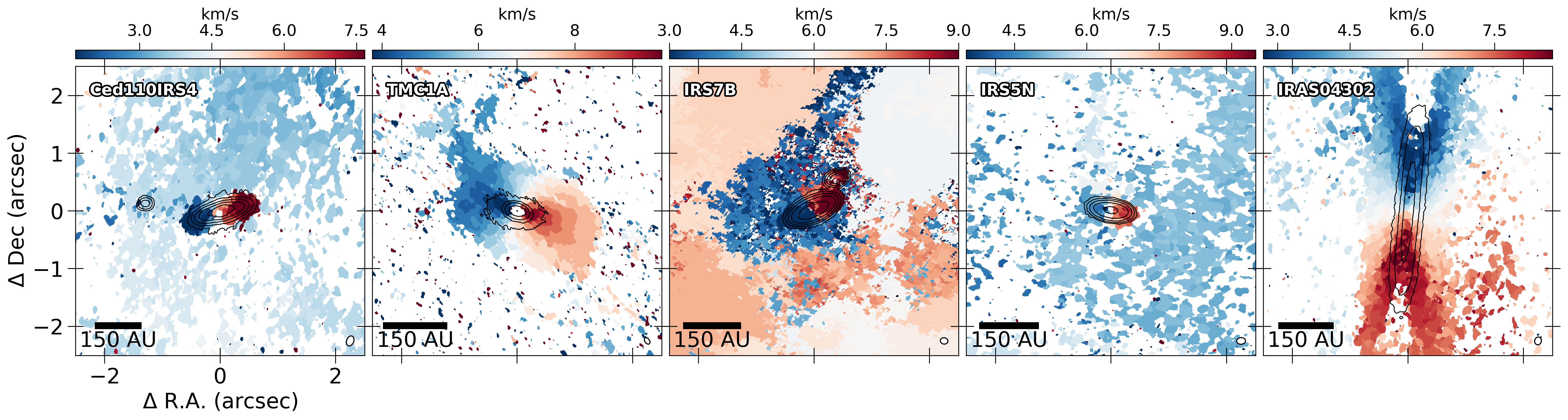}
\caption{Zoomed-in moment 8 maps (\textit{top}) and moment 9 maps (\textit{bottom}) of \ceteno~($J$=2--1) created using $\geq3\sigma$ emissions toward five representative eDisk sources. The contours display the continuum emission at the same levels as Figure~\ref{fig:12co_mom8}. The scale bar is located at the bottom left, and the synthesized beam is indicated in white at the bottom-right corner of each image.\label{fig:c18o_mom8_mom9}}
\end{figure*}

In line with the observations of other CO isotopologues, \ceteno~(2--1) emissions are detected in all sources within the eDisk sample. Figure~\ref{fig:c18o_mom8_mom9} shows the zoomed-in moment 8 maps (top) and moment 9 maps (bottom) of the \ceteno~emissions toward five eDisk sources. The moment 8 maps show that the emissions are mostly concentrated in the inner envelope and the disk region, similar to that of the \thrco~emissions. For most sources, the \ceteno~emission appears to have an absorption profile at the protostar position. This feature likely results from the optically thick and cold foreground clouds absorbing the warmer continuum emissions at low velocities, causing the continuum oversubtraction. This absorption also occurs toward the emissions of \tlvco~and \thrco~but due to their relatively high brightness, is less apparent and important in those molecules. Furthermore, bright and optically thick continuum emissions at small radii can block the relatively weak \ceteno~emissions, resulting in continuum oversubtraction during data reduction. A distinct separation between the blue- and the red-shifted emission can be seen in the moment 9 maps along the major axis of the continuum. PV analyses of the \ceteno~emission conducted as part of the first look results of eDisk sources have observed Keplerian disks in 14 of the 19 eDisk sources \citep{Flores_2023,Hoff_2023,Kido_2023,Lin_2023,Ohashi_2023,Sai_2023,Sharma_2023,Thieme_2023,Yamato_2023,Encalada_2024,Santamaria_2024,Han_2025,Phuong_2025}. 

Large-scale \ceteno~emissions extending over 3--4$\arcsec$ are also observed toward a couple of sources (see Figure~\ref{appendix:c18o_mom8_mom9}). In IRAS15398, IRAS16253, and L1527, the large-scale emission faintly traces an outline of the outflow cavity walls. In IRAS16544 and IRAS04169, the extended \ceteno~emission likely traces an accretion streamer that transports material from the outer envelope to the inner envelope and disk nonaxisymmetrically (see Section~\ref{subsubsection:streamer}), whereas the arc-like structure seen toward the north of Ced110IRS4 is most likely tracing a shocked shell caused by a large-scale outflow in the past \citep{Sai_2023}. As with the \tlvco~and \thrco~emission maps, the large-scale \ceteno~emission seen toward IRS5N and IRS7B most likely originates from the complex environment of the Coronet region (see Figure~\ref{appendix:c18o_mom8_mom9}). However, they also exhibit compact emissions that overlap the continuum with a much different velocity profile than the surrounding emissions and likely trace the disk rotation, as shown in the bottom panels of Figure~\ref{fig:c18o_mom8_mom9}.

\subsection{SiO}

\begin{figure}[ht!]
    \includegraphics[width=0.95\linewidth]{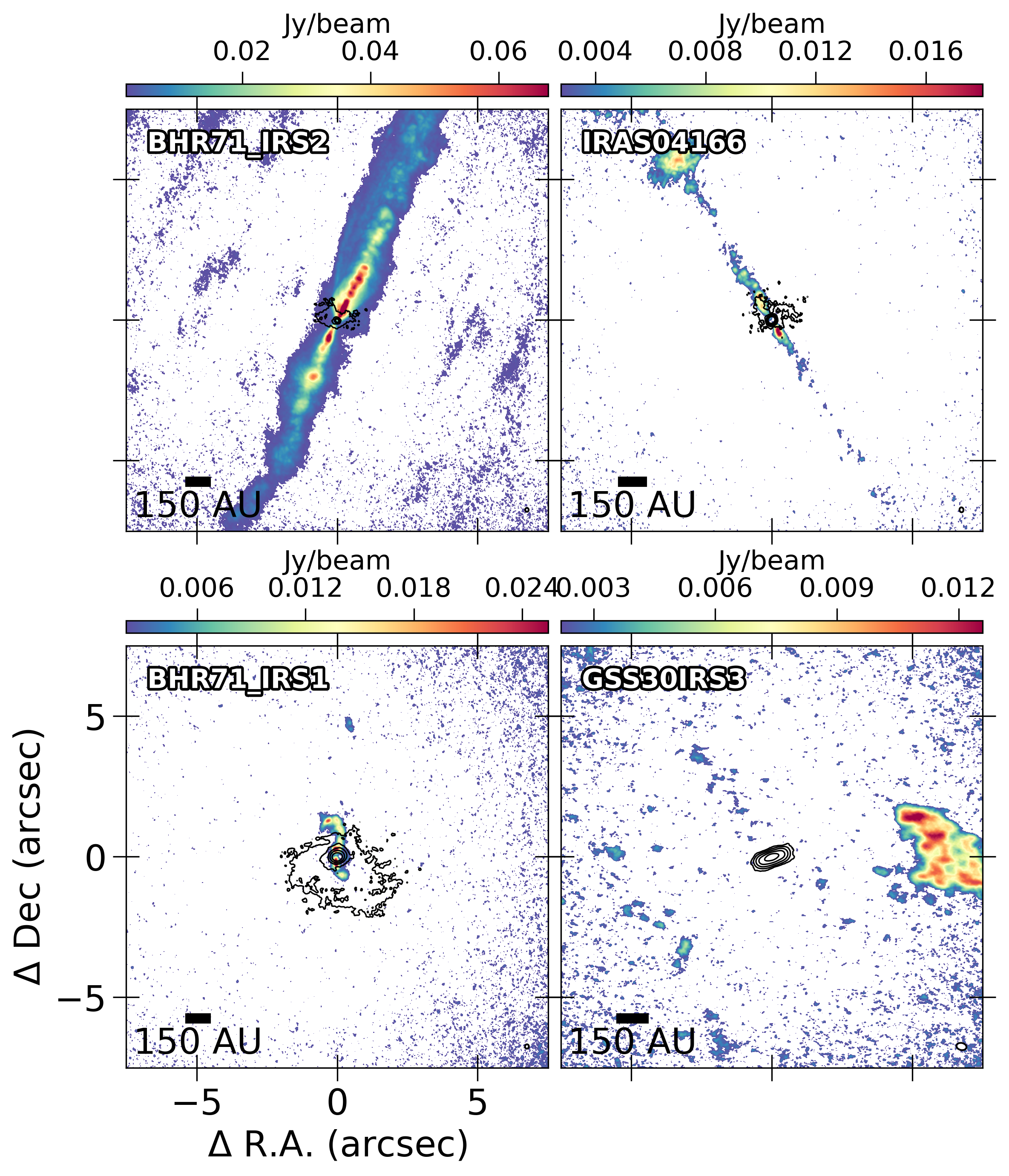}
\caption{Moment 8 maps of SiO~($J$=5--4) created using $\geq3\sigma$ emissions toward four representative eDisk sources. The contours display the continuum emission at the same levels as Figure~\ref{fig:12co_mom8}. The scale bar is located at the bottom left, and the synthesized beam is indicated in white at the bottom-right corner of each image. \label{fig:sio_mom8}}
\end{figure}

SiO (5--4) emission is only clearly detected toward eight eDisk sources, namely BHR71 IRS1/2, IRAS16544, IRAS15398, IRAS04166, GSS30IRS3, L1527, and IRS7B. Figure~\ref{fig:sio_mom8} shows the moment 8 maps of the SiO emission toward four systems. SiO is a well-known tracer of protostellar jets, and, unlike other molecules, the emission from SiO has generally been observed to be much more compact and collimated, containing knots and clumps \citep{Jhan_2022,Takahashi_2024}. The moment 8 maps for BHR71 IRS2 and IRAS 04166 show multiple knots throughout the elongated SiO emission, closely aligned with their outflows. The emissions toward BHR71 IRS1, IRAS16544, and L1527, while much less prominent and only observed near the protostar, also appear in the same direction as the outflow. Isolated clumps of SiO emission are observed $\sim$4$\arcsec$ away from the protostar position toward GSS30IRS3, IRAS15398, and IRS7B in the west, southwest, and southeast, respectively (see Figure~\ref{appendix:sio_mom8_mom9}). The clumps observed in GSS30IRS3 and IRS7B are observed along the direction of the major axis of the continuum, whereas the clump seen in IRAS15398 is oriented along the outflow direction. These clumps may be created by other shocks taking place in these regions. Nevertheless, it should be acknowledged that both sources are situated in complex regions, and some of the SiO emissions detected might be unrelated to the sources.

\subsection{SO}

\begin{figure*}[ht!]
    \includegraphics[width=1.0\textwidth]{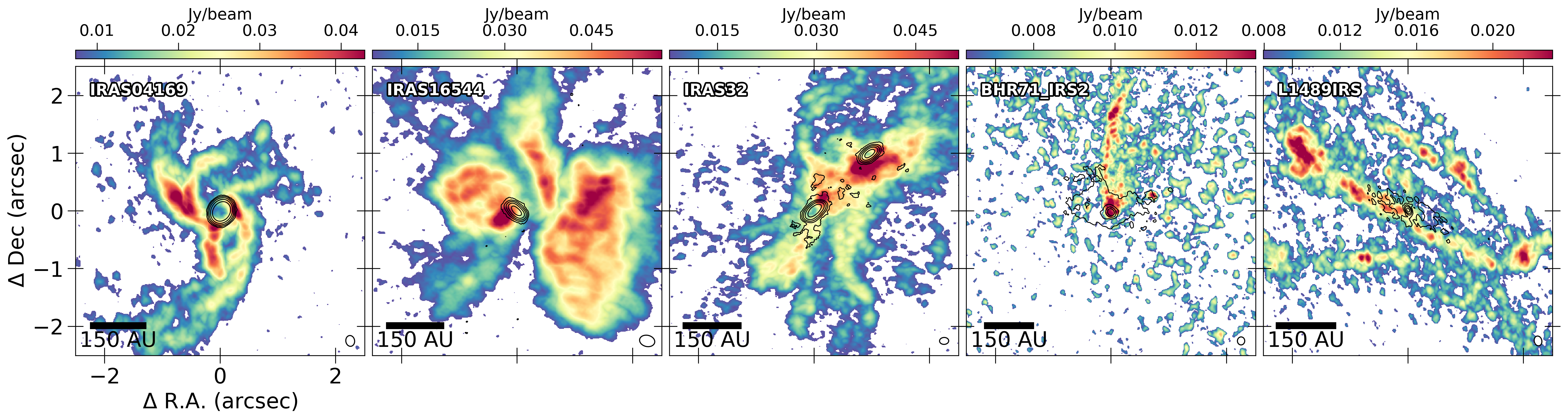}
\caption{Moment 8 maps of SO~($J$=$6_5$--$5_4$) created using $\geq3\sigma$ emissions toward five representative eDisk sources. The contours display the continuum emission at the same levels as Figure~\ref{fig:12co_mom8}. The scale bar is located at the bottom left, and the synthesized beam is indicated in white at the bottom-right corner of each image. \label{fig:so_mom8_zoomed}}
\end{figure*}

Emissions from SO (6$_5$--5$_4$) are detected in all eDisk sources. However, in the case of IRS5N, all of the emission only appears toward the edges of the map and is most likely associated with the surrounding environment and not with the protostar itself. Figure~\ref{fig:so_mom8_zoomed} shows the zoomed-in moment 8 maps of SO (6$_5$--5$_4$) emission observed toward five systems. The emission typically peaks within a few tens of astronomical units from the protostar, though the exact morphology varies among the sources. These emissions appear to trace the inner envelope and disk regions. In IRAS04169 and OphIRS63, the emissions reveal spiral structures that connect to the inner envelope and disk regions. These structures likely represent accretion streamers that funnel material nonaxisymmetrically to the inner envelope and the disk regions \citep{Pineda_2020,Flores_2023,Lee_2023,Lee_2024,Phuong_2025}. The moment 9 maps of the SO emission toward most sources show that near the continuum, the emission shows clear signs of rotation (see Figure~\ref{appendix:so_mom8_mom9}). 

Large-scale emissions are also observed in 8 of the 19 sources. These extended emissions appear to trace different components among the different sources. In BHR71 IRS2 and L1527, the extended SO emissions faintly trace a section of the outflow cavity wall, whereas in IRAS32, extended emissions are seen perpendicular to the outflow direction. A clump of blue-shifted SO emission is also seen southwest of IRAS15398 at the same location as the SiO clump. In IRAS16544 and Ced110IRS4, extended emissions seen toward the north of the protostar weakly trace the streamer and the shocked shell due to outflow, respectively, as observed in \ceteno~emission. The extended emissions observed toward GSS30IRS3 and IRS7B appear to be affected by the environment. However, for these two sources, the emissions near the protostar exhibit distinct velocity profiles in the moment 9 maps, suggesting an association with the protostar (see Figure~\ref{appendix:so_mom8_mom9_zoomed}). 

\subsection{DCN}

\begin{figure}[ht!]
    \includegraphics[width=0.95\linewidth]{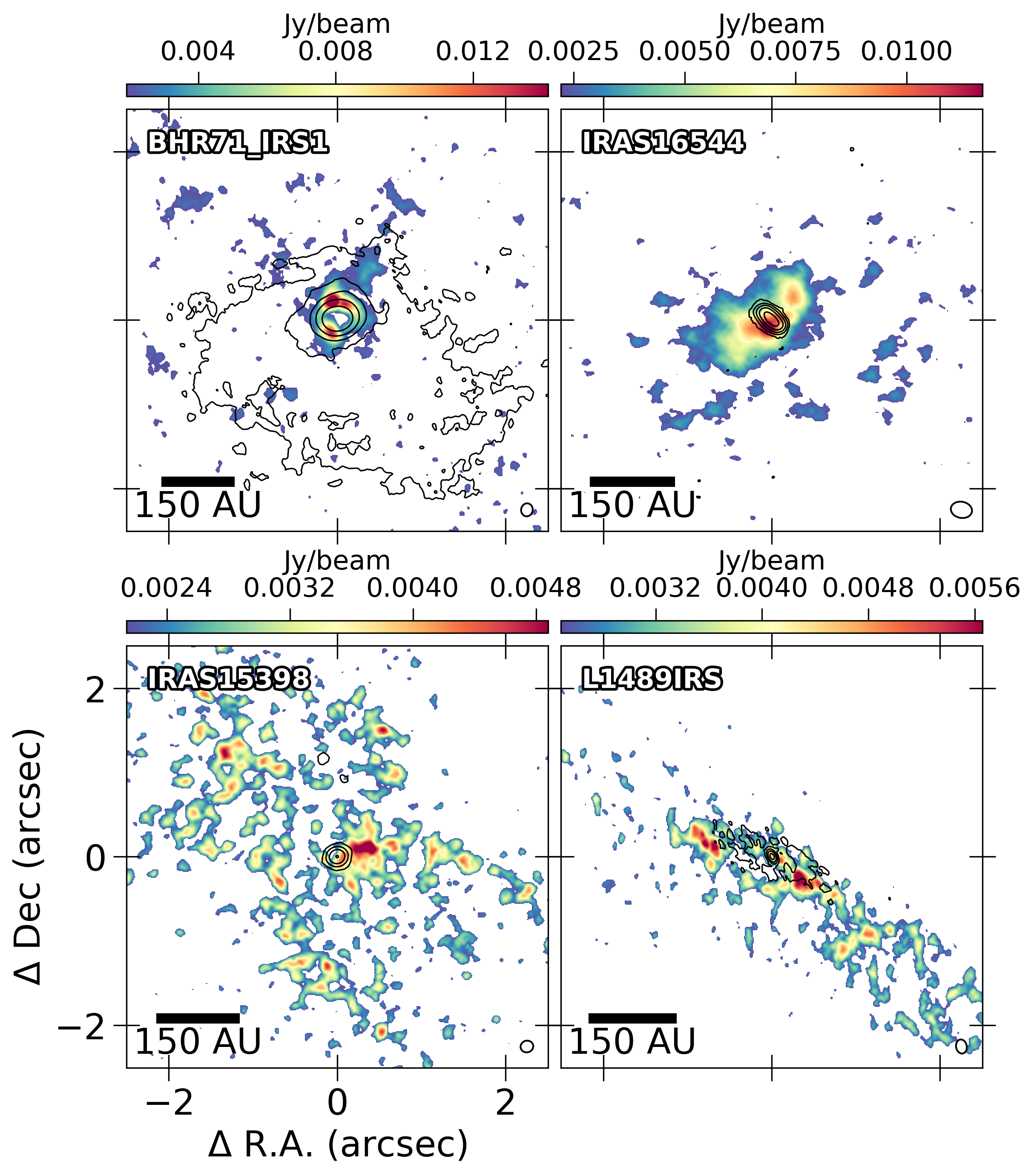}
\caption{Zoomed-in moment 8 maps of DCN~($J$=3--2) created using $\geq3\sigma$ emissions toward four representative eDisk sources. The contours display the continuum emission at the same levels as Figure~\ref{fig:12co_mom8}. The scale bar is located at the bottom left, and the synthesized beam is indicated in white at the bottom-right corner of each image. \label{fig:dcn_mom8_zoomed}}
\end{figure}

Emissions from DCN (3--2) are clearly detected in nine sources, and in all cases, the bulk of the emission appears to originate in the inner warm envelope and disk region of the protostar. Figure \ref{fig:dcn_mom8_zoomed} shows the zoomed-in moment 8 maps of DCN emission observed toward four eDisk sources. Emissions in BHR71 IRS1/2, IRAS16544, and L1489 are concentrated toward the disk region and show signs of rotation in the moment 9 maps (see Figure~\ref{appendix:dcn_mom8_mom9_zoomed}). Notably, the emissions in BHR71 IRS1 and L1489 show ring-shaped structures with an absorption profile in the middle, similar to the ones seen in \ceteno. Extended emission that is clearly associated with the source is only observed toward IRAS15398, which shows an elongated structure extending from the northeast to the southwest, along the direction of the outflow, and has a similar velocity structure as shown by the \thrco~emission (see Figure~\ref{appendix:dcn_mom8_mom9}).

DCN emission is also detected toward L1527, IRAS04166, IRAS32, and IRAS16253 (see Figure~\ref{appendix:dcn_mom8_mom9}). However, these detections lack clear association with specific source structures. In IRAS32, knotty emission is detected between the two continuum sources of the binary, accompanied by two sets of elongated emissions. One emission is slightly blue-shifted in the north-south direction, while the other one is slightly red-shifted in the northeast-southwest direction. Patches of emission are seen toward the south and the northwest of the continuum in IRAS16253, while tentative and relatively weak emissions are also seen near the continuum positions of L1527 and IRAS04166. These sources are also affected by strong absorption at the position of their continuum. In addition to the nine sources mentioned above, DCN emission is also seen in IRS7B. Although the emissions toward IRS7B again appear to be affected by the surrounding environment, as seen in observations of other molecules, a distinct patch of highly blue-shifted emission is present toward the southeast, near the edge of the major axis of the continuum. This distinct velocity structure of this patch compared to surrounding emissions suggests a separate origin, likely associated with IRS7B itself, similar to the emissions in \ceteno (see Figures~\ref{fig:c18o_mom8_mom9},\ref{appendix:dcn_mom8_mom9_zoomed}).

\subsection{\ch3oh}\label{results:ch3oh}

\begin{figure}[ht!]
    \includegraphics[width=0.95\linewidth]{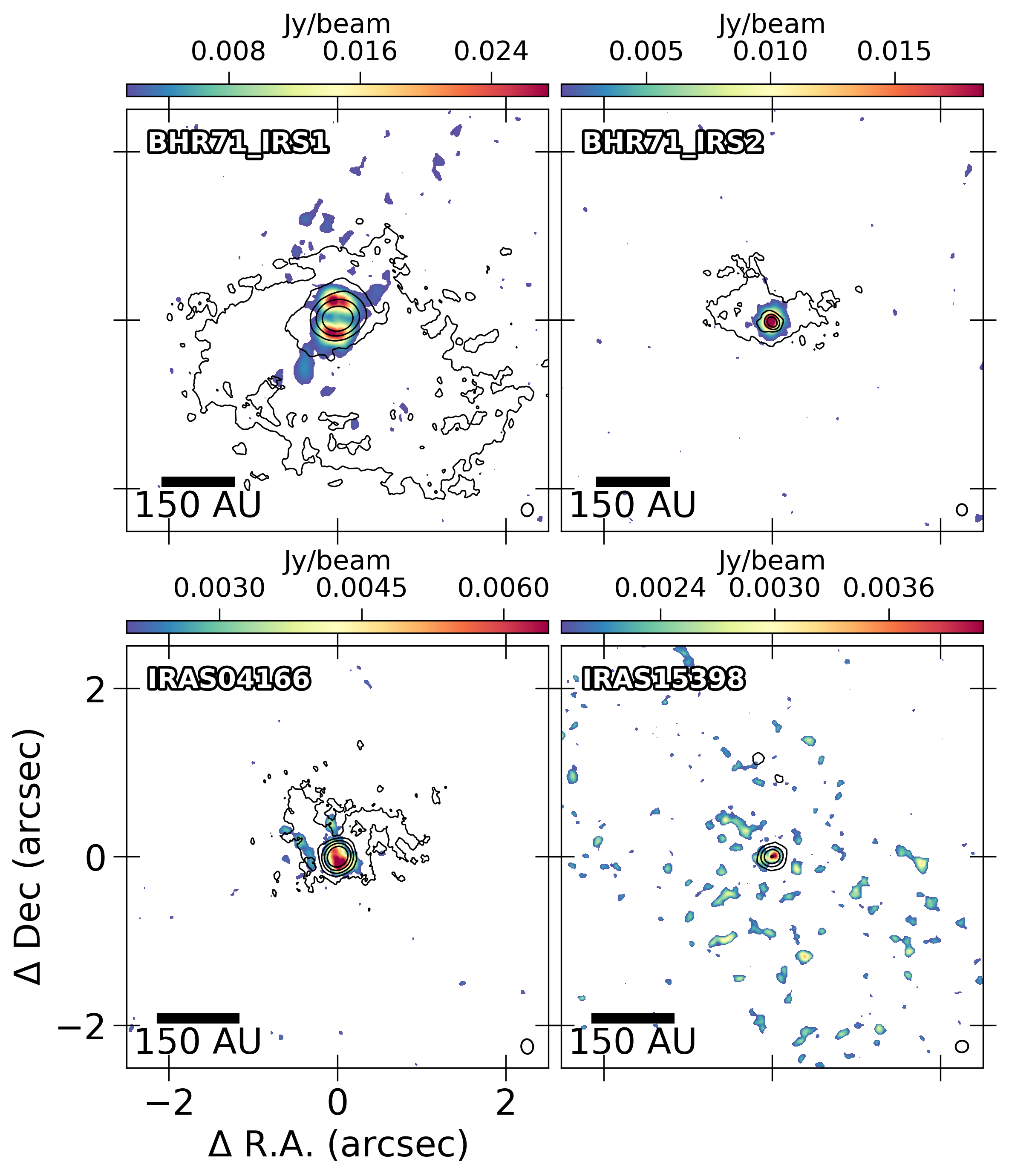}
\caption{Zoomed-in moment 8 maps of \ch3oh~($J$=$4_2$--$3_1$) created using $\geq3\sigma$ emissions toward four representative eDisk sources. The contours display the continuum emissions at the same levels as Figure~\ref{fig:12co_mom8}. The scale bar is located at the bottom left, and the synthesized beam is indicated in white at the bottom-right corner of each image. \label{fig:ch3oh_mom8_zoomed}}
\end{figure}

Emissions in \ch3oh (4$_2$--3$_1$) are clearly detected toward seven sources. Figure~\ref{fig:ch3oh_mom8_zoomed} shows the zoomed-in moment 8 maps illustrating \ch3oh emissions toward four eDisk sources. Toward BHR71 IRS 1/2, IRAS16544, and IRAS04166, these emissions are mostly compact and concentrated in the inner envelope and disk region, with their peaks coinciding with the positions of the corresponding protostars. In contrast, the emission toward IRAS15398 peaks near the position of the protostar, but additional patchy emissions are also seen in the direction of the outflow.

In addition, patches of \ch3oh emission can be seen in the large-scale maps of three sources (see Figure~\ref{appendix:ch3oh_mom8_mom9}). Toward IRAS15398 and GSS30IRS3, a knot-like feature can be seen in the northeast direction, which aligns with the direction of their respective outflow. IRAS04166 also exhibits a similar but slightly more elongated emission in the direction of its outflow. These knots might suggest previous outbursts in these sources \citep{Kim_2024,Lee_2024}. A compact, likely unresolved blue-shifted patch of emission is observed near the primary source of IRAS32, potentially indicating molecular jets that have recently emerged from the source (see Figure~\ref{appendix:ch3oh_mom8_mom9_zoomed}; but for a better view, consult the web application). The IRAS15398 map again reveals a clump of \ch3oh~emission in the large-scale map at the same location as the SiO and SO emissions.

\subsection{\h2co} \label{results:h2co}

\begin{figure*}[ht!]
    \includegraphics[width=1\linewidth]{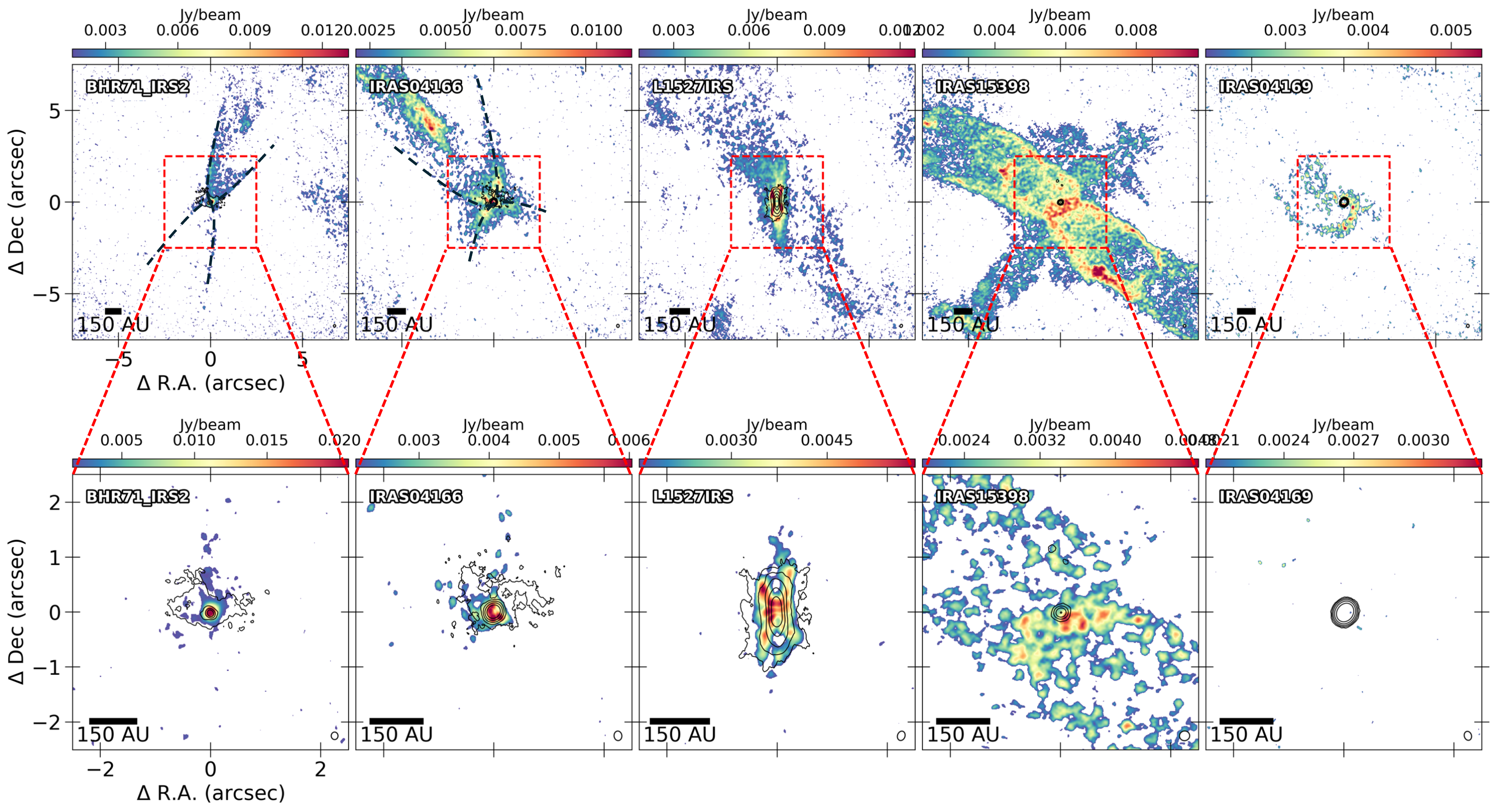}
\caption{Moment 8 maps of \h2co emissions in \htwocolow~(\textit{top}) and \htwocohigh~(\textit{bottom}) created using $\geq3\sigma$ emissions toward five of the eDisk sources. Dashed lines show the outflow cavity walls seen toward the \htwocolow~emissions. The contours display the continuum emission at the same levels as Figure~\ref{fig:12co_mom8}. The scale bar is located at the bottom left, and the synthesized beam is indicated in white at the bottom-right corner of each image. \label{fig:h2co_mom8_zoomed_combined}}
\end{figure*}

The spectral windows of the eDisk observations also cover three transitions of \h2co: \htwocolow, \htwocomid, and \htwocohigh. Figure~\ref{fig:h2co_mom8_zoomed_combined} displays the zoomed-in moment 8 maps of the \h2co emissions for the \htwocolow~transition with $E_{\mathrm{up}}$ = 21 K and the \htwocohigh~transition $E_{\mathrm{up}}$ = 68 K toward five eDisk sources. The plots show that, in most cases, the different transitions exhibit similar morphologies for a given source, with stronger and more extended emissions seen in the lower-energy transition. Emissions in \h2co are observed in all 19 eDisk sources for the lower-energy \htwocolow~transition, with emissions mostly concentrating near the position of the protostar. These emissions likely trace the inner envelope and disk regions and, in some cases, also trace the base of the outflow cavity walls, forming a U-shaped morphology on either side of the minor axis of the continuum. In the large-scale map of a few sources, such as BHR71 IRS2, IRAS15398, and IRAS04166, clumpy emissions are also seen in the direction of the outflow, which is suggestive of outflow jets and can be released in shocks \citep{Tychoniec_2019}. 

Additionally, the moment maps of the \h2co emission also trace the inner envelope and disks and accretion streamers toward some sources (see Figure~\ref{appendix:h2co_3_03-2_02_mom8_mom9}). In L1489, L1527, IRAS16253, and OphIRS43, the emission is elongated along the major axis of the continuum. In IRAS04169, the emission traces the spiral structure around the source, while in Ced110IRS4, the emission traces the extended structure seen toward the north of the source. These components have also been observed in \ceteno~and SO molecules and are likely trace accretion streamers in IRAS04169 and outflow shocked shells in Ced110IRS4. In IRAS32, the \h2co emission is observed mostly in the mutual envelope material between the two binary sources. The emission toward OphIRS43 peaks toward the west of the source and is extended in the east-west direction. In GSS30IRS3, an elongated band of \h2co emission in the north-south direction is seen toward the northeast of the source. IRS7B once again has two small patches of \h2co emission with vastly different velocity structures with respect to the surrounding emission (see Figure~\ref{appendix:h2co_3_03-2_02_mom8_mom9_zoomed}).

\subsection{\C3h2}
\begin{figure*}[ht!]
    \includegraphics[width=1\linewidth]{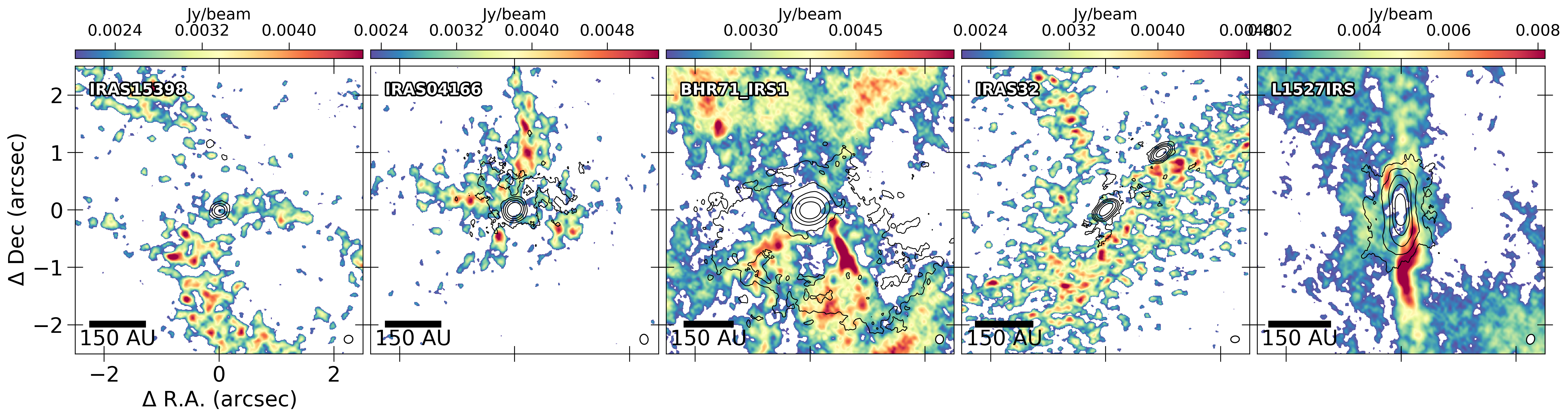}
\caption{Zoomed-in moment 8 maps of blended \C3h2~$J$=6$_{0,6}$--5$_{1,5}$ and $J$=6$_{1,6}$--5$_{0,5}$ transitions created using $\geq3\sigma$ emissions toward five representative eDisk sources. The contours display the continuum emissions at the same levels as Figure~\ref{fig:12co_mom8}. The scale bar is located at the bottom left, and the synthesized beam is indicated in white at the bottom-right corner of each image. \label{fig:c3h2_mom8_zoomed}}
\end{figure*}

The spectral windows of the eDisk observations also cover four transitions of \C3h2 emissions: 6$_{0,6}$--5$_{1,5}$, 6$_{1,6}$--5$_{0,5}$, 5$_{1,4}$--4$_{2,3}$, and 5$_{2,4}$--4$_{1,3}$. Emissions in \C3h2 are detected toward twelve eDisk sources. Figure~\ref{fig:c3h2_mom8_zoomed} shows the moment 8 maps made with the blended 6$_{0,6}$--5$_{1,5}$ and 6$_{1,6}$--5$_{0,5}$ transitions toward four eDisk sources. The emissions of \C3h2 appear to primarily trace the outflow cavity walls near the position of the protostar in many sources, such as BHR71 IRS1/2, IRAS15398, IRAS04166, IRAS16253, and IRAS16544. Emissions are also seen toward the envelope in L1527, L1489, IRAS32, IRAS16253, and IRAS16544. These emissions generally have constant velocities that are close to the systemic velocity ($v_{sys}$) of the source (see Figure~\ref{appendix:c3h2_217.82_mom8_mom9_zoomed}). Likewise, the emission toward Ced110IRS4 traces the outflow-shocked shell that is also seen in other molecules, and IRS7B once again shows large-scale emissions from the surroundings.

The emissions from the remaining two transitions also trace similar structures to the two blended lines toward the different sources. One notable exception to this is the source BHR71 IRS2, where each transition displays a different feature. Emissions are seen tracing the cavity walls in the blended lines, no emission is seen in the 5$_{1,4}$--4$_{2,3}$ transition, and a patch of highly red-shifted emission is seen right at the center in the 5$_{2,4}$--4$_{1,3}$ line (see Figures~\ref{appendix:c3h2_217.82_mom8_mom9_zoomed}, ~\ref{appendix:c3h2_217.94_mom8_mom9_zoomed}, and ~\ref{appendix:c3h2_218.16_mom8_mom9_zoomed}; but for a better view, consult the web application).

\section{Discussion}\label{sec:discussion}

\begin{table*}[]
\centering
\caption{Summary of molecules observed toward the various components.} \label{tab:molecular_components}
\begin{tabular}{cccccccccc}
\hline\hline
Components & \tlvco & \thrco & \ceteno & SiO & SO & DCN & \ch3oh & \h2co & \C3h2 \\
 \hline
Outflows & \checkmark & \checkmark &  &  &  &  & \checkmark & \checkmark &  \\
High velocity jets & \checkmark &  &  & \checkmark &  &  &  &  &  \\
Outflow cavity walls & \checkmark & \checkmark &  &  &  &  &  & \checkmark & \checkmark \\
Inner envelope and disk &  & \checkmark & \checkmark &  & \checkmark & \checkmark & \checkmark &  & \checkmark  \\
Accretion streamers &  &  & \checkmark &  & \checkmark &  &  & \checkmark & \\
 \hline
\end{tabular}
\end{table*}

\subsection{Chemical morphology of the embedded sources}
The high resolution and sensitivity of the eDisk observations provide a unique opportunity to study the spatial distribution of the molecules in the inner envelope and disk regions of embedded sources. Table~\ref{tab:molecular_components} summarizes the principal molecular lines detected toward the various morphological components identified in this study. In this section, we only discuss these primary molecules targeted by the spectral window of the eDisk observations. For the sources, BHR71 IRS1/2, as well as IRAS16544, line emissions of several other COMs are seen beyond the \ch3oh emissions presented in Sect.~\ref{results:ch3oh}. The analysis of these more complex species observed lies beyond the scope of this paper and will be presented in a forthcoming study.

\subsubsection{Outflowing material}
Outflows and jets play a crucial role in transferring the excess angular momentum back to the molecular cloud and are essential feedback processes in protostellar systems \citep{Offner_2014}. Outflows generally consist of low-velocity components, with gases moving at velocities up to 20 km s$^{-1}$ relative to the $v_{sys}$ of the source. The threshold of 20 km s$^{-1}$ has been selected based on the channel maps of the molecular emissions, and similar values have been adapted in previous studies \citep[e.g.][]{Arce_2007,Frank_2014}. In eDisk observations, they are commonly traced by low-velocity channels of \tlvco~and \thrco~emissions.
All eDisk sources display a certain extent of \tlvco~emission in the direction of the outflow (perpendicular to the major axis of the continuum). 15 of the 19 sources clearly trace large-scale bipolar outflows in \tlvco~emission, whereas the distinction is not as clear in the remaining sources (see Figure~\ref{appendix:12co_mom8_mom9}). The low-velocity \tlvco~ emission typically displays a parabolic shape with a wide-angle morphology. Our observations show that such outflows originate within the inner disk region, very close to the position of the protostar. Similar structures are also observed in the \thrco~emission toward some eDisk sources (see Figure~\ref{appendix:13co_mom8_mom9}). The \thrco~emission, however, is associated with denser regions of a protostar, as it is much less abundant than \tlvco~with the abundance ratio of \tlvco/\thrco~$\sim$77 \citep{Wilson_1994}. 

\begin{figure}[ht!]
    \includegraphics[width=0.95\linewidth]{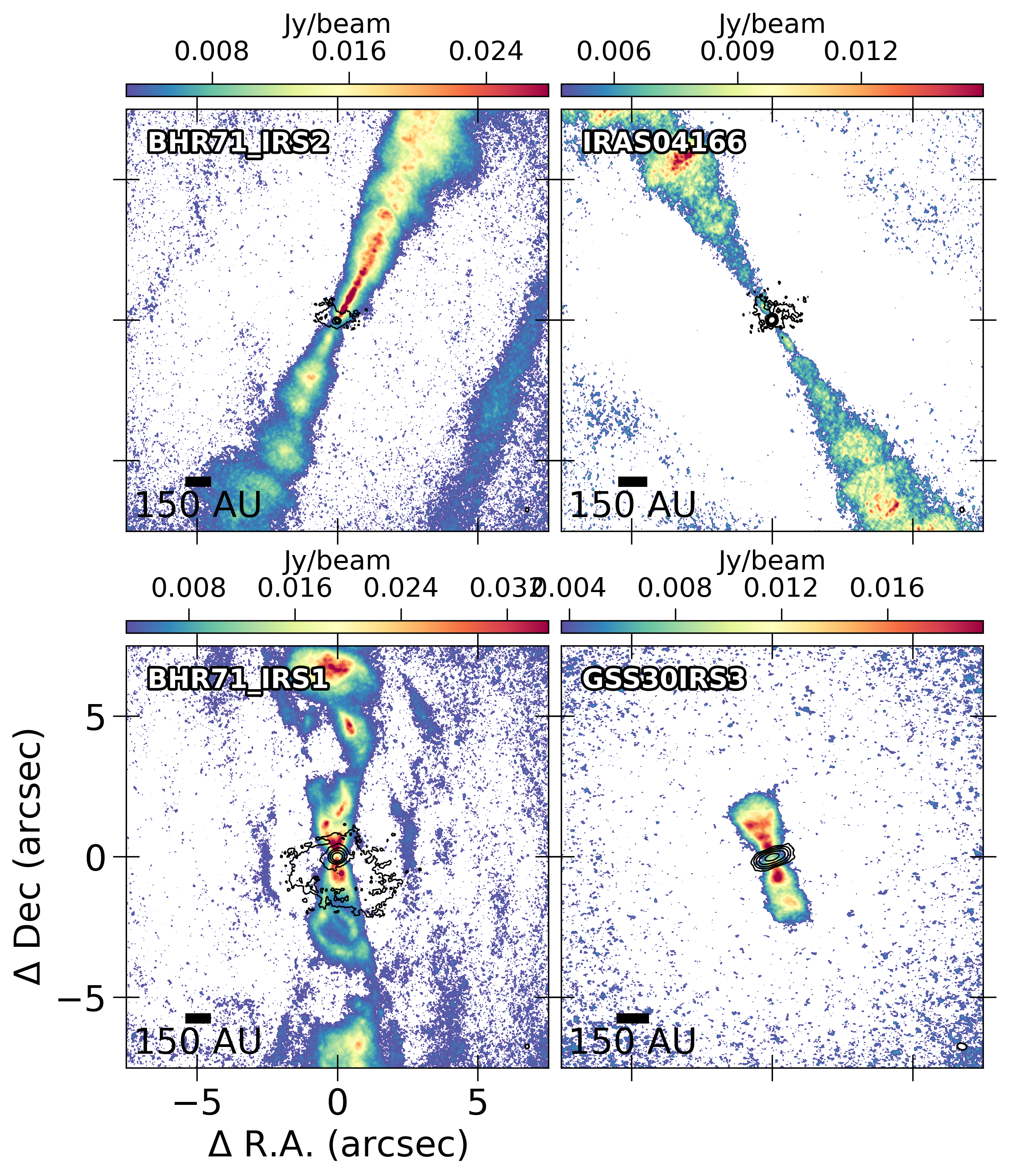}
\caption{Moment 8 maps of high-velocity jets seen in \tlvco~emission toward five of the eDisk sources. The maps are created using the velocity channels where $|v_{sys} - v_{chan}| > 20$ km s$^{-1}$. These are the same sources used for the SiO images in Figure~\ref{fig:sio_mom8}, allowing for a comparison with the SiO emissions. The contours display the continuum emission at the same levels as Figure~\ref{fig:12co_mom8}. The scale bar is located at the bottom left, and the synthesized beam is indicated in white at the bottom-right corner of each image. \label{fig:12co_high_vel_mom8}}
\end{figure}

In contrast, protostellar jets consist of high-velocity components, with gases moving at speeds greater than 20 km s$^{-1}$ relative to $v_{sys}$ of the source and can reach upwards of 100 km s$^{-1}$ \citep[e.g.,][]{Bachiller_1996,Arce_2007}. These jets are often seen in high-velocity channels of \tlvco~and SiO emissions. Figure~\ref{fig:12co_high_vel_mom8} shows the moment 8 maps of high-velocity \tlvco~emission seen toward four eDisk sources. Compared to low-velocity outflows with $|v_{sys} - v_{chan}| < 20$ km s$^{-1}$, these jets are much more collimated, have narrower opening angles, and originate much closer to the protostar. Furthermore, they typically show clumpy emission toward very young protostars, which is believed to be indicative of past episodic accretion \citep{Plunkett_2015,Vorobyov_2018,Sharma_2020}. Except in GSS30IRS3, all other sources that display high-velocity \tlvco~emission also display corresponding jets in SiO. These SiO emissions also exhibit similar structures to their high-velocity \tlvco~counterparts. In contrast, \tlvco~emissions in L1527 do not show any high-velocity components, consistent with the CARMA observations of \tlvco~(1--0) reported by \citealt{Rivera_2021}, and the only indication of a jet is a patch of mostly unresolved high-velocity SiO emission close to the protostar \citep{Hoff_2023}. 

The absence of SiO emission in GSS30IRS3 as a complement to its high-velocity \tlvco~emission could be attributed to the lack of sensitivity of the observation to detect the SiO present in the jets or to the jets themselves not being energetic enough to generate shocks to release SiO into gas, as SiO is a well-known shock tracer \citep{Pintado_1992,Bergin_1998}. Although SiO emissions in IRAS16544 align well with the outflow direction, their velocities are close to $v_{sys}$ of the source, suggesting that these emissions may result from other shock processes such as a shock created by the interaction of the infalling streamer depositing material onto the disk \citep{Garufi_2022,Kido_2023} rather than from shocks due to high-velocity jets. The general lack of SiO emission in the outflow toward most eDisk sources is in contrast to the findings of \citet{Tychoniec_2021} where jets in SiO were seen toward six of the seven Class 0 sources observed. The Class 0 targets of \citet{Tychoniec_2021} are much more luminous, averaging $\sim$ 19 $L_{\odot}$. This could reflect that these sources are potentially more actively accreting and driving more energetic outflows than those targeted as part of eDisk. However, both samples are still relatively small for any more conclusive statements about this.

As mentioned in Sect.\ref{sec:results}, sporadic emissions of other molecules such as \ch3oh~and \h2co~are also seen toward the outflow direction in some sources. These emissions are consistent with the release of molecules from dust grain surfaces via sputtering due to slow shocks, as observed by \citet{Tychoniec_2021}. These slow shocks can raise the temperatures in these regions and thermally desorb the frozen molecules off the dust grains. We discuss these molecules in greater detail in Sect.\ref{subsubsection:envelope} and Sect.\ref{discussion:h2co}.

\subsubsection{Outflows cavity walls}\label{subsubsection:cavity_walls}
The boundary where the infalling envelope meets the cavity opened up by the outflows and jets is referred to as the outflow cavity wall. These walls typically exhibit increased chemical complexity due to the interaction between the infalling envelope and the outflows, and also due to their exposure to UV radiation from the protostar \citep{Arce_2006,Visser_2012,Murillo_2018,Tychoniec_2021}. Emissions of \tlvco, \thrco, and \C3h2 are commonly seen toward the outflow cavity walls in the eDisk sources. A few sources also show emissions in \h2co and SO in the cavity walls. We discuss these two molecules separately in Sect.~\ref{subsection:h2co_so}.

In addition to tracing the outflows, both \tlvco~and \thrco~emissions delineate the parabolic shape of the outflow cavity walls. The emissions seen toward the cavity walls generally have velocities close to the $v_{sys}$ of the source and likely originate from the sublimation of these molecules from the dust grains caused by the heating of the material by UV radiation and high-velocity jets interacting with the infalling envelope. For the protostars IRAS04166, IRAS15398, IRAS16253, and B335, most of the \thrco~emission in the direction of the outflow is concentrated primarily in the cavity walls. This indicates that the cavity walls are typically denser than both the outflow and the surrounding envelope.

Hydrocarbons such as \C3h2 are well-known tracers of photodissociation regions (PDRs), which are dominated by UV radiation \citep[e.g.,][]{vanderWiel_2009,Guzman_2015}. Exposure of the cavity walls to UV radiation from the source creates the PDR-like environment necessary for the release of atomic carbon, enabling the production of \C3h2. The emission in \C3h2 is concentrated in the cavity walls close to the protostar in many sources (see Figures~\ref{appendix:c3h2_217.82_mom8_mom9_zoomed},\ref{appendix:c3h2_217.94_mom8_mom9_zoomed},\ref{appendix:c3h2_218.16_mom8_mom9_zoomed}). This is anticipated, as the walls nearest to the protostar are exposed to more intense UV radiation from the protostar. Over time, this concentration of \C3h2 in the outflow cavity shifts toward the inner envelope and disk as more of the envelope is accreted, allowing UV radiation to penetrate further \citep{Drozdovskaya_2015}. This can likely explain some of the \C3h2 emissions observed toward the envelope regions of some of the protostars.

\cite{Tychoniec_2021} reported detections of \C3h2 emissions in 4 of their 16 sources. In three of these cases, the \C3h2 emission traces the walls of the UV-irradiated outflow cavities. For TMC1, however, the emission is observed along the extended envelope region, possibly reflecting the case where UV radiation penetrates further as the envelope dissipates \citep{Drozdovskaya_2015}. Our observations of \C3h2 emissions are consistent with these findings, further supporting that hydrocarbons such as \C3h2 are enhanced in PDR-like regions created by UV radiation from the protostar.

\subsubsection{Inner envelope and disk}\label{subsubsection:envelope}
The inner envelope consists of regions within several hundred au of the protostar and is characterized by higher temperatures ($T \gtrsim$ 30 K) and densities ($n \gtrsim 10^7$ cm$^{-3}$) compared to the outer envelope \citep{Jorgensen_2002}. This leads to a rich chemistry in this region, including the formation of COMs \cite[see reviews by][]{Jorgensen_2020,Oberg_2023}. Infalling material from the inner envelope to the protostar funnels onto the circumstellar disk, which not only regulates the material accreted onto the protostar but also serves as a site for planet formation. Among the molecules targeted in the eDisk observations, emissions of \thrco, \ceteno, DCN, and \ch3oh are commonly seen toward the inner envelope and disk regions of the sources.

The CO isotopologues \thrco~and \ceteno~are well-known tracers of the gas distribution and kinematics in the inner envelope and disk region of protostars. This is because CO is ubiquitous in regions where temperatures rise above 20 K, and due to its low critical densities of $\sim$10$^4$ cm$^{-3}$, it can be readily excited even in the low-density conditions typical of outer envelope regions. As a result, CO emissions do not require active chemistry to be observed at spatial scales probed by the eDisk observations. Moment 9 maps from both these molecules at small and intermediate scales ($\lesssim$ 500 au) display a velocity gradient along the major axis of the continuum, consistent with the rotation of the disk or infalling envelope (see Figures~\ref{appendix:13co_mom8_mom9_zoomed}, \ref{appendix:c18o_mom8_mom9_zoomed}). \citet{Hoff_2018} found that emissions from \thrco~and \ceteno~in the inner envelope and disk regions of embedded sources are mostly optically thick. This suggests that temperatures in these regions are at least high enough to cause the CO molecules to sublimate ($T \geq 20$ K).  

Unlike \thrco~and \ceteno, the DCN emissions observed toward the eDisk sources are much more compact and generally peak near the central protostar (see Figure~\ref{appendix:dcn_mom8_mom9_zoomed}). DCN is produced by two main reaction pathways that are active in protostellar disks. At temperatures below 30 K, it forms via D-atom transfer with H$_2$D$^+$, and at temperatures above 30 K, it forms through reactions with deuterated hydrocarbons such as C$_2$HD$^+$ \citep{Millar_1989,Turner_2001,Willacy_2007, Aikawa_2018}. The low-temperature pathway is primarily active in the outer disk midplane, just inside the CO snowline, while the warmer formation pathway is prevalent at disk surface layers where the temperatures are elevated. The observation of DCN emissions in the inner disk region, very close to the protostar toward BHR71 IRS1/2, IRAS16544, IRAS04166, and IRAS15398, suggests that the warmer temperature pathway predominantly drives DCN production in these sources. Alternatively, these emissions could also arise from DCN molecules that sublimate off dust grains in envelope regions that are close to the protostar, where temperatures are high. The sublimation temperature of DCN should be similar to that of HCN ($\sim$80 K, \citealt{Bergner_2022}). Consequently, DCN formed in the gas phase can freeze out onto the dust grains in regions of the envelope where temperatures are lower. The extended emissions seen toward some sources likely trace the DCN produced through the lower temperature route.  

An interesting result obtained in our observations is that, except for L1489 and maybe toward the disk of IRS7B, DCN emissions are observed only toward the Class 0 sources. This non-detection toward Class I sources can simply be attributed to the lack of sensitivity in our observations of the less abundant DCN emissions toward the Class I sources. This result would be a natural consequence of the alternative pathway for the DCN emission mentioned above. It is likely that DCN shares similar chemistry in the envelopes of both Class 0 and Class I sources; however, since the masses of the envelopes of the Class I sources are smaller, detecting them becomes more challenging. Alternatively, as accretion proceeds, DCN gets broken down readily by reactions with H$^+$ and H$_3$$^+$ \citep{Albertsson_2013}, leading to a reduction in the amount of DCN detectable at the provided sensitivity. In either case, DCN might be a more reliable tracer of the earlier, younger stages of protostellar evolution.

Similar to DCN, emissions from \ch3oh~are also only observed toward Class 0 sources. These emissions are also very compact and generally peak near the position of the central protostar, likely tracing the hot core region, as also inferred from the lower resolution ($\sim$100 au) data by \citet{Tychoniec_2021}. The emitting regions of \ch3oh~are contained within the angular extent of DCN, suggesting that they trace higher temperature regions than the DCN emissions (see Figure~\ref{appendix:ch3oh_mom8_mom9_zoomed}). \ch3oh~primarily forms on grain surfaces through successive hydrogenation of CO-rich ices \citep{Hiraoka_1994,Watanabe_2002,Fuchs_2009,Simons_2020} or surface chain reaction in H$_2$O-rich ices \citep{Bergner_2017,Qasim_2018} and sublimate at high temperatures \citep[$\sim$100 K;][]{Brown_2007,Kristensen_2010,Penteado_2017}. The presence of \ch3oh~in the inner envelope and disk regions of BHR71 IRS1/2, IRAS16544, and likely IRAS04166 indicates that these sources are ``hot corinos'', where the temperatures exceed 100 K in the innermost ($\leq 100$ au) dense regions around protostars \citep{Ceccarelli_2004,Ceccarelli_2007}. These hot corinos are regions rich in COMs, suggesting that the production of complex molecules is well underway in the early stages of star formation.

\subsubsection{Accretion streamers} \label{subsubsection:streamer}

\begin{figure*}[ht!]
    \includegraphics[width=1.0\linewidth]{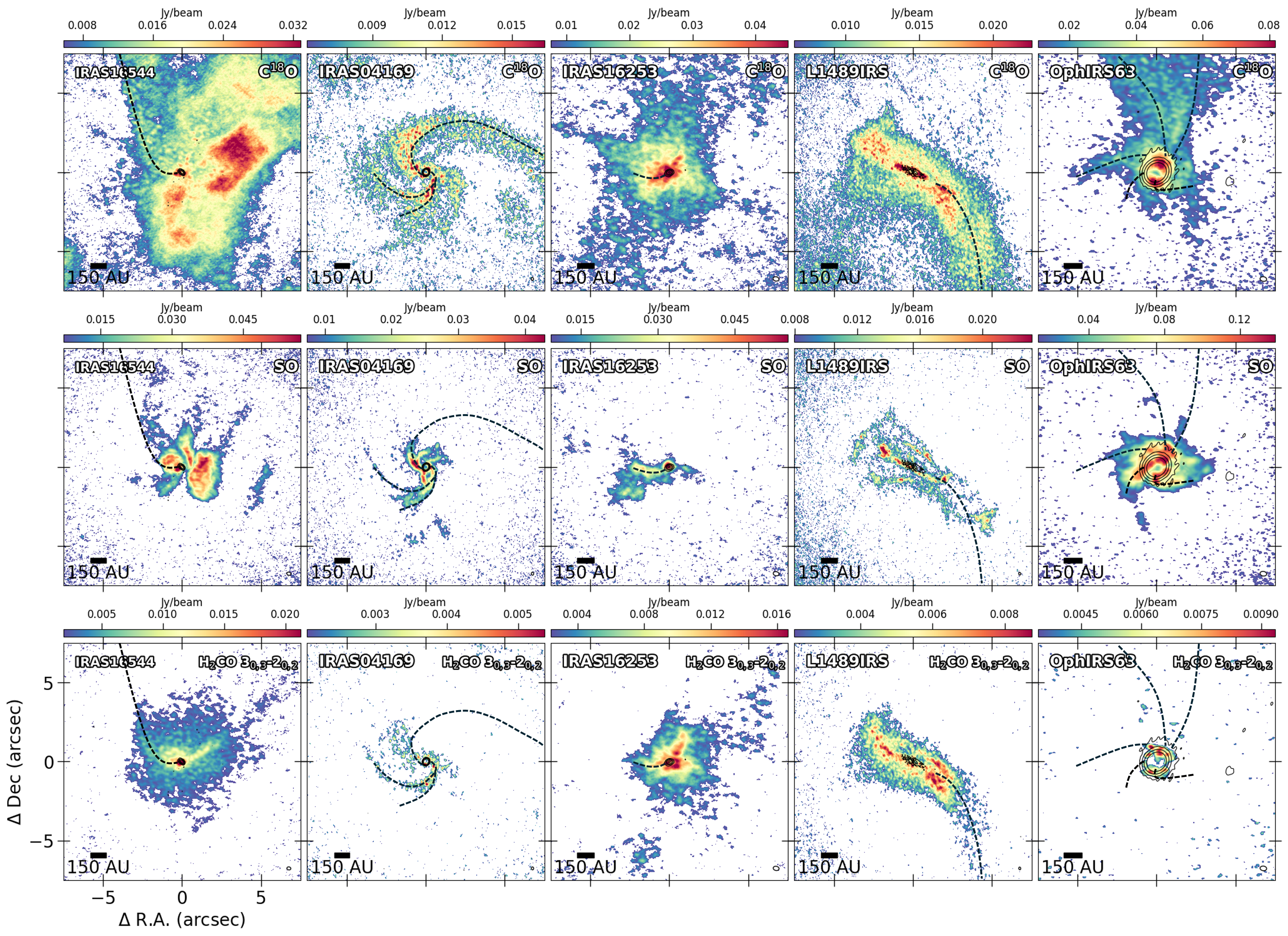}
\caption{Moment 8 maps depicting the streamers observed toward the eDisk sources. The streamers seen toward each source are traced by the dashed lines. The contours display the continuum emission at the same levels as Figure~\ref{fig:12co_mom8}. The scale bar is located at the bottom left, and the synthesized beam is indicated in white at the bottom-right corner of each image. \label{fig:streamers}}
\end{figure*}

The recent increase in the detection of streamers toward YSOs suggests that these structures likely play a crucial role in the star and planet formation process \citep[see][for an overview]{Pineda_2023}. Accretion streamers channel material from the broader surrounding environment, such as filaments and fibers, directly onto the protostar's inner disk-forming region \citep[e.g.][]{Pineda_2020,Valdivia-Mena_2024}. They serve as a critical means to replenish the mass of the system, often having infall rates that exceed the accretion rates of protostars \citep{Hsieh_2019,Kido_2023,Lee_2023,Flores_2023}. This influx of material can change the temperatures and densities of the system and can also generate shocks and instabilities at regions where material is deposited \citep{Lee_2024}. This can lead to the formation of dust traps, the development of substructures, the initiation of outbursts, and changes in the chemical makeup of the system \citep{Pineda_2023}. 

Streamers are identified toward IRAS16544, IRAS04169, IRAS16253, OphIRS63, and L1489, observed in emissions of \ceteno, SO, and \h2co. Figure~\ref{fig:streamers} presents the moment 8 maps of \ceteno, SO, and \h2co \htwocolow~emissions, with the streamers marked by dashed lines. For IRAS16544, IRAS04169, and L1489, all three molecules evidently trace the same streamer structures, whereas the distinction is less clear in IRAS16253 and OphIRS63. The \ceteno~emissions reveal large-scale arc-like streamers in IRAS16544 and L1489, and spiral streamers in IRAS04169. These structures might extend beyond the MRS of our observations and likely trace regions of elevated column densities where temperatures exceed 20 K, causing the CO to be released from the ice. In contrast, the SO and \h2co emissions trace the same structures but are confined within the inner 3\arcsec -- 5\arcsec region of the protostar, with enhanced emissions at the landing point of where the streamer meets the inner disk-envelope region. This enhancement is likely due to the shock resulting from the streamer interacting with the disk-envelope region. In IRAS16253, a small streamer is detected toward the east of the source in SO but appears to be obscured by emissions from the envelope in \ceteno~and \h2co. 

For OphIRS63, the \ceteno~emission shows a long arc-like streamer to the north of the protostar and a shorter spiral streamer toward the east. The SO emission does not show the extended streamer toward the north but shows three short spiral streamers arriving from the east, south, and west. The spiral arriving from the east likely traces the same streamer seen toward the east in the \ceteno~emission. The \h2co emissions from eDisk observations do not appear to trace any of the streamers observed in \ceteno~and SO. However, a more sensitive observation conducted by the ALMA Large Program Fifty AU STudy of the chemistry in the disk/envelope system of solar-like protostars (FAUST) detected a streamer to the north of the source, likely tracing the same streamer seen toward the north in our \ceteno~emissions \citep{Podio_2024}.

The molecules \ceteno, \h2co, and SO observed in this work are well-known tracers of infall and shocks in protostellar systems \citep[e.g.,][]{vanDishoeck_1998,Oya_2014,Sakai_2014b,Yen_2017}.
Thus, these molecules are useful for identifying such asymmetric infalling structures and have been employed in several studies to identify accretion streamers. \ceteno~emissions commonly trace dense structures in low-temperature regions of 20--40 K due to their relatively low optical depth and were used to identify streamers in several embedded sources, including Lupus 3-MMS \citep{Thieme_2022}, Per-emb-50 \citep{Valdivia-Mena_2022}, VLA 1623-2417W \citep{Mercimek_2023}, and Per-emb-8 \citep{Lin_2024}. SO and \h2co~emissions, on the other hand, are often enhanced at elevated temperatures and shocked regions, making them useful tracers of impact zones where these infalling streamers deposit their material, typically in the inner envelope and disk regions of protostars \citep[e.g.,][]{Garufi_2022,Valdivia-Mena_2022,Valdivia-Mena_2023,Lee_2023,Lee_2024}.

In addition to \ceteno, \h2co, and SO, several other molecular tracers, including HC$_3$N, N$_2$H$^+$, CCS, CS, HCO$^+$, and SO$_2$, have been used to identify streamers in protostellar systems \citep[e.g.,][]{Yen_2019,Pineda_2020,Garufi_2022,Murillo_2022}. A recent systematic survey of the NGC 1333 star-forming region conducted by \citet{Valdivia-Mena_2024} found evidence of streamers using emissions in HC$_3$N and N$_2$H$^+$ in nearly 60\% of the embedded sources. Within the eDisk survey, streamers are detected in $\sim$30\% of the sources (5 out of 19). However, it is important to note that the eDisk observations were not optimized for detecting large-scale streamers. The NGC1333 survey of \citet{Valdivia-Mena_2024}, covering an area of approximately 150\arcsec $\times$ 150\arcsec~with spatial resolutions of $\sim$5\arcsec, was specifically designed to capture large-scale accretion structures and filaments over extended envelopes. In comparison, the eDisk survey probed considerably smaller spatial scales, with spatial resolutions of $\sim$0\farcs1 and maximum recoverable scales of only $\sim$3\arcsec. As a result, the eDisk survey is primarily sensitive to the warm, shock-influenced inner regions of streamers. Therefore, the molecular tracers used in each study are complementary and highlight the multiscale nature of accretion streamers and the importance of combining both large- and small-scale observations in order to obtain a comprehensive understanding of the accretion process in protostellar systems.
Consequently, streamers may have been missed in other sources, especially if they originate from much larger scales. Notably, streamers are present in both OphIRS63 and L1489, the only two sources in our sample that also show clear signs of substructures, suggesting a possible link between the presence of streamers and the development of these substructures.

\subsection{SO and \h2co} \label{subsection:h2co_so}

The emissions of SO and \h2co exhibit complex morphology and trace disparate spatial distributions within the eDisk sources. These apparent variations suggest multiple concurrent physical processes at play in protostars, the understanding of which can provide valuable insights into the star formation process.

\subsubsection{Morphology of SO emission}
The SO emissions exhibit peaks near the outer edge of the continuum position in several eDisk sources (see Figure~\ref{appendix:so_mom8_mom9_zoomed}). SO is considered to be a reliable tracer of shocked regions in protostellar systems \citep[e.g.,][]{Wakelam_2005,Tafalla_2010,Podio_2015}. In the cases of IRAS04169, IRAS16544, L1489, and OphIRS63, this enhancement is probably due to accretion shocks triggered by material from streamers landing onto the inner-disk envelope region \citep{Flores_2023,Kido_2023,Yamato_2023,Han_2025}. Likewise, the SO emissions show ring-like structures in L1527, OphIRS63, IRAS16253 (only visible in the robust = 0.5 image, see~\citealt{Aso_2023}), and possibly in BHR71 IRS1. The emission in L1527 was found to be likely originating from the disk surface layer and outflow cavity walls \citep{Hoff_2023}, and this might be the case for the other sources as well. Additionally, such ring-like morphology is also believed to result from accretion shocks at the centrifugal barrier, where the infalling material from the envelope or streamers interacts with the rotating disk \citep{Ohashi_2014,Sakai_2014,Sakai_2017,Aso_2023}. 

In the innermost regions ($\lesssim0\farcs4$, $\lesssim$70 au) of BHR71 IRS2, IRAS04166, IRAS15398, IRAS32, and OphIRS43, the SO emissions peak on or close to the peak of the continuum and, in most cases, are asymmetric. The emissions at these distances might be associated with the hot core region \citep{Drozdovskaya_2018}, where SO can sublimate off the disk surface, as temperatures can reach higher than its sublimation temperature of $\sim$50 K. Indeed, moment 9 maps show a velocity gradient along the major axis in most sources, which suggests that the emission close to the protostar likely originates from the innermost envelope and disk region. These emissions could represent the thermal sublimation of SO itself or its precursor. Similar findings have been noted in the outbursting protostar V883 Ori, where the central sublimated component of the SO emissions displays similar size and kinematic properties to those of the COMs \citep{Lee_2024}. However, this scenario does not account for the asymmetry observed toward many sources. The asymmetrical peaks in these sources are oriented in the direction of the outflow, which suggests that they probably result from localized enhancement of SO caused by disk winds or bow shocks from outflows and jets \citep{Tabone_2017}. This enhancement likely complements the SO emissions from the inner envelope and disk surface. Additionally, the large-scale SO emissions seen in a few sources likely also trace shocked regions caused by the outflows (see Figure~\ref{appendix:so_mom8_mom9}). A large arc-like structure is seen in the northeast of Ced110IRS4, and a small blob of emission is seen in the southwest of IRAS15398. Both structures are seen in the direction of the outflow and probably result from SO emissions from gas-phase reactions due to slow-moving outflows ($\sim$3 km s$^{-1}$) or sublimation of SO from ices due to fast-moving outflows ($\gtrsim$ 4 km s$^{-1}$; \citealt{vanGelder_2021}). In BHR71 IRS2 and L1527, large-scale emissions are seen toward the outflow cavities and likely also originate from the sublimation of SO from the ices due to increased temperatures in these regions.

\citet{Tychoniec_2021} discovered that SO and SiO emissions exhibit similar morphologies in most sources in their study. However, we only observe the similarity between the two molecules in IRAS16544, where both molecules peak near the protostar toward the southeast, and in IRAS15398, where a clumpy emission is seen $\sim$1000 au southwest in the direction of the outflow. Corresponding moment 9 maps of both molecules show that the emissions show velocity structures that are close to the $v_{sys}$ of the corresponding source. This suggests that only the shocks produced by the low-velocity outflows with velocities $\lesssim$20 km s$^{-1}$ are able to induce SO emissions either chemically or due to excitation. Likewise, \citet{Tychoniec_2021} observed a clear decrease in emissions in SO from Class 0 sources to Class I sources. However, we do not see a discernible difference between the sources. It is crucial to note that \citet{Tychoniec_2021} detected SO emissions in only five of the sixteen sources observed, with only one being a Class I source. Because of this small sample size, caution must be exercised in generalizing this as a trend. Nevertheless, this apparent decrease likely results from the observations lacking the sensitivity needed to detect the lower column densities of the SO emissions in Class I sources due to their thinner envelopes.

\subsubsection{Morphology of \h2co emission} \label{discussion:h2co}
\h2co was the first organic molecule identified in the ISM that includes elements other than carbon and hydrogen \citep{Snyder_1969}. It has multiple reaction pathways and can form both in warm regions through gas-phase reactions \citep[e.g.,][]{Fockenberg_2002,Atkinson_2006,Marel_2014,Loomis_2015} and cold regions on grain surfaces through the hydrogenation of CO ices \citep[e.g.,][]{Watanabe_2002,Cuppen_2009,Fuchs_2009}. Its importance goes beyond being a simple organic molecule, as it is considered a key intermediate molecule in the synthesis pathways of numerous COMs and is often detected in protostellar systems \citep[e.g.,][]{Maret_2004,Jorgensen_2005,Oberg_2017}. The multiple \h2co transitions covered by the eDisk observations provide a valuable opportunity to probe the physical conditions of the protostellar environments where these emissions are observed. 

The zoomed-in \htwocolow~maps reveal that emissions in the inner envelope and disk regions are observed toward all eDisk sources (see Figure~\ref{appendix:h2co_3_03-2_02_mom8_mom9_zoomed}). Emissions from higher energy transitions are also concentrated in these regions toward BHR71 IRS1/2, IRAS16544, IRAS04166, IRAS15398, IRAS16253, IRAS32, and L1527 (see Figures~\ref{appendix:h2co_3_21-2_20_mom8_mom9_zoomed},~\ref{appendix:h2co_3_22-2_21_mom8_mom9_zoomed}). These emissions likely trace the disk surface and inner envelope regions where temperatures exceed the excitation temperatures required to populate the high-energy states. The detection of emissions in high-energy lines and the fact that CO cannot remain frozen in dust grains in these areas suggest that gas-phase reactions are dominant in these regions \citep{Loomis_2015,Oberg_2017}. Similarly, clusters of \h2co emissions are detected across all transitions toward GSS30IRS3, IRAS04169, and IRAS15398 at larger distances in the outflow direction. These emissions are also probably dominated by gas-phase reactions, with shock-induced heating from outflows raising the temperatures in these regions.

Although the \h2co emissions from both the low- and high-energy transitions trace similar morphologies in the overall moment 8 maps, some differences are observed between these transitions, especially within the individual channel maps. For instance, for the source L1527, emissions in the \htwocohigh~transition were found to originate mostly from the disk surface layer, while significant contributions from the envelope are seen in \htwocolow~transition (see \citealt{Hoff_2023} for further details). Such discrepancies indicate that emissions from different physical layers can contribute to different transitions.

Aside from BHR71 IRS1, IRAS15398, and OphIRS43, extended \h2co emissions are detected only in the \htwocolow~transition. This absence of extended emissions in the higher energy transitions indicates that these emissions primarily trace the relatively colder, outer regions of the protostellar envelope, where the temperatures are not quite warm enough to excite \h2co molecules to higher energy states. Another possibility is that the quantity of molecules excited to the higher energy states in these outer regions is insufficient to produce emissions that exceed the brightness temperature thresholds sensitive to the eDisk observations. In either case, the emissions in these regions are likely dominated by the grain surface hydrogenation pathway, with \h2co remaining frozen out until temperatures rise above the \h2co~sublimation temperatures of $\sim$40 -- 70 K \citep{Aikawa_1997,Noble_2012,Fedoseev_2015}. A more comprehensive analysis using chemical models incorporating both formation pathways is necessary to fully untangle the chemistry of \h2co emissions in protostellar sources.

The ratios of different $K_a$ ladders of \h2co transitions, such as $J_{K_aK_c}$ = \htwocolow, and $J_{K_aK_c}$ = \htwocohigh~are often considered to be good tracers of the kinetic temperatures of dense regions \citep[e.g.,][]{Mangum_1993,Tang_2017,Artur_2019,Hoff_2023}. As mentioned in sections~\ref{results:h2co} and \ref{discussion:h2co}, the eDisk observations feature several sources where emissions are detected in both \h2co (\htwocolow) and (\htwocohigh) transitions. However, on the scales of the eDisk sources, the temperatures increase above 70 K and extinctions are $\gtrsim$100 mag, corresponding to H$_2$ column densities of $\gtrsim 10^{23}$ cm$^{-2}$. A simple non-local thermal equilibrium (non-LTE) calculation using RADEX \citep{Tak_2007} demonstrates that the \htwocolow~transition becomes optically thick under these conditions, even for a conservative \h2co abundance of 10$^{-9}$, which is why this method is not applicable on these scales. Additional molecular tracers that are optically thin and sensitive to higher temperatures, such as CH$_3$CN or H$_3$CN \citep[e.g.,][]{Bergner_2018,Hsieh_2023,Hoff_2024}, as well as more sophisticated radiative transfer modeling, are necessary to accurately determine the temperature structures on these scales of protostellar systems.

Emissions from \h2co were also identified in various morphologies of a protostellar system, including envelope, disk, outflow, and jets, toward the sources studied by \citet{Tychoniec_2021}, consistent with the interpretation that \h2co has a multicomponent origin in the eDisk observations. However, a notable distinction between the two studies is that in our observations, we detect \h2co emissions toward the walls of the outflow cavity in some sources, including BHR71 IRS1 and IRAS04166 (see Figure~\ref{fig:h2co_mom8_zoomed_combined}), which was not reported by \citet{Tychoniec_2021}. This discrepancy most likely arises from the difference in the spatial resolution of the two observations. The eDisk observations, with resolutions of $\sim$15 au, are significantly more sensitive to emissions from small-scale structures compared to the $\gtrsim$100 au resolution in \citet{Tychoniec_2021}. Thus, any emissions arising from the narrow outflow cavity walls in \h2co may have been unresolved or spatially filtered out in the previous study.

\begin{figure}[ht!]
\centering
    \includegraphics[width=1.0\linewidth]{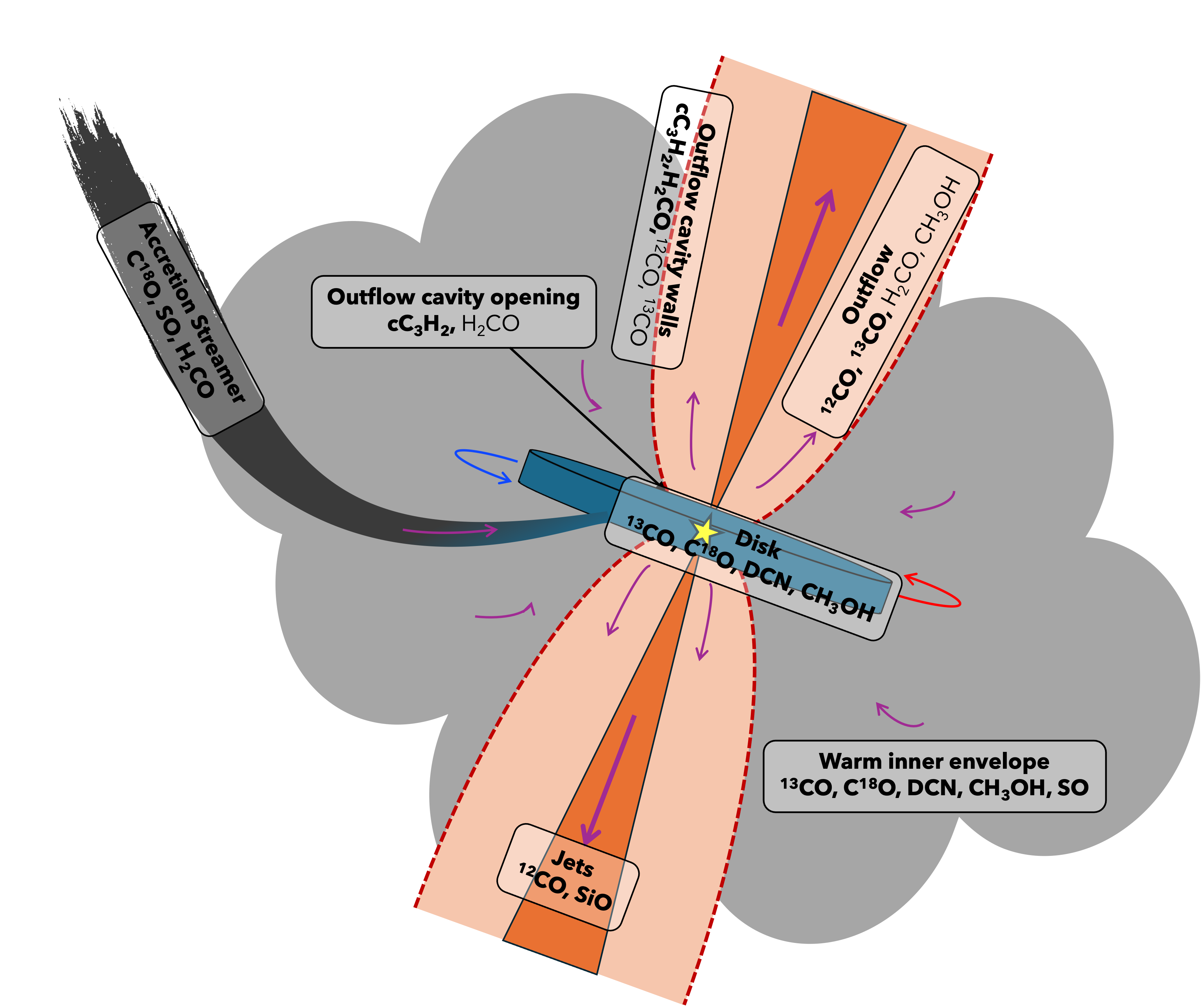}
\caption{Illustration depicting the various molecules identified in different components of this study. Molecules highlighted in bold are tracers typically found in these components, whereas those in regular font indicate tracers that are occasionally identified in some sources. \label{fig:embedded_source_filled}}
\end{figure}

\section{Conclusions}\label{sec:conclusion}

We have presented a comprehensive analysis of high-angular resolution ($\sim$0$\farcs1$ or $\sim$15 au) spectral line emissions detected toward 19 nearby Class 0/I protostars in an attempt to characterize the various morphologies traced by different molecules and their physical and chemical implications. Figure~\ref{fig:embedded_source_filled} provides a summary of our results by showing the molecules detected in each spatial region, illustrated schematically in Figure~\ref{fig:embedded_source}. Our main findings are as follows:

   \begin{enumerate}
      \item Protostellar outflows are observed in 15 of the 19 sources and are primarily traced by large-scale emissions in \tlvco~and \thrco. Additionally, we also detect high-velocity molecular jets with $|v_{sys} - v_{chan}| > 20$ km s$^{-1}$, in emissions of \tlvco~and SiO toward four sources.
      \item The walls of the outflow cavities have high densities and are delineated by large-scale \thrco~emissions. Exposure to UV radiation and outflow shocks leads to increased temperatures in the cavity walls, leading to enhanced emissions from \C3h2 and \h2co in these regions.
      \item Rotation in the inner envelope and disk regions is traced by emissions in \thrco~and \ceteno, with Keplerian rotation identified in 14 of the 19 sources. Emissions in all nine sources where DCN is detected, and in five of the seven sources where \ch3oh is detected, are confined to the innermost regions. These molecules are most likely tracing the hot core regions of these sources.
      \item Accretion streamers are identified in five sources. These appear as spirals or elongated arc-like structures in emissions of \ceteno, SO, and \h2co.
      \item Emissions in SO and \h2co~display complex morphologies and trace disparate structures across the different sources. Both molecules have multiple reaction pathways that likely contribute to the overall emissions observed toward the eDisk sources.
   \end{enumerate}

\begin{acknowledgements}

The authors would like to thank the anonymous referee for the helpful comments and suggestions that improved the quality of the manuscript. This paper makes use of the following ALMA data: ADS/ JAO.ALMA\#2019.1.00261.L, ADS/ JAO.ALMA\#2019.A.00034.S. ALMA is a partnership of ESO (representing its member states), NSF (USA), and NINS (Japan), together with NRC (Canada), MOST and ASIAA (Taiwan), and KASI (Republic of Korea), in cooperation with the Republic of Chile. The Joint ALMA Observatory is operated by ESO, AUI/NRAO, and NAOJ. 
The National Radio Astronomy Observatory is a facility of the National Science Foundation operated under cooperative agreement by Associated Universities, Inc. 
R.S, J.K.J, and S.G. acknowledge support from the Independent Research Fund Denmark (grant No. 0135-00123B). 
J.J.T. acknowledges support from NASA XRP 80NSSC22K1159. 
L.W.L. acknowledges support from NSF AST-2108794.
Y.-L.Y. acknowledges support from Grant-in-Aid from the Ministry of Education, Culture, Sports, Science, and Technology of Japan (20H05845, 20H05844, 22K20389), and a pioneering project in RIKEN (Evolution of Matter in the Universe). Y.A acknowledges support from Grain-in-Aid from the Ministry of Education, Culture, Sports, Science, and Technology of Japan (24K00674, 21H04495, 20H05847).
J.-E.L. was supported by the National Research Foundation of Korea (NRF) grant funded by the Korea government (MSIT) (grant numbers 2021R1A2C1011718 and RS-2024-00416859).
S.N. acknowledges support from NSF GRFP grant No. 2236415 and the P.E.O. Scholar Award. Z.-Y.L. is supported in part by NASA 80NSSC20K0533
and NSF AST-2307199.
N.O. and M.N. acknowledge support from National Science and Technology Council (NSTC 113-2112-M-001-037) and the Academia Sinica Investigator Project Grant (AS-IV-114-M02).

\end{acknowledgements}

\bibliographystyle{bibtex/aa}
\bibliography{references}

\begin{thebibliography}{130}
\expandafter\ifx\csname natexlab\endcsname\relax\def\natexlab#1{#1}\fi

\bibitem[{{Aikawa} {et~al.}(2018){Aikawa}, {Furuya}, {Hincelin}, \& {Herbst}}]{Aikawa_2018}
{Aikawa}, Y., {Furuya}, K., {Hincelin}, U., \& {Herbst}, E. 2018, \apj, 855, 119

\bibitem[{{Aikawa} {et~al.}(1997){Aikawa}, {Umebayashi}, {Nakano}, \& {Miyama}}]{Aikawa_1997}
{Aikawa}, Y., {Umebayashi}, T., {Nakano}, T., \& {Miyama}, S.~M. 1997, \apjl, 486, L51

\bibitem[{{Albertsson} {et~al.}(2013){Albertsson}, {Semenov}, {Vasyunin}, {Henning}, \& {Herbst}}]{Albertsson_2013}
{Albertsson}, T., {Semenov}, D.~A., {Vasyunin}, A.~I., {Henning}, T., \& {Herbst}, E. 2013, \apjs, 207, 27

\bibitem[{{ALMA Partnership} {et~al.}(2015){ALMA Partnership}, {Brogan}, {P{\'e}rez}, {Hunter}, {Dent}, {Hales}, {Hills}, {Corder}, {Fomalont}, {Vlahakis}, {Asaki}, {Barkats}, {Hirota}, {Hodge}, {Impellizzeri}, {Kneissl}, {Liuzzo}, {Lucas}, {Marcelino}, {Matsushita}, {Nakanishi}, {Phillips}, {Richards}, {Toledo}, {Aladro}, {Broguiere}, {Cortes}, {Cortes}, {Espada}, {Galarza}, {Garcia-Appadoo}, {Guzman-Ramirez}, {Humphreys}, {Jung}, {Kameno}, {Laing}, {Leon}, {Marconi}, {Mignano}, {Nikolic}, {Nyman}, {Radiszcz}, {Remijan}, {Rod{\'o}n}, {Sawada}, {Takahashi}, {Tilanus}, {Vila Vilaro}, {Watson}, {Wiklind}, {Akiyama}, {Chapillon}, {de Gregorio-Monsalvo}, {Di Francesco}, {Gueth}, {Kawamura}, {Lee}, {Nguyen Luong}, {Mangum}, {Pietu}, {Sanhueza}, {Saigo}, {Takakuwa}, {Ubach}, {van Kempen}, {Wootten}, {Castro-Carrizo}, {Francke}, {Gallardo}, {Garcia}, {Gonzalez}, {Hill}, {Kaminski}, {Kurono}, {Liu}, {Lopez}, {Morales}, {Plarre}, {Schieven}, {Testi}, {Videla}, {Villard}, {Andreani}, {Hibbard}, \&
  {Tatematsu}}]{Alma_2015}
{ALMA Partnership}, {Brogan}, C.~L., {P{\'e}rez}, L.~M., {et~al.} 2015, \apjl, 808, L3

\bibitem[{{Andr{\`e}} {et~al.}(1993){Andr{\`e}}, {Ward-Thompson}, \& {Barsony}}]{Andre_1993}
{Andr{\`e}}, P., {Ward-Thompson}, D., \& {Barsony}, M. 1993, \apj, 406, 122

\bibitem[{{Andr{\`e}} {et~al.}(2000){Andr{\`e}}, {Ward-Thompson}, \& {Barsony}}]{Andre_2000}
{Andr{\`e}}, P., {Ward-Thompson}, D., \& {Barsony}, M. 2000, in Protostars and Planets IV, ed. V.~{Mannings}, A.~P. {Boss}, \& S.~S. {Russell}, 59

\bibitem[{Andrews {et~al.}(2018)Andrews, Huang, Pérez, Isella, Dullemond, Kurtovic, Guzmán, Carpenter, Wilner, Zhang, Zhu, Birnstiel, Bai, Benisty, Hughes, Öberg, \& Ricci}]{Andrews_2018}
Andrews, S.~M., Huang, J., Pérez, L.~M., {et~al.} 2018, The Astrophysical Journal Letters, 869, L41

\bibitem[{{Arce} \& {Sargent}(2006)}]{Arce_2006}
{Arce}, H.~G. \& {Sargent}, A.~I. 2006, \apj, 646, 1070

\bibitem[{{Arce} {et~al.}(2007){Arce}, {Shepherd}, {Gueth}, {Lee}, {Bachiller}, {Rosen}, \& {Beuther}}]{Arce_2007}
{Arce}, H.~G., {Shepherd}, D., {Gueth}, F., {et~al.} 2007, in Protostars and Planets V, ed. B.~{Reipurth}, D.~{Jewitt}, \& K.~{Keil}, 245

\bibitem[{{Artur de la Villarmois} {et~al.}(2019){Artur de la Villarmois}, {Kristensen}, \& {J{\o}rgensen}}]{Artur_2019}
{Artur de la Villarmois}, E., {Kristensen}, L.~E., \& {J{\o}rgensen}, J.~K. 2019, \aap, 627, A37

\bibitem[{{Aso} {et~al.}(2023){Aso}, {Kwon}, {Ohashi}, {J{\o}rgensen}, {Tobin}, {Aikawa}, {de Gregorio-Monsalvo}, {Han}, {Kido}, {Koch}, {Lai}, {Lee}, {Lee}, {Li}, {Lin}, {Looney}, {Narayanan}, {Phuong}, {Sai}, {Saigo}, {Santamar{\'\i}a-Miranda}, {Sharma}, {Takakuwa}, {Thieme}, {Tomida}, {Williams}, \& {Yen}}]{Aso_2023}
{Aso}, Y., {Kwon}, W., {Ohashi}, N., {et~al.} 2023, \apj, 954, 101

\bibitem[{{Atkinson} {et~al.}(2006){Atkinson}, {Baulch}, {Cox}, {Crowley}, {Hampson}, {Hynes}, {Jenkin}, {Rossi}, {Troe}, \& {Subcommittee}}]{Atkinson_2006}
{Atkinson}, R., {Baulch}, D.~L., {Cox}, R.~A., {et~al.} 2006, Atmospheric Chemistry \& Physics, 6, 3625

\bibitem[{{Bachiller}(1996)}]{Bachiller_1996}
{Bachiller}, R. 1996, \araa, 34, 111

\bibitem[{{Bergin} {et~al.}(1998){Bergin}, {Melnick}, \& {Neufeld}}]{Bergin_1998}
{Bergin}, E.~A., {Melnick}, G.~J., \& {Neufeld}, D.~A. 1998, \apj, 499, 777

\bibitem[{{Bergner} {et~al.}(2018){Bergner}, {Guzm{\'a}n}, {{\"O}berg}, {Loomis}, \& {Pegues}}]{Bergner_2018}
{Bergner}, J.~B., {Guzm{\'a}n}, V.~G., {{\"O}berg}, K.~I., {Loomis}, R.~A., \& {Pegues}, J. 2018, \apj, 857, 69

\bibitem[{{Bergner} {et~al.}(2017){Bergner}, {{\"O}berg}, \& {Rajappan}}]{Bergner_2017}
{Bergner}, J.~B., {{\"O}berg}, K.~I., \& {Rajappan}, M. 2017, \apj, 845, 29

\bibitem[{{Bergner} {et~al.}(2022){Bergner}, {Rajappan}, \& {{\"O}berg}}]{Bergner_2022}
{Bergner}, J.~B., {Rajappan}, M., \& {{\"O}berg}, K.~I. 2022, \apj, 933, 206

\bibitem[{{Bjerkeli} {et~al.}(2019){Bjerkeli}, {Ramsey}, {Harsono}, {Calcutt}, {Kristensen}, {van der Wiel}, {J{\o}rgensen}, {Muller}, \& {Persson}}]{Bjerkeli_2019}
{Bjerkeli}, P., {Ramsey}, J.~P., {Harsono}, D., {et~al.} 2019, \aap, 631, A64

\bibitem[{{Bjerkeli} {et~al.}(2016){Bjerkeli}, {van der Wiel}, {Harsono}, {Ramsey}, \& {J{\o}rgensen}}]{Bjerkeli_2016}
{Bjerkeli}, P., {van der Wiel}, M. H.~D., {Harsono}, D., {Ramsey}, J.~P., \& {J{\o}rgensen}, J.~K. 2016, \nat, 540, 406

\bibitem[{{Boogert} {et~al.}(2015){Boogert}, {Gerakines}, \& {Whittet}}]{Boogert_2015}
{Boogert}, A.~C.~A., {Gerakines}, P.~A., \& {Whittet}, D. C.~B. 2015, \araa, 53, 541

\bibitem[{{Brown} \& {Bolina}(2007)}]{Brown_2007}
{Brown}, W.~A. \& {Bolina}, A.~S. 2007, \mnras, 374, 1006

\bibitem[{{Ceccarelli}(2004)}]{Ceccarelli_2004}
{Ceccarelli}, C. 2004, in Astronomical Society of the Pacific Conference Series, Vol. 323, Star Formation in the Interstellar Medium: In Honor of David Hollenbach, ed. D.~{Johnstone}, F.~C. {Adams}, D.~N.~C. {Lin}, D.~A. {Neufeeld}, \& E.~C. {Ostriker}, 195

\bibitem[{{Ceccarelli} {et~al.}(2007){Ceccarelli}, {Caselli}, {Herbst}, {Tielens}, \& {Caux}}]{Ceccarelli_2007}
{Ceccarelli}, C., {Caselli}, P., {Herbst}, E., {Tielens}, A.~G.~G.~M., \& {Caux}, E. 2007, in Protostars and Planets V, ed. B.~{Reipurth}, D.~{Jewitt}, \& K.~{Keil}, 47

\bibitem[{{Ceccarelli} {et~al.}(2023){Ceccarelli}, {Codella}, {Balucani}, {Bockelee-Morvan}, {Herbst}, {Vastel}, {Caselli}, {Favre}, {Lefloch}, {Oberg}, \& {Yamamoto}}]{Ceccarelli_2023}
{Ceccarelli}, C., {Codella}, C., {Balucani}, N., {et~al.} 2023, in Astronomical Society of the Pacific Conference Series, Vol. 534, Protostars and Planets VII, ed. S.~{Inutsuka}, Y.~{Aikawa}, T.~{Muto}, K.~{Tomida}, \& M.~{Tamura}, 379

\bibitem[{{Collings} {et~al.}(2004){Collings}, {Anderson}, {Chen}, {Dever}, {Viti}, {Williams}, \& {McCoustra}}]{Collings_2004}
{Collings}, M.~P., {Anderson}, M.~A., {Chen}, R., {et~al.} 2004, \mnras, 354, 1133

\bibitem[{{Cuppen} {et~al.}(2009){Cuppen}, {van Dishoeck}, {Herbst}, \& {Tielens}}]{Cuppen_2009}
{Cuppen}, H.~M., {van Dishoeck}, E.~F., {Herbst}, E., \& {Tielens}, A.~G.~G.~M. 2009, \aap, 508, 275

\bibitem[{{Drozdovskaya} {et~al.}(2018){Drozdovskaya}, {van Dishoeck}, {J{\o}rgensen}, {Calmonte}, {van der Wiel}, {Coutens}, {Calcutt}, {M{\"u}ller}, {Bjerkeli}, {Persson}, {Wampfler}, \& {Altwegg}}]{Drozdovskaya_2018}
{Drozdovskaya}, M.~N., {van Dishoeck}, E.~F., {J{\o}rgensen}, J.~K., {et~al.} 2018, \mnras, 476, 4949

\bibitem[{{Drozdovskaya} {et~al.}(2015){Drozdovskaya}, {Walsh}, {Visser}, {Harsono}, \& {van Dishoeck}}]{Drozdovskaya_2015}
{Drozdovskaya}, M.~N., {Walsh}, C., {Visser}, R., {Harsono}, D., \& {van Dishoeck}, E.~F. 2015, \mnras, 451, 3836

\bibitem[{{Encalada} {et~al.}(2024){Encalada}, {Looney}, {Takakuwa}, {Tobin}, {Ohashi}, {J{\o}rgensen}, {Li}, {Aikawa}, {Aso}, {Koch}, {Kwon}, {Lai}, {Lee}, {Lin}, {Santamar{\'\i}a-Miranda}, {de Gregorio-Monsalvo}, {Phuong}, {Plunkett}, {Sai (Insa Choi)}, {Sharma}, {Yen}, \& {Han}}]{Encalada_2024}
{Encalada}, F.~J., {Looney}, L.~W., {Takakuwa}, S., {et~al.} 2024, \apj, 966, 32

\bibitem[{{Fedoseev} {et~al.}(2015){Fedoseev}, {Cuppen}, {Ioppolo}, {Lamberts}, \& {Linnartz}}]{Fedoseev_2015}
{Fedoseev}, G., {Cuppen}, H.~M., {Ioppolo}, S., {Lamberts}, T., \& {Linnartz}, H. 2015, \mnras, 448, 1288

\bibitem[{{Feeney-Johansson} {et~al.}(2025){Feeney-Johansson}, {Aikawa}, {Takakuwa}, {Ohashi}, {Plunkett}, {J\o rgensen}, {Shang}, {Li}, {Sharma}, {Kwon}, {Lee}, {Looney}, {Yang}, {Narang}, \& {de Gregorio-Monsalvo}}]{Feeney-Johansson_prep}
{Feeney-Johansson}, A., {Aikawa}, Y., {Takakuwa}, S., {et~al.} 2025, Submitted to \apj

\bibitem[{{Flores} {et~al.}(2023){Flores}, {Ohashi}, {Tobin}, {J{\o}rgensen}, {Takakuwa}, {Li}, {Lin}, {van't Hoff}, {Plunkett}, {Yamato}, {Sai (Insa Choi)}, {Koch}, {Yen}, {Aikawa}, {Aso}, {de Gregorio-Monsalvo}, {Kido}, {Kwon}, {Lee}, {Lee}, {Looney}, {Santamar{\'\i}a-Miranda}, {Sharma}, {Thieme}, {Williams}, {Han}, {Narayanan}, \& {Lai}}]{Flores_2023}
{Flores}, C., {Ohashi}, N., {Tobin}, J.~J., {et~al.} 2023, \apj, 958, 98

\bibitem[{{Flores-Rivera} {et~al.}(2021){Flores-Rivera}, {Terebey}, {Willacy}, {Isella}, {Turner}, \& {Flock}}]{Rivera_2021}
{Flores-Rivera}, L., {Terebey}, S., {Willacy}, K., {et~al.} 2021, \apj, 908, 108

\bibitem[{{Fockenberg} \& {Preses}(2002)}]{Fockenberg_2002}
{Fockenberg}, C. \& {Preses}, J.~M. 2002, Journal of Physical Chemistry A, 106, 2924

\bibitem[{{Frank} {et~al.}(2014){Frank}, {Ray}, {Cabrit}, {Hartigan}, {Arce}, {Bacciotti}, {Bally}, {Benisty}, {Eisl{\"o}ffel}, {G{\"u}del}, {Lebedev}, {Nisini}, \& {Raga}}]{Frank_2014}
{Frank}, A., {Ray}, T.~P., {Cabrit}, S., {et~al.} 2014, in Protostars and Planets VI, ed. H.~{Beuther}, R.~S. {Klessen}, C.~P. {Dullemond}, \& T.~{Henning}, 451--474

\bibitem[{{Fuchs} {et~al.}(2009){Fuchs}, {Cuppen}, {Ioppolo}, {Romanzin}, {Bisschop}, {Andersson}, {van Dishoeck}, \& {Linnartz}}]{Fuchs_2009}
{Fuchs}, G.~W., {Cuppen}, H.~M., {Ioppolo}, S., {et~al.} 2009, \aap, 505, 629

\bibitem[{{Garufi} {et~al.}(2022){Garufi}, {Podio}, {Codella}, {Segura-Cox}, {Vander Donckt}, {Mercimek}, {Bacciotti}, {Fedele}, {Kasper}, {Pineda}, {Humphreys}, \& {Testi}}]{Garufi_2022}
{Garufi}, A., {Podio}, L., {Codella}, C., {et~al.} 2022, \aap, 658, A104

\bibitem[{{Gavino} {et~al.}(2024){Gavino}, {J{\o}rgensen}, {Sharma}, {Yang}, {Li}, {Tobin}, {Ohashi}, {Takakuwa}, {Plunkett}, {Kwon}, {de Gregorio-Monsalvo}, {Lin}, {Santamar{\'\i}a-Miranda}, {Aso}, {Sai}, {Aikawa}, {Tomida}, {Koch}, {Lee}, {Lee}, {Lai}, {Looney}, {Narayanan}, {Phuong}, {Thieme}, {van't Hoff}, {Williams}, \& {Yen}}]{Gavino_2024}
{Gavino}, S., {J{\o}rgensen}, J.~K., {Sharma}, R., {et~al.} 2024, \apj, 974, 21

\bibitem[{{Guzm{\'a}n} {et~al.}(2015){Guzm{\'a}n}, {Pety}, {Goicoechea}, {Gerin}, {Roueff}, {Gratier}, \& {{\"O}berg}}]{Guzman_2015}
{Guzm{\'a}n}, V.~V., {Pety}, J., {Goicoechea}, J.~R., {et~al.} 2015, \apjl, 800, L33

\bibitem[{{Han} {et~al.}(2025){Han}, {Kwon}, {Aso}, {Ohashi}, {Tobin}, {J{\o}rgensen}, {Takakuwa}, {Looney}, {Aikawa}, {Flores}, {de Gregorio-Monsalvo}, {Koch}, {Lee}, {Lee}, {Li}, {Lin}, {Sai}, {Thieme}, {Williams}, {Gavino}, {Kido}, {Lai}, {Phuong}, {Santamar{\'\i}a-Miranda}, \& {Yen}}]{Han_2025}
{Han}, I., {Kwon}, W., {Aso}, Y., {et~al.} 2025, \apj~in press, arXiv:2506.16569

\bibitem[{{Harsono} {et~al.}(2018){Harsono}, {Bjerkeli}, {van der Wiel}, {Ramsey}, {Maud}, {Kristensen}, \& {J{\o}rgensen}}]{Harsono_2018}
{Harsono}, D., {Bjerkeli}, P., {van der Wiel}, M. H.~D., {et~al.} 2018, Nature Astronomy, 2, 646

\bibitem[{{Herbst} \& {van Dishoeck}(2009)}]{Herbst_2009}
{Herbst}, E. \& {van Dishoeck}, E.~F. 2009, \araa, 47, 427

\bibitem[{{Hiraoka} {et~al.}(1994){Hiraoka}, {Ohashi}, {Kihara}, {Yamamoto}, {Sato}, \& {Yamashita}}]{Hiraoka_1994}
{Hiraoka}, K., {Ohashi}, N., {Kihara}, Y., {et~al.} 1994, Chemical Physics Letters, 229, 408

\bibitem[{{Hsieh} {et~al.}(2019){Hsieh}, {Murillo}, {Belloche}, {Hirano}, {Walsh}, {van Dishoeck}, {J{\o}rgensen}, \& {Lai}}]{Hsieh_2019}
{Hsieh}, T.-H., {Murillo}, N.~M., {Belloche}, A., {et~al.} 2019, \apj, 884, 149

\bibitem[{{Hsieh} {et~al.}(2023){Hsieh}, {Segura-Cox}, {Pineda}, {Caselli}, {Bouscasse}, {Neri}, {Lopez-Sepulcre}, {Valdivia-Mena}, {Maureira}, {Henning}, {Smirnov-Pinchukov}, {Semenov}, {M{\"o}ller}, {Cunningham}, {Fuente}, {Marino}, {Dutrey}, {Tafalla}, {Chapillon}, {Ceccarelli}, \& {Zhao}}]{Hsieh_2023}
{Hsieh}, T.~H., {Segura-Cox}, D.~M., {Pineda}, J.~E., {et~al.} 2023, \aap, 669, A137

\bibitem[{{Jhan} {et~al.}(2022){Jhan}, {Lee}, {Johnstone}, {Liu}, {Liu}, {Hirano}, {Tatematsu}, {Dutta}, {Moraghan}, {Shang}, {Lee}, {Li}, {Liu}, {Hsu}, {Kwon}, {Sahu}, {Liu}, {Kim}, {Luo}, {Qin}, {Sanhueza}, {Bronfman}, {Qizhou}, {Eden}, {Traficante}, {Lee}, \& {Almasop Team}}]{Jhan_2022}
{Jhan}, K.-S., {Lee}, C.-F., {Johnstone}, D., {et~al.} 2022, \apjl, 931, L5

\bibitem[{{J{\o}rgensen} {et~al.}(2020){J{\o}rgensen}, {Belloche}, \& {Garrod}}]{Jorgensen_2020}
{J{\o}rgensen}, J.~K., {Belloche}, A., \& {Garrod}, R.~T. 2020, \araa, 58, 727

\bibitem[{{J{\o}rgensen} {et~al.}(2002){J{\o}rgensen}, {Sch{\"o}ier}, \& {van Dishoeck}}]{Jorgensen_2002}
{J{\o}rgensen}, J.~K., {Sch{\"o}ier}, F.~L., \& {van Dishoeck}, E.~F. 2002, \aap, 389, 908

\bibitem[{{J{\o}rgensen} {et~al.}(2005){J{\o}rgensen}, {Sch{\"o}ier}, \& {van Dishoeck}}]{Jorgensen_2005}
{J{\o}rgensen}, J.~K., {Sch{\"o}ier}, F.~L., \& {van Dishoeck}, E.~F. 2005, \aap, 437, 501

\bibitem[{{Kido} {et~al.}(2023){Kido}, {Takakuwa}, {Saigo}, {Ohashi}, {Tobin}, {J{\o}rgensen}, {Aikawa}, {Aso}, {Encalada}, {Flores}, {Gavino}, {de Gregorio-Monsalvo}, {Han}, {Hirano}, {Koch}, {Kwon}, {Lai}, {Lee}, {Lee}, {Li}, {Lin}, {Looney}, {Mori}, {Narayanan}, {Plunkett}, {Phuong}, {(Insa Choi)}, {Santamar{\'\i}a-Miranda}, {Sharma}, {Sheehan}, {Thieme}, {Tomida}, {van't Hoff}, {Williams}, {Yamato}, \& {Yen}}]{Kido_2023}
{Kido}, M., {Takakuwa}, S., {Saigo}, K., {et~al.} 2023, \apj, 953, 190

\bibitem[{{Kim} {et~al.}(2024){Kim}, {Lee}, {Pe{\~n}a}, {Johnstone}, {Herczeg}, {Tobin}, \& {Evans}}]{Kim_2024}
{Kim}, C.-H., {Lee}, J.-E., {Pe{\~n}a}, C.~C., {et~al.} 2024, \apj, 961, 108

\bibitem[{{Kristensen} {et~al.}(2010){Kristensen}, {van Dishoeck}, {van Kempen}, {Cuppen}, {Brinch}, {J{\o}rgensen}, \& {Hogerheijde}}]{Kristensen_2010}
{Kristensen}, L.~E., {van Dishoeck}, E.~F., {van Kempen}, T.~A., {et~al.} 2010, \aap, 516, A57

\bibitem[{{Lada} \& {Wilking}(1984)}]{Lada_1984}
{Lada}, C.~J. \& {Wilking}, B.~A. 1984, \apj, 287, 610

\bibitem[{{Lee} {et~al.}(2024){Lee}, {Kim}, {Lee}, {Lee}, {Baek}, {Yun}, {Aikawa}, {Johnstone}, {Herczeg}, \& {Cieza}}]{Lee_2024}
{Lee}, J.-E., {Kim}, C.-H., {Lee}, S., {et~al.} 2024, \apj, 966, 119

\bibitem[{{Lee} {et~al.}(2023){Lee}, {Matsumoto}, {Kim}, {Lee}, {Harsono}, {Bae}, {Evans}, {Inutsuka}, {Choi}, {Tatematsu}, {Lee}, \& {Jaffe}}]{Lee_2023}
{Lee}, J.-E., {Matsumoto}, T., {Kim}, H.-J., {et~al.} 2023, \apj, 953, 82

\bibitem[{{Lee} {et~al.}(2015){Lee}, {Lee}, \& {Bergin}}]{Lee_2015}
{Lee}, S., {Lee}, J.-E., \& {Bergin}, E.~A. 2015, \apjs, 217, 30

\bibitem[{{Lee} {et~al.}(2014){Lee}, {Lee}, {Bergin}, \& {Park}}]{Lee_2014}
{Lee}, S., {Lee}, J.-E., {Bergin}, E.~A., \& {Park}, Y.-S. 2014, \apjs, 213, 33

\bibitem[{{Lin} {et~al.}(2024){Lin}, {Yen}, \& {Lai}}]{Lin_2024}
{Lin}, S.-J., {Yen}, H.-W., \& {Lai}, S.-P. 2024, \aj, 168, 107

\bibitem[{{Lin} {et~al.}(2023){Lin}, {Li}, {Tobin}, {Ohashi}, {J{\o}rgensen}, {Looney}, {Aso}, {Takakuwa}, {Aikawa}, {van't Hoff}, {de Gregorio-Monsalvo}, {Encalada}, {Flores}, {Gavino}, {Han}, {Kido}, {Koch}, {Kwon}, {Lai}, {Lee}, {Lee}, {Phuong}, {Sai}, {Sharma}, {Sheehan}, {Thieme}, {Williams}, {Yamato}, \& {Yen}}]{Lin_2023}
{Lin}, Z.-Y.~D., {Li}, Z.-Y., {Tobin}, J.~J., {et~al.} 2023, \apj, 951, 9

\bibitem[{{Loomis} {et~al.}(2015){Loomis}, {Cleeves}, {{\"O}berg}, {Guzman}, \& {Andrews}}]{Loomis_2015}
{Loomis}, R.~A., {Cleeves}, L.~I., {{\"O}berg}, K.~I., {Guzman}, V.~V., \& {Andrews}, S.~M. 2015, \apjl, 809, L25

\bibitem[{{Mangum} \& {Wootten}(1993)}]{Mangum_1993}
{Mangum}, J.~G. \& {Wootten}, A. 1993, \apjs, 89, 123

\bibitem[{{Maret} {et~al.}(2004){Maret}, {Ceccarelli}, {Caux}, {Tielens}, {J{\o}rgensen}, {van Dishoeck}, {Bacmann}, {Castets}, {Lefloch}, {Loinard}, {Parise}, \& {Sch{\"o}ier}}]{Maret_2004}
{Maret}, S., {Ceccarelli}, C., {Caux}, E., {et~al.} 2004, \aap, 416, 577

\bibitem[{{Martin-Pintado} {et~al.}(1992){Martin-Pintado}, {Bachiller}, \& {Fuente}}]{Pintado_1992}
{Martin-Pintado}, J., {Bachiller}, R., \& {Fuente}, A. 1992, \aap, 254, 315

\bibitem[{{Maury} {et~al.}(2019){Maury}, {Andr{\'e}}, {Testi}, {Maret}, {Belloche}, {Hennebelle}, {Cabrit}, {Codella}, {Gueth}, {Podio}, {Anderl}, {Bacmann}, {Bontemps}, {Gaudel}, {Ladjelate}, {Lef{\`e}vre}, {Tabone}, \& {Lefloch}}]{Maury_2019}
{Maury}, A.~J., {Andr{\'e}}, P., {Testi}, L., {et~al.} 2019, \aap, 621, A76

\bibitem[{{McMullin} {et~al.}(2007){McMullin}, {Waters}, {Schiebel}, {Young}, \& {Golap}}]{Mcmullin_2007}
{McMullin}, J.~P., {Waters}, B., {Schiebel}, D., {Young}, W., \& {Golap}, K. 2007, in Astronomical Society of the Pacific Conference Series, Vol. 376, Astronomical Data Analysis Software and Systems XVI, ed. R.~A. {Shaw}, F.~{Hill}, \& D.~J. {Bell}, 127

\bibitem[{{Mercimek} {et~al.}(2023){Mercimek}, {Podio}, {Codella}, {Chahine}, {L{\'o}pez-Sepulcre}, {Ohashi}, {Loinard}, {Johnstone}, {Menard}, {Cuello}, {Caselli}, {Zamponi}, {Aikawa}, {Bianchi}, {Busquet}, {Pineda}, {Bouvier}, {De Simone}, {Zhang}, {Sakai}, {Chandler}, {Ceccarelli}, {Alves}, {Dur{\'a}n}, {Fedele}, {Murillo}, {Jim{\'e}nez-Serra}, \& {Yamamoto}}]{Mercimek_2023}
{Mercimek}, S., {Podio}, L., {Codella}, C., {et~al.} 2023, \mnras, 522, 2384

\bibitem[{{Millar} {et~al.}(1989){Millar}, {Bennett}, \& {Herbst}}]{Millar_1989}
{Millar}, T.~J., {Bennett}, A., \& {Herbst}, E. 1989, \apj, 340, 906

\bibitem[{{Murillo} {et~al.}(2022){Murillo}, {van Dishoeck}, {Hacar}, {Harsono}, \& {J{\o}rgensen}}]{Murillo_2022}
{Murillo}, N.~M., {van Dishoeck}, E.~F., {Hacar}, A., {Harsono}, D., \& {J{\o}rgensen}, J.~K. 2022, \aap, 658, A53

\bibitem[{{Murillo} {et~al.}(2018){Murillo}, {van Dishoeck}, {van der Wiel}, {J{\o}rgensen}, {Drozdovskaya}, {Calcutt}, \& {Harsono}}]{Murillo_2018}
{Murillo}, N.~M., {van Dishoeck}, E.~F., {van der Wiel}, M.~H.~D., {et~al.} 2018, \aap, 617, A120

\bibitem[{{Nakamura} \& {Li}(2014)}]{Nakamura_2014}
{Nakamura}, F. \& {Li}, Z.-Y. 2014, \apj, 783, 115

\bibitem[{{Narayanan} {et~al.}(2023){Narayanan}, {Williams}, {Tobin}, {J{\o}rgensen}, {Ohashi}, {Lin}, {van't Hoff}, {Li}, {Plunkett}, {Looney}, {Takakuwa}, {Yen}, {Aso}, {Flores}, {Lee}, {Lai}, {Kwon}, {de Gregorio-Monsalvo}, {Sharma}, \& {Lee}}]{Narayanan_2023}
{Narayanan}, S., {Williams}, J.~P., {Tobin}, J.~J., {et~al.} 2023, \apj, 958, 20

\bibitem[{{Noble} {et~al.}(2012){Noble}, {Theule}, {Mispelaer}, {Duvernay}, {Danger}, {Congiu}, {Dulieu}, \& {Chiavassa}}]{Noble_2012}
{Noble}, J.~A., {Theule}, P., {Mispelaer}, F., {et~al.} 2012, \aap, 543, A5

\bibitem[{{{\"O}berg} \& {Bergin}(2021)}]{Oberg_2021}
{{\"O}berg}, K.~I. \& {Bergin}, E.~A. 2021, \physrep, 893, 1

\bibitem[{{{\"O}berg} {et~al.}(2023){{\"O}berg}, {Facchini}, \& {Anderson}}]{Oberg_2023}
{{\"O}berg}, K.~I., {Facchini}, S., \& {Anderson}, D.~E. 2023, \araa, 61, 287

\bibitem[{{{\"O}berg} {et~al.}(2017){{\"O}berg}, {Guzm{\'a}n}, {Merchantz}, {Qi}, {Andrews}, {Cleeves}, {Huang}, {Loomis}, {Wilner}, {Brinch}, \& {Hogerheijde}}]{Oberg_2017}
{{\"O}berg}, K.~I., {Guzm{\'a}n}, V.~V., {Merchantz}, C.~J., {et~al.} 2017, \apj, 839, 43

\bibitem[{{Offner} \& {Arce}(2014)}]{Offner_2014}
{Offner}, S. S.~R. \& {Arce}, H.~G. 2014, \apj, 784, 61

\bibitem[{{Ohashi} {et~al.}(2014){Ohashi}, {Saigo}, {Aso}, {Aikawa}, {Koyamatsu}, {Machida}, {Saito}, {Takahashi}, {Takakuwa}, {Tomida}, {Tomisaka}, \& {Yen}}]{Ohashi_2014}
{Ohashi}, N., {Saigo}, K., {Aso}, Y., {et~al.} 2014, \apj, 796, 131

\bibitem[{{Ohashi} {et~al.}(2023){Ohashi}, {Tobin}, {J{\o}rgensen}, {Takakuwa}, {Sheehan}, {Aikawa}, {Li}, {Looney}, {Williams}, {Aso}, {Sharma}, {Sai}, {Yamato}, {Lee}, {Tomida}, {Yen}, {Encalada}, {Flores}, {Gavino}, {Kido}, {Han}, {Lin}, {Narayanan}, {Phuong}, {Santamar{\'\i}a-Miranda}, {Thieme}, {van't Hoff}, {de Gregorio-Monsalvo}, {Koch}, {Kwon}, {Lai}, {Lee}, {Plunkett}, {Saigo}, {Hirano}, {Lam}, \& {Mori}}]{Ohashi_2023}
{Ohashi}, N., {Tobin}, J.~J., {J{\o}rgensen}, J.~K., {et~al.} 2023, \apj, 951, 8

\bibitem[{{Oya} {et~al.}(2014){Oya}, {Sakai}, {Sakai}, {Watanabe}, {Hirota}, {Lindberg}, {Bisschop}, {J{\o}rgensen}, {van Dishoeck}, \& {Yamamoto}}]{Oya_2014}
{Oya}, Y., {Sakai}, N., {Sakai}, T., {et~al.} 2014, \apj, 795, 152

\bibitem[{{Penteado} {et~al.}(2017){Penteado}, {Walsh}, \& {Cuppen}}]{Penteado_2017}
{Penteado}, E.~M., {Walsh}, C., \& {Cuppen}, H.~M. 2017, \apj, 844, 71

\bibitem[{{Phuong} {et~al.}(2025){Phuong}, {Lee}, {Tobin}, {Ohashi}, {J{\o}rgensen}, {Takakuwa}, {Aikawa}, {Aso}, {Li}, {Koch}, {Williams}, {Gavino}, {Lin}, {Tomida}, {Kwon}, {Looney}, {Han}, {Santamar{\i}a-Miranda}, {Lai}, {Hsi-Wei}, {Thieme}, {Sai}, \& {Flores}}]{Phuong_2025}
{Phuong}, N.~T., {Lee}, C.~W., {Tobin}, J.~J., {et~al.} 2025, \apj~in press, arXiv:2508.07212

\bibitem[{{Pineda} {et~al.}(2023){Pineda}, {Arzoumanian}, {Andre}, {Friesen}, {Zavagno}, {Clarke}, {Inoue}, {Chen}, {Lee}, {Soler}, \& {Kuffmeier}}]{Pineda_2023}
{Pineda}, J.~E., {Arzoumanian}, D., {Andre}, P., {et~al.} 2023, in Astronomical Society of the Pacific Conference Series, Vol. 534, Protostars and Planets VII, ed. S.~{Inutsuka}, Y.~{Aikawa}, T.~{Muto}, K.~{Tomida}, \& M.~{Tamura}, 233

\bibitem[{{Pineda} {et~al.}(2020){Pineda}, {Segura-Cox}, {Caselli}, {Cunningham}, {Zhao}, {Schmiedeke}, {Maureira}, \& {Neri}}]{Pineda_2020}
{Pineda}, J.~E., {Segura-Cox}, D., {Caselli}, P., {et~al.} 2020, Nature Astronomy, 4, 1158

\bibitem[{{Plunkett} {et~al.}(2015){Plunkett}, {Arce}, {Mardones}, {van Dokkum}, {Dunham}, {Fern{\'a}ndez-L{\'o}pez}, {Gallardo}, \& {Corder}}]{Plunkett_2015}
{Plunkett}, A.~L., {Arce}, H.~G., {Mardones}, D., {et~al.} 2015, \nat, 527, 70

\bibitem[{{Podio} {et~al.}(2024){Podio}, {Ceccarelli}, {Codella}, {Sabatini}, {Segura-Cox}, {Balucani}, {Rimola}, {Ugliengo}, {Chandler}, {Sakai}, {Svoboda}, {Pineda}, {De Simone}, {Bianchi}, {Caselli}, {Isella}, {Aikawa}, {Bouvier}, {Caux}, {Chahine}, {Charnley}, {Cuello}, {Dulieu}, {Evans}, {Fedele}, {Feng}, {Fontani}, {Hama}, {Hanawa}, {Herbst}, {Hirota}, {Jim{\'e}nez-Serra}, {Johnstone}, {Lefloch}, {Le Gal}, {Loinard}, {Liu}, {L{\'o}pez-Sepulcre}, {Maud}, {Maureira}, {Menard}, {Miotello}, {Moellenbrock}, {Nomura}, {Oba}, {Ohashi}, {Okoda}, {Oya}, {Sakai}, {Shirley}, {Testi}, {Vastel}, {Viti}, {Watanabe}, {Watanabe}, {Zhang}, {Zhang}, \& {Yamamoto}}]{Podio_2024}
{Podio}, L., {Ceccarelli}, C., {Codella}, C., {et~al.} 2024, \aap, 688, L22

\bibitem[{{Podio} {et~al.}(2015){Podio}, {Codella}, {Gueth}, {Cabrit}, {Bachiller}, {Gusdorf}, {Lee}, {Lefloch}, {Leurini}, {Nisini}, \& {Tafalla}}]{Podio_2015}
{Podio}, L., {Codella}, C., {Gueth}, F., {et~al.} 2015, \aap, 581, A85

\bibitem[{{Pontoppidan} {et~al.}(2014){Pontoppidan}, {Salyk}, {Bergin}, {Brittain}, {Marty}, {Mousis}, \& {{\"O}berg}}]{Pontoppidan_2014}
{Pontoppidan}, K.~M., {Salyk}, C., {Bergin}, E.~A., {et~al.} 2014, in Protostars and Planets VI, ed. H.~{Beuther}, R.~S. {Klessen}, C.~P. {Dullemond}, \& T.~{Henning}, 363--385

\bibitem[{{Qasim} {et~al.}(2018){Qasim}, {Chuang}, {Fedoseev}, {Ioppolo}, {Boogert}, \& {Linnartz}}]{Qasim_2018}
{Qasim}, D., {Chuang}, K.~J., {Fedoseev}, G., {et~al.} 2018, \aap, 612, A83

\bibitem[{{Sai} {et~al.}(2023){Sai}, {Yen}, {Ohashi}, {Tobin}, {J{\o}rgensen}, {Takakuwa}, {Saigo}, {Aso}, {Lin}, {Koch}, {Aikawa}, {Flores}, {de Gregorio-Monsalvo}, {Han}, {Kido}, {Kwon}, {Lai}, {Lee}, {Lee}, {Li}, {Looney}, {Mori}, {Phuong}, {Santamar{\'\i}a-Miranda}, {Sharma}, {Thieme}, {Tomida}, \& {Williams}}]{Sai_2023}
{Sai}, J., {Yen}, H.-W., {Ohashi}, N., {et~al.} 2023, \apj, 954, 67

\bibitem[{{Sakai} {et~al.}(2017){Sakai}, {Oya}, {Higuchi}, {Aikawa}, {Hanawa}, {Ceccarelli}, {Lefloch}, {L{\'o}pez-Sepulcre}, {Watanabe}, {Sakai}, {Hirota}, {Caux}, {Vastel}, {Kahane}, \& {Yamamoto}}]{Sakai_2017}
{Sakai}, N., {Oya}, Y., {Higuchi}, A.~E., {et~al.} 2017, \mnras, 467, L76

\bibitem[{{Sakai} {et~al.}(2014{\natexlab{a}}){Sakai}, {Oya}, {Sakai}, {Watanabe}, {Hirota}, {Ceccarelli}, {Kahane}, {Lopez-Sepulcre}, {Lefloch}, {Vastel}, {Bottinelli}, {Caux}, {Coutens}, {Aikawa}, {Takakuwa}, {Ohashi}, {Yen}, \& {Yamamoto}}]{Sakai_2014b}
{Sakai}, N., {Oya}, Y., {Sakai}, T., {et~al.} 2014{\natexlab{a}}, \apjl, 791, L38

\bibitem[{{Sakai} {et~al.}(2014{\natexlab{b}}){Sakai}, {Sakai}, {Hirota}, {Watanabe}, {Ceccarelli}, {Kahane}, {Bottinelli}, {Caux}, {Demyk}, {Vastel}, {Coutens}, {Taquet}, {Ohashi}, {Takakuwa}, {Yen}, {Aikawa}, \& {Yamamoto}}]{Sakai_2014}
{Sakai}, N., {Sakai}, T., {Hirota}, T., {et~al.} 2014{\natexlab{b}}, \nat, 507, 78

\bibitem[{{Santamar{\'\i}a-Miranda} {et~al.}(2024){Santamar{\'\i}a-Miranda}, {de Gregorio-Monsalvo}, {Ohashi}, {Tobin}, {Sai}, {J{\o}rgensen}, {Aso}, {Daniel Lin}, {Flores}, {Kido}, {Koch}, {Kwon}, {Lee}, {Li}, {Looney}, {Plunkett}, {Takakuwa}, {R van't Hoff}, {Williams}, \& {Yen}}]{Santamaria_2024}
{Santamar{\'\i}a-Miranda}, A., {de Gregorio-Monsalvo}, I., {Ohashi}, N., {et~al.} 2024, \aap, 690, A46

\bibitem[{{Sharma} {et~al.}(2023){Sharma}, {J{\o}rgensen}, {Gavino}, {Ohashi}, {Tobin}, {Lin}, {Li}, {Takakuwa}, {Lee}, {Sai (Insa Choi)}, {Kwon}, {de Gregorio-Monsalvo}, {Santamar{\'\i}a-Miranda}, {Yen}, {Aikawa}, {Aso}, {Lai}, {Lee}, {Looney}, {Phuong}, {Thieme}, \& {Williams}}]{Sharma_2023}
{Sharma}, R., {J{\o}rgensen}, J.~K., {Gavino}, S., {et~al.} 2023, \apj, 954, 69

\bibitem[{{Sharma} {et~al.}(2020){Sharma}, {Tobin}, {Sheehan}, {Megeath}, {Fischer}, {J{\o}rgensen}, {Safron}, \& {Nagy}}]{Sharma_2020}
{Sharma}, R., {Tobin}, J.~J., {Sheehan}, P.~D., {et~al.} 2020, \apj, 904, 78

\bibitem[{{Simons} {et~al.}(2020){Simons}, {Lamberts}, \& {Cuppen}}]{Simons_2020}
{Simons}, M.~A.~J., {Lamberts}, T., \& {Cuppen}, H.~M. 2020, \aap, 634, A52

\bibitem[{{Snyder} {et~al.}(1969){Snyder}, {Buhl}, {Zuckerman}, \& {Palmer}}]{Snyder_1969}
{Snyder}, L.~E., {Buhl}, D., {Zuckerman}, B., \& {Palmer}, P. 1969, \prl, 22, 679

\bibitem[{{Tabone} {et~al.}(2017){Tabone}, {Cabrit}, {Bianchi}, {Ferreira}, {Pineau des For{\^e}ts}, {Codella}, {Gusdorf}, {Gueth}, {Podio}, \& {Chapillon}}]{Tabone_2017}
{Tabone}, B., {Cabrit}, S., {Bianchi}, E., {et~al.} 2017, \aap, 607, L6

\bibitem[{{Tafalla} {et~al.}(2010){Tafalla}, {Santiago-Garc{\'\i}a}, {Hacar}, \& {Bachiller}}]{Tafalla_2010}
{Tafalla}, M., {Santiago-Garc{\'\i}a}, J., {Hacar}, A., \& {Bachiller}, R. 2010, \aap, 522, A91

\bibitem[{{Takahashi} {et~al.}(2024){Takahashi}, {Machida}, {Omura}, {Johnstone}, {Saigo}, {Harada}, {Tomisaka}, {Ho}, {Zapata}, {Mairs}, {Herczeg}, {Taniguchi}, {Liu}, \& {Sato}}]{Takahashi_2024}
{Takahashi}, S., {Machida}, M.~N., {Omura}, M., {et~al.} 2024, \apj, 964, 48

\bibitem[{{Takakuwa} {et~al.}(2024){Takakuwa}, {Saigo}, {Kido}, {Ohashi}, {Tobin}, {J{\o}rgensen}, {Aikawa}, {Aso}, {Gavino}, {Han}, {Koch}, {Kwon}, {Lee}, {Lee}, {Li}, {Lin}, {Looney}, {Mori}, {Sai}, {Sharma}, {Sheehan}, {Tomida}, {Williams}, {Yamato}, \& {Yen}}]{Takakuwa_2024}
{Takakuwa}, S., {Saigo}, K., {Kido}, M., {et~al.} 2024, \apj, 964, 24

\bibitem[{{Tang} {et~al.}(2017){Tang}, {Henkel}, {Chen}, {Menten}, {Indebetouw}, {Zheng}, {Esimbek}, {Zhou}, {Yuan}, {Li}, \& {He}}]{Tang_2017}
{Tang}, X.~D., {Henkel}, C., {Chen}, C. H.~R., {et~al.} 2017, \aap, 600, A16

\bibitem[{{Testi} {et~al.}(2014){Testi}, {Birnstiel}, {Ricci}, {Andrews}, {Blum}, {Carpenter}, {Dominik}, {Isella}, {Natta}, {Williams}, \& {Wilner}}]{Testi_2014}
{Testi}, L., {Birnstiel}, T., {Ricci}, L., {et~al.} 2014, in Protostars and Planets VI, ed. H.~{Beuther}, R.~S. {Klessen}, C.~P. {Dullemond}, \& T.~{Henning}, 339--361

\bibitem[{{Thieme} {et~al.}(2022){Thieme}, {Lai}, {Lin}, {Cheong}, {Lee}, {Yen}, {Li}, {Lam}, \& {Zhao}}]{Thieme_2022}
{Thieme}, T.~J., {Lai}, S.-P., {Lin}, S.-J., {et~al.} 2022, \apj, 925, 32

\bibitem[{{Thieme} {et~al.}(2023){Thieme}, {Lai}, {Ohashi}, {Tobin}, {J{\o}rgensen}, {Sai}, {Aso}, {Williams}, {Yamato}, {Aikawa}, {de Gregorio-Monsalvo}, {Han}, {Kwon}, {Lee}, {Lee}, {Li}, {Lin}, {Looney}, {Narayanan}, {Phuong}, {Plunkett}, {Santamar{\'\i}a-Miranda}, {Sharma}, {Takakuwa}, \& {Yen}}]{Thieme_2023}
{Thieme}, T.~J., {Lai}, S.-P., {Ohashi}, N., {et~al.} 2023, \apj, 958, 60

\bibitem[{{Turner}(2001)}]{Turner_2001}
{Turner}, B.~E. 2001, \apjs, 136, 579

\bibitem[{{Tychoniec} {et~al.}(2019){Tychoniec}, {Hull}, {Kristensen}, {Tobin}, {Le Gouellec}, \& {van Dishoeck}}]{Tychoniec_2019}
{Tychoniec}, {\L}., {Hull}, C. L.~H., {Kristensen}, L.~E., {et~al.} 2019, \aap, 632, A101

\bibitem[{{Tychoniec} {et~al.}(2021){Tychoniec}, {van Dishoeck}, {van't Hoff}, {van Gelder}, {Tabone}, {Chen}, {Harsono}, {Hull}, {Hogerheijde}, {Murillo}, \& {Tobin}}]{Tychoniec_2021}
{Tychoniec}, {\L}., {van Dishoeck}, E.~F., {van't Hoff}, M. L.~R., {et~al.} 2021, \aap, 655, A65

\bibitem[{{Valdivia-Mena} {et~al.}(2024){Valdivia-Mena}, {Pineda}, {Caselli}, {Segura-Cox}, {Schmiedeke}, {Spezzano}, {Offner}, {Ivlev}, {Kuffmeier}, {Cunningham}, {Neri}, \& {Maureira}}]{Valdivia-Mena_2024}
{Valdivia-Mena}, M.~T., {Pineda}, J.~E., {Caselli}, P., {et~al.} 2024, \aap, 687, A71

\bibitem[{{Valdivia-Mena} {et~al.}(2022){Valdivia-Mena}, {Pineda}, {Segura-Cox}, {Caselli}, {Neri}, {L{\'o}pez-Sepulcre}, {Cunningham}, {Bouscasse}, {Semenov}, {Henning}, {Pi{\'e}tu}, {Chapillon}, {Dutrey}, {Fuente}, {Guilloteau}, {Hsieh}, {Jim{\'e}nez-Serra}, {Marino}, {Maureira}, {Smirnov-Pinchukov}, {Tafalla}, \& {Zhao}}]{Valdivia-Mena_2022}
{Valdivia-Mena}, M.~T., {Pineda}, J.~E., {Segura-Cox}, D.~M., {et~al.} 2022, \aap, 667, A12

\bibitem[{{Valdivia-Mena} {et~al.}(2023){Valdivia-Mena}, {Pineda}, {Segura-Cox}, {Caselli}, {Schmiedeke}, {Choudhury}, {Offner}, {Neri}, {Goodman}, \& {Fuller}}]{Valdivia-Mena_2023}
{Valdivia-Mena}, M.~T., {Pineda}, J.~E., {Segura-Cox}, D.~M., {et~al.} 2023, \aap, 677, A92

\bibitem[{{van der Marel} {et~al.}(2014){van der Marel}, {van Dishoeck}, {Bruderer}, \& {van Kempen}}]{Marel_2014}
{van der Marel}, N., {van Dishoeck}, E.~F., {Bruderer}, S., \& {van Kempen}, T.~A. 2014, \aap, 563, A113

\bibitem[{{van der Tak} {et~al.}(2007){van der Tak}, {Black}, {Sch{\"o}ier}, {Jansen}, \& {van Dishoeck}}]{Tak_2007}
{van der Tak}, F.~F.~S., {Black}, J.~H., {Sch{\"o}ier}, F.~L., {Jansen}, D.~J., \& {van Dishoeck}, E.~F. 2007, \aap, 468, 627

\bibitem[{{van der Wiel} {et~al.}(2009){van der Wiel}, {van der Tak}, {Ossenkopf}, {Spaans}, {Roberts}, {Fuller}, \& {Plume}}]{vanderWiel_2009}
{van der Wiel}, M.~H.~D., {van der Tak}, F.~F.~S., {Ossenkopf}, V., {et~al.} 2009, \aap, 498, 161

\bibitem[{{van Dishoeck} \& {Blake}(1998)}]{vanDishoeck_1998}
{van Dishoeck}, E.~F. \& {Blake}, G.~A. 1998, \araa, 36, 317

\bibitem[{{van Gelder} {et~al.}(2021){van Gelder}, {Tabone}, {van Dishoeck}, \& {Godard}}]{vanGelder_2021}
{van Gelder}, M.~L., {Tabone}, B., {van Dishoeck}, E.~F., \& {Godard}, B. 2021, \aap, 653, A159

\bibitem[{{van 't Hoff} {et~al.}(2018){van 't Hoff}, {Tobin}, {Harsono}, \& {van Dishoeck}}]{Hoff_2018}
{van 't Hoff}, M. L.~R., {Tobin}, J.~J., {Harsono}, D., \& {van Dishoeck}, E.~F. 2018, \aap, 615, A83

\bibitem[{{van't Hoff} {et~al.}(2024){van't Hoff}, {Bergin}, {Riley}, {Mittal}, {J{\o}rgensen}, \& {Tobin}}]{Hoff_2024}
{van't Hoff}, M. L.~R., {Bergin}, E.~A., {Riley}, P., {et~al.} 2024, \apj, 970, 138

\bibitem[{{van't Hoff} {et~al.}(2023){van't Hoff}, {Tobin}, {Li}, {Ohashi}, {J{\o}rgensen}, {Lin}, {Aikawa}, {Aso}, {de Gregorio-Monsalvo}, {Gavino}, {Han}, {Koch}, {Kwon}, {Lee}, {Lee}, {Looney}, {Narayanan}, {Plunkett}, {Sai Insa Choi}, {Santamar{\'\i}a-Miranda}, {Sharma}, {Sheehan}, {Takakuwa}, {Thieme}, {Williams}, {Lai}, {Phuong}, \& {Yen}}]{Hoff_2023}
{van't Hoff}, M. L.~R., {Tobin}, J.~J., {Li}, Z.-Y., {et~al.} 2023, \apj, 951, 10

\bibitem[{{Visser} {et~al.}(2012){Visser}, {Kristensen}, {Bruderer}, {van Dishoeck}, {Herczeg}, {Brinch}, {Doty}, {Harsono}, \& {Wolfire}}]{Visser_2012}
{Visser}, R., {Kristensen}, L.~E., {Bruderer}, S., {et~al.} 2012, \aap, 537, A55

\bibitem[{{Vorobyov} {et~al.}(2018){Vorobyov}, {Elbakyan}, {Plunkett}, {Dunham}, {Audard}, {Guedel}, \& {Dionatos}}]{Vorobyov_2018}
{Vorobyov}, E.~I., {Elbakyan}, V.~G., {Plunkett}, A.~L., {et~al.} 2018, \aap, 613, A18

\bibitem[{{Wakelam} {et~al.}(2005){Wakelam}, {Ceccarelli}, {Castets}, {Lefloch}, {Loinard}, {Faure}, {Schneider}, \& {Benayoun}}]{Wakelam_2005}
{Wakelam}, V., {Ceccarelli}, C., {Castets}, A., {et~al.} 2005, \aap, 437, 149

\bibitem[{{Wang} {et~al.}(2004){Wang}, {Mundt}, {Henning}, \& {Apai}}]{Wang_2004}
{Wang}, H., {Mundt}, R., {Henning}, T., \& {Apai}, D. 2004, \apj, 617, 1191

\bibitem[{{Watanabe} \& {Kouchi}(2002)}]{Watanabe_2002}
{Watanabe}, N. \& {Kouchi}, A. 2002, \apjl, 571, L173

\bibitem[{{Willacy}(2007)}]{Willacy_2007}
{Willacy}, K. 2007, \apj, 660, 441

\bibitem[{{Wilson} \& {Rood}(1994)}]{Wilson_1994}
{Wilson}, T.~L. \& {Rood}, R. 1994, \araa, 32, 191

\bibitem[{{Yamato} {et~al.}(2023){Yamato}, {Aikawa}, {Ohashi}, {Tobin}, {J{\o}rgensen}, {Takakuwa}, {Aso}, {Sai}, {Flores}, {de Gregorio-Monsalvo}, {Hirano}, {Han}, {Kido}, {Koch}, {Kwon}, {Lai}, {Lee}, {Lee}, {Li}, {Lin}, {Looney}, {Mori}, {Narayanan}, {Phuong}, {Saigo}, {Santamar{\'\i}a-Miranda}, {Sharma}, {Thieme}, {Tomida}, {van't Hoff}, \& {Yen}}]{Yamato_2023}
{Yamato}, Y., {Aikawa}, Y., {Ohashi}, N., {et~al.} 2023, \apj, 951, 11

\bibitem[{{Yen} {et~al.}(2019){Yen}, {Gu}, {Hirano}, {Koch}, {Lee}, {Liu}, \& {Takakuwa}}]{Yen_2019}
{Yen}, H.-W., {Gu}, P.-G., {Hirano}, N., {et~al.} 2019, \apj, 880, 69

\bibitem[{{Yen} {et~al.}(2017){Yen}, {Koch}, {Takakuwa}, {Krasnopolsky}, {Ohashi}, \& {Aso}}]{Yen_2017}
{Yen}, H.-W., {Koch}, P.~M., {Takakuwa}, S., {et~al.} 2017, \apj, 834, 178

\bibitem[{{Yen} {et~al.}(2015){Yen}, {Takakuwa}, {Koch}, {Aso}, {Koyamatsu}, {Krasnopolsky}, \& {Ohashi}}]{Yen_2015}
{Yen}, H.-W., {Takakuwa}, S., {Koch}, P.~M., {et~al.} 2015, \apj, 812, 129

\end{thebibliography}

\appendix
\onecolumn
\section{Large-scale moment 8 and moment 9 maps}\label{appendix:large_scale_emissions}
\begin{figure*}[htbp]
  \includegraphics[width=1.0\linewidth]{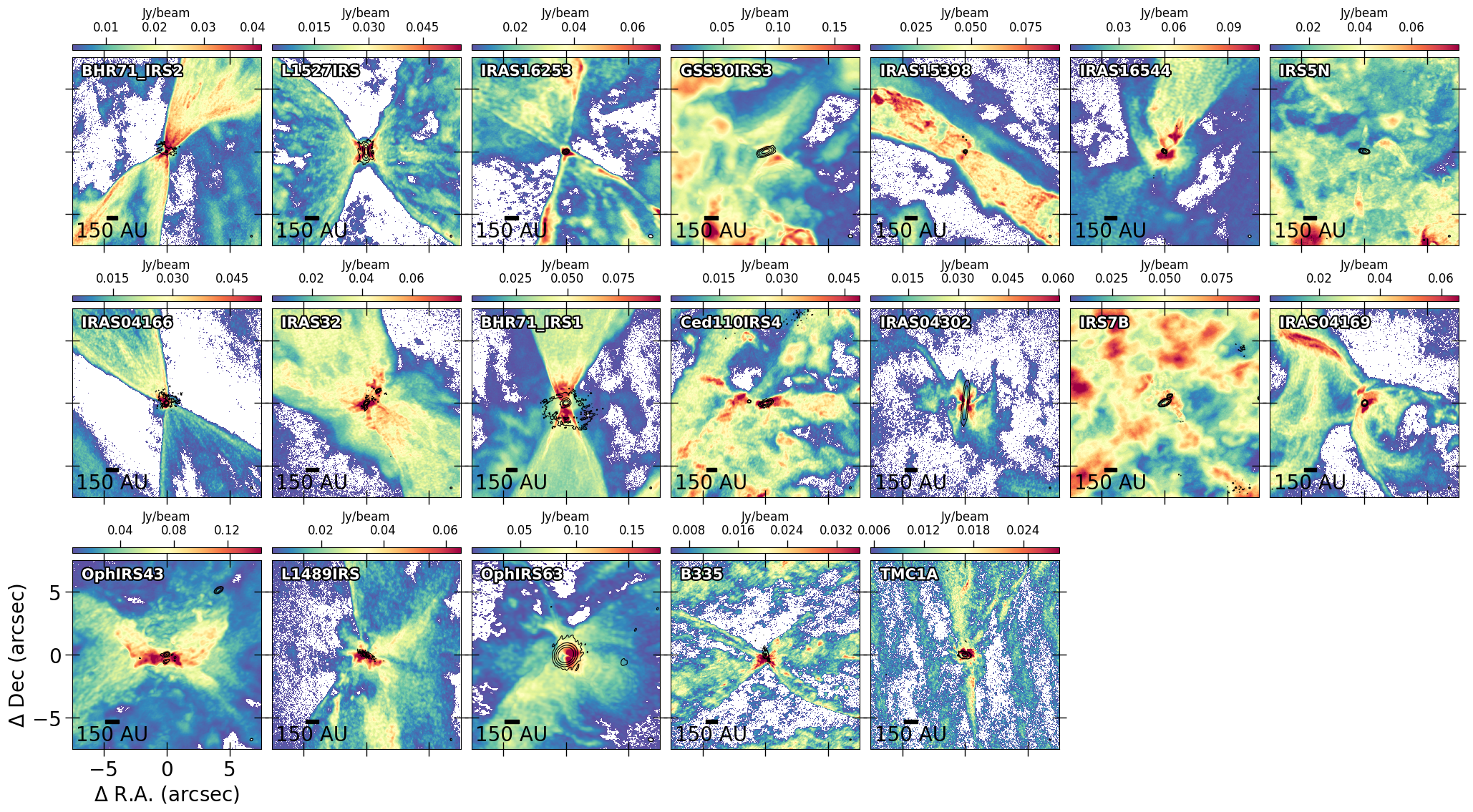}
  \includegraphics[width=1.0\linewidth]{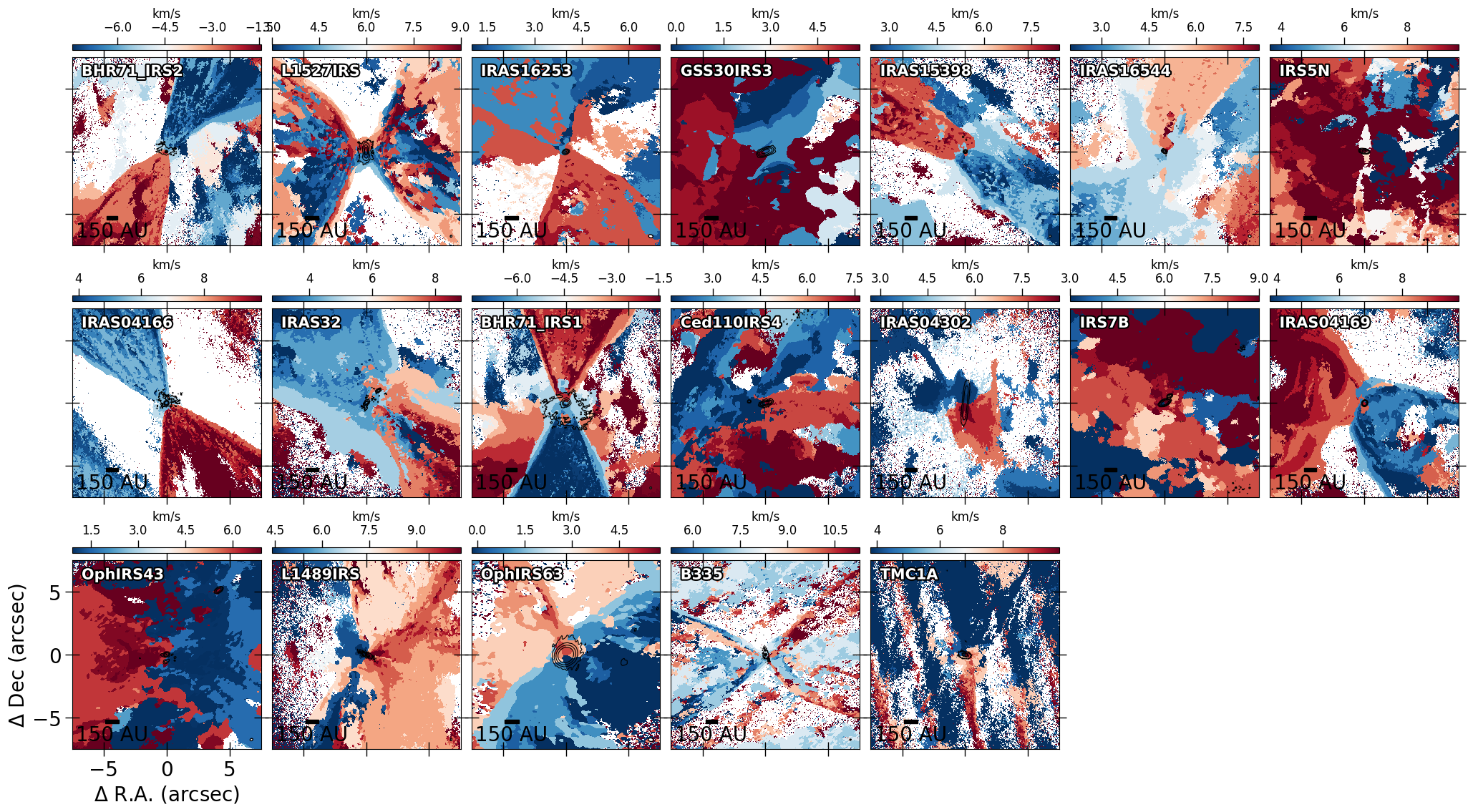} 
  \caption{Moment 8 (\textit{top}) and moment 9 (\textit{bottom}) maps depicting the \tlvco~(2--1) emission in the inner 15\asec region of the nineteen eDisk sources. Moment maps were generated by integrating the regions where $I_{\nu} > 3\sigma$, where $\sigma$ is the rms per channel. The sources are arranged in ascending order of $L_{bol}$ except for B335 and TMC1A, the two sources taken from the archive. The contour lines display the continuum emission at thresholds of 5$\sigma$, 20$\sigma$, 80$\sigma$, and 320$\sigma$ for each source. The scale bar located at the bottom left shows the 150 au scale in each source, and the synthesized beam is indicated in white at the bottom-right corner of each image. \label{appendix:12co_mom8_mom9}}
\end{figure*}
\newpage

\begin{figure*}[ht]
  \includegraphics[width=1.0\linewidth]{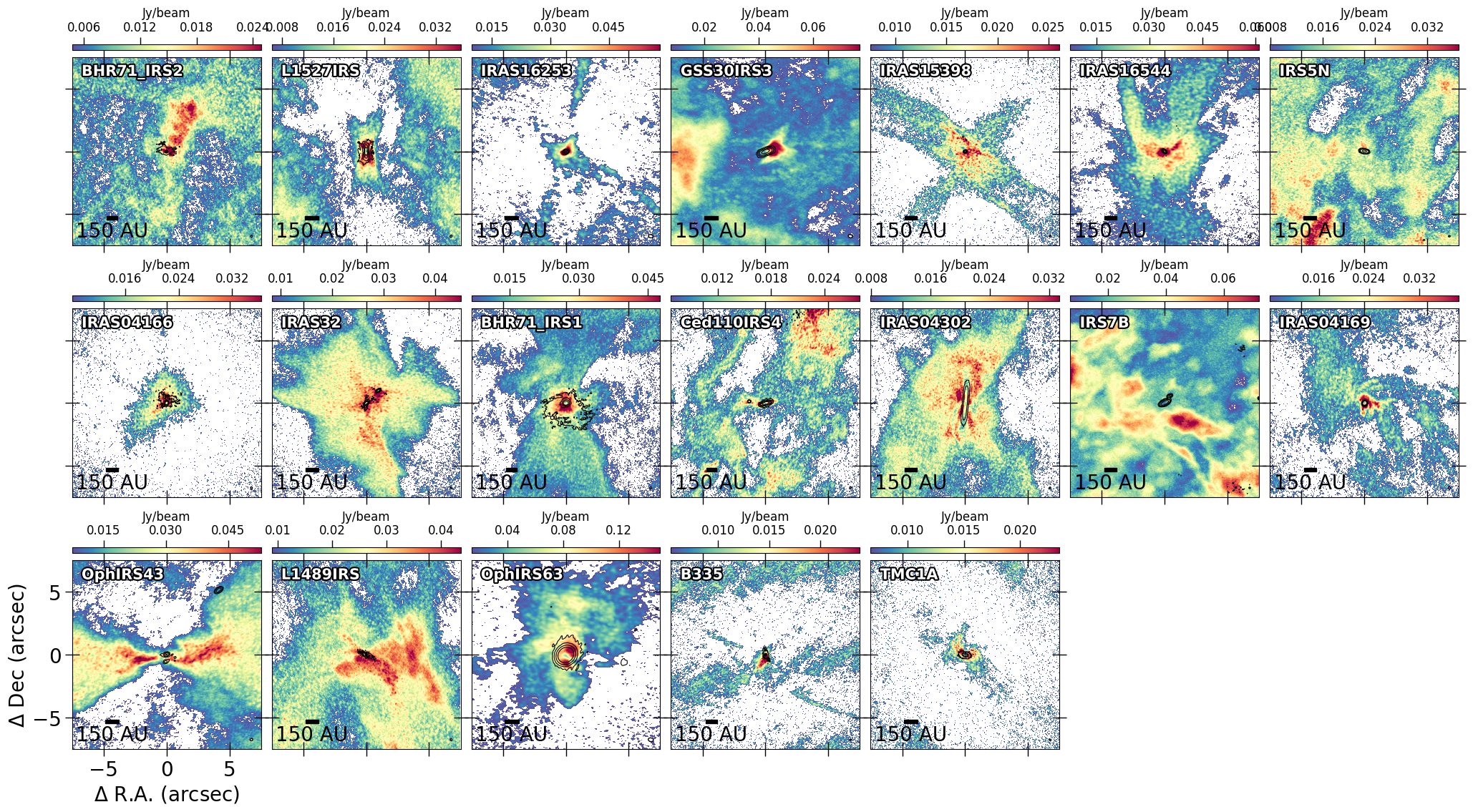}
  \includegraphics[width=1.0\linewidth]{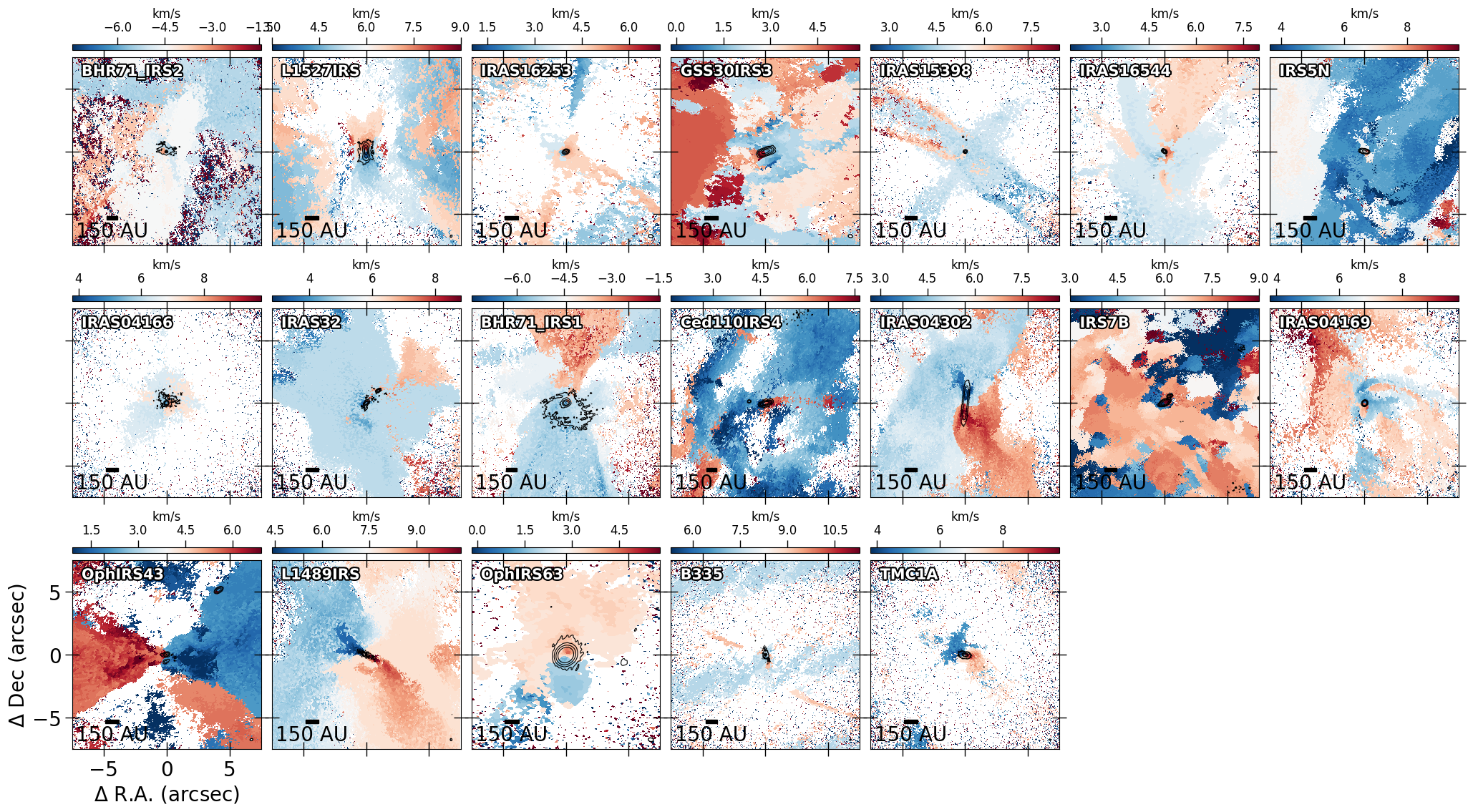}
  \caption{Same as Figure~\ref{appendix:12co_mom8_mom9} but for \thrco~($2$--$1$) instead. \label{appendix:13co_mom8_mom9}}
\end{figure*}
\newpage

\begin{figure*}[htbp]
  \includegraphics[width=1.0\linewidth]{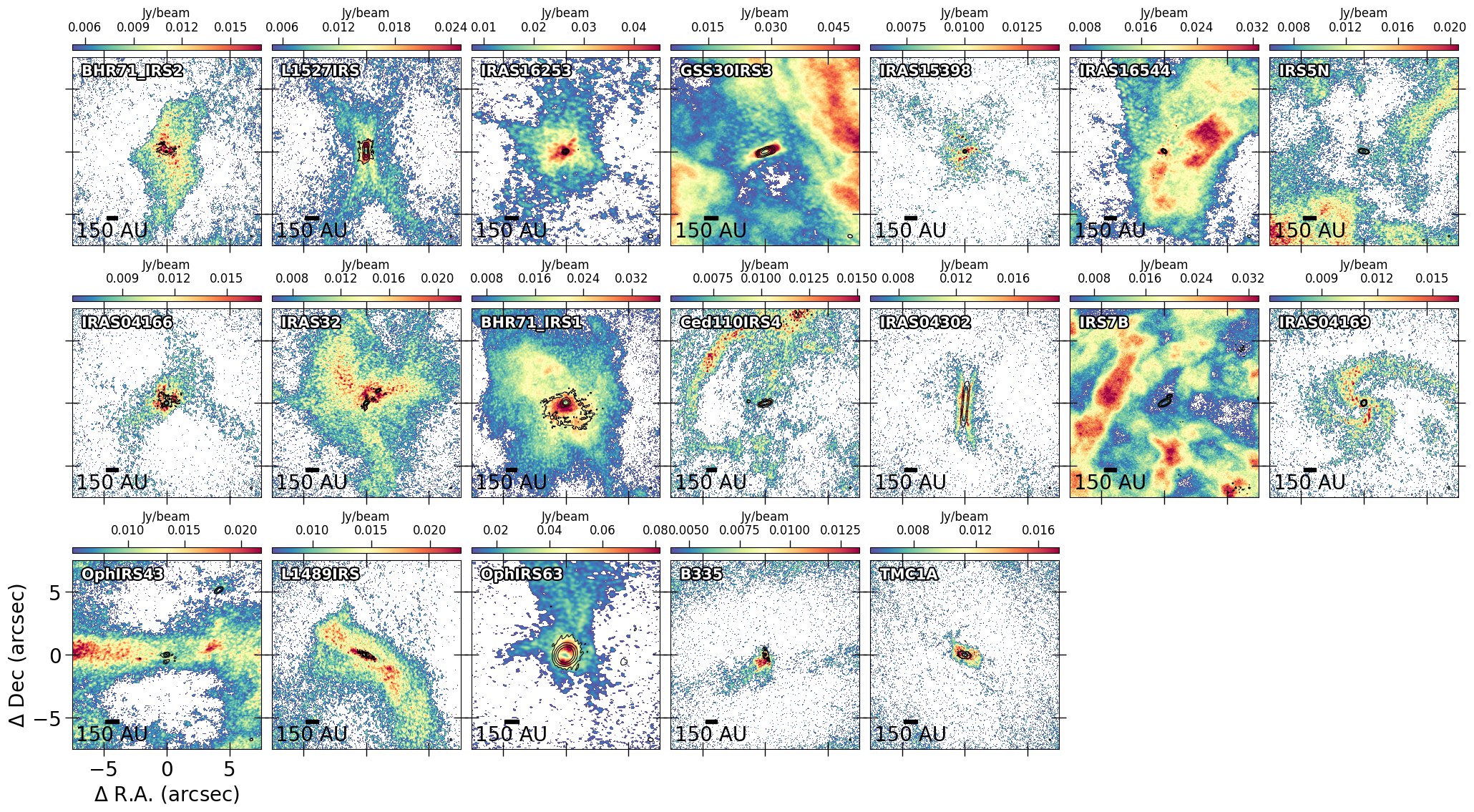}
  \includegraphics[width=1.0\linewidth]{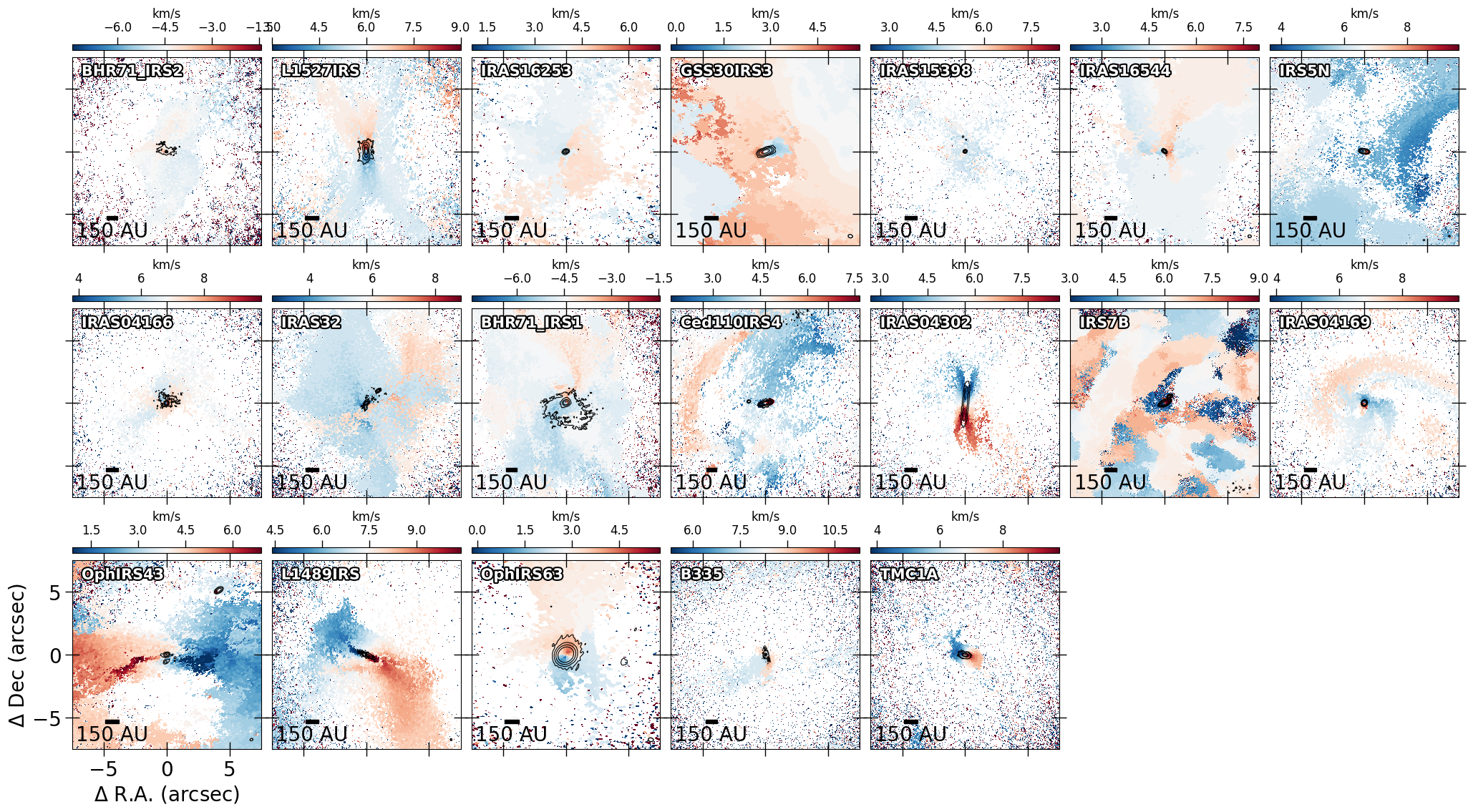}
  \caption{Same as Figure~\ref{appendix:12co_mom8_mom9} but for \ceteno~($2$--$1$) instead. \label{appendix:c18o_mom8_mom9}}
\end{figure*}
\newpage

\begin{figure*}[htbp]
  \includegraphics[width=1.0\linewidth]{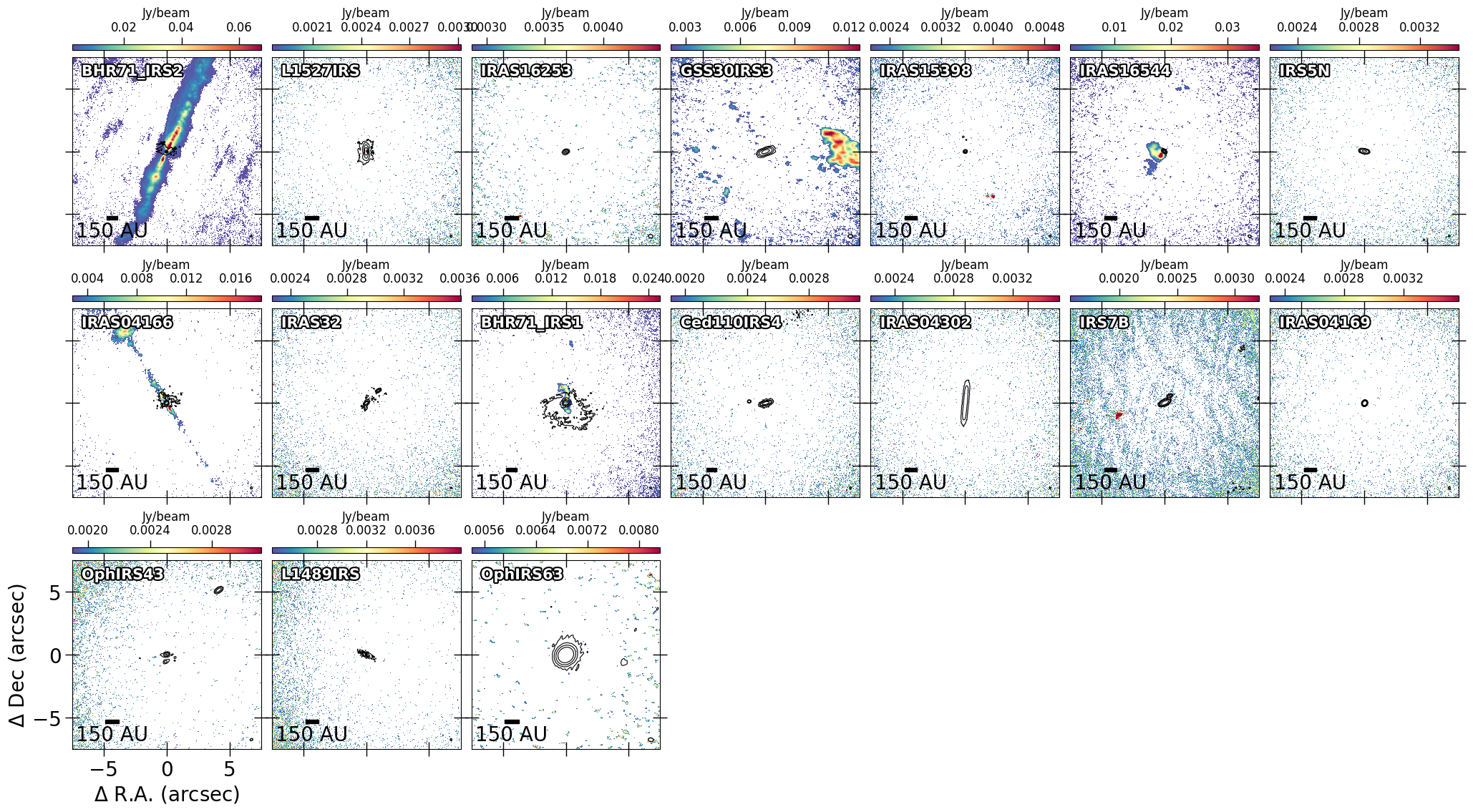} 
  \includegraphics[width=1.0\linewidth]{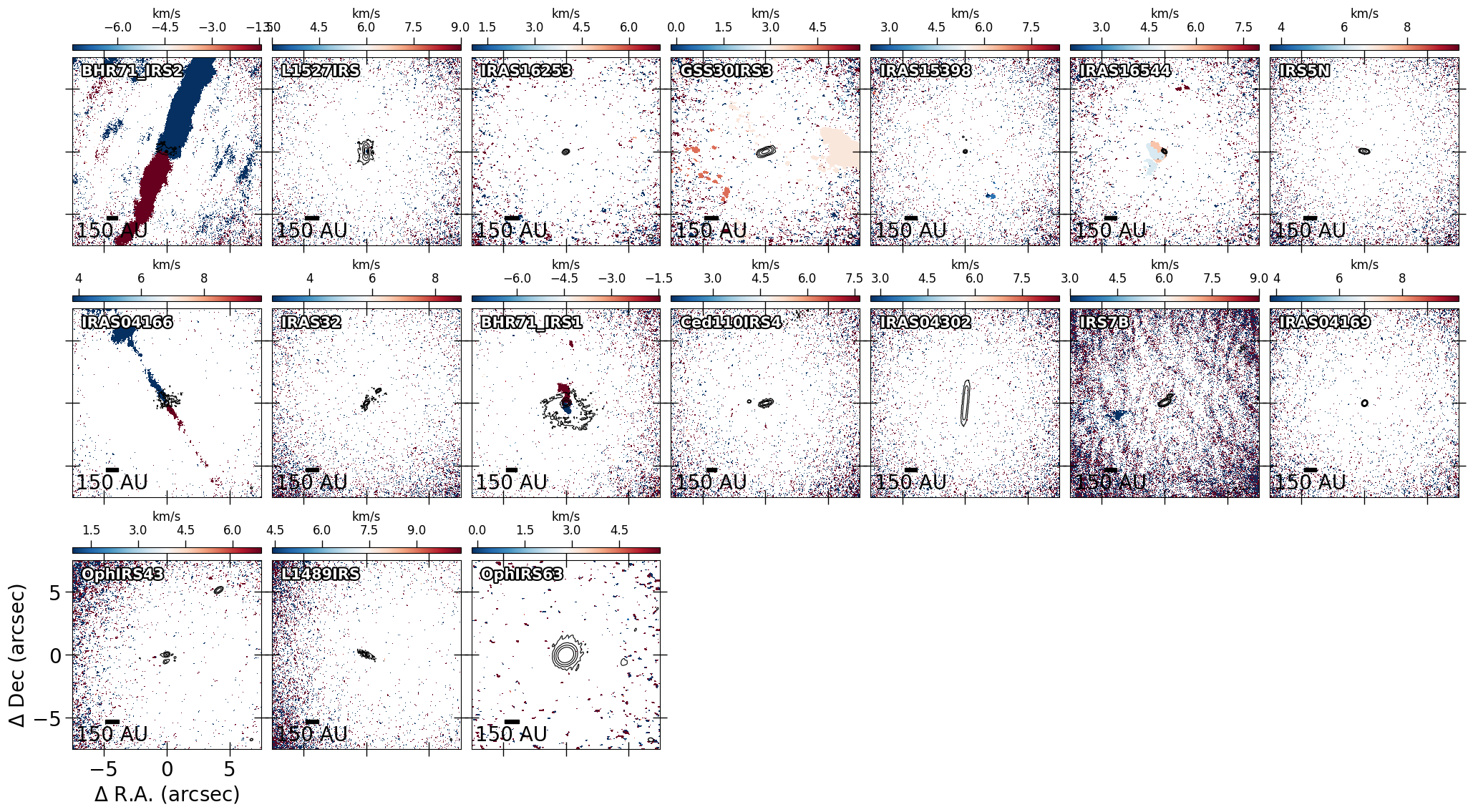}
  \caption{Same as Figure~\ref{appendix:12co_mom8_mom9} but for SiO~(5--4) instead. \label{appendix:sio_mom8_mom9}}
\end{figure*}
\newpage

\begin{figure*}[htbp]
  \includegraphics[width=1.0\linewidth]{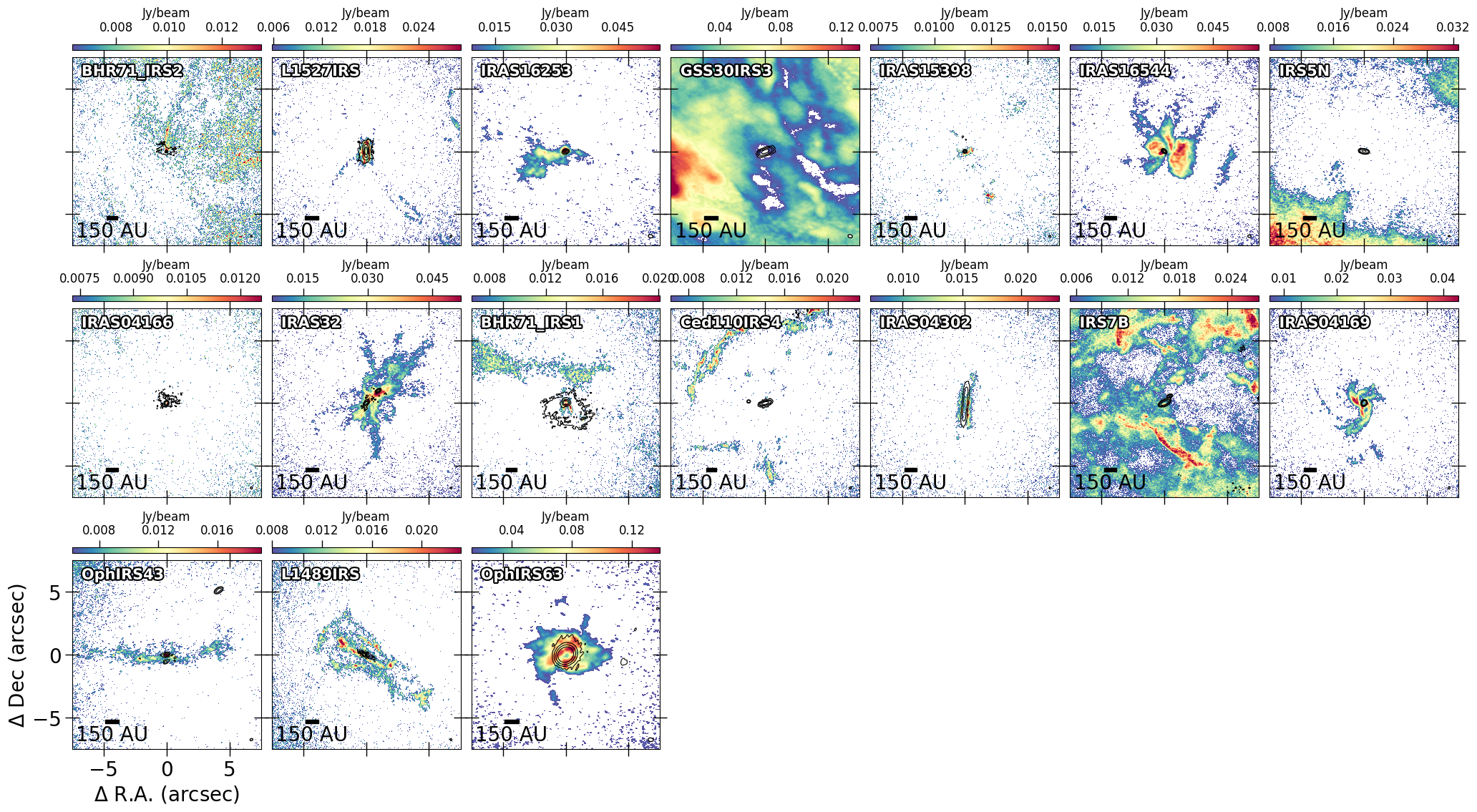}
  \includegraphics[width=1.0\linewidth]{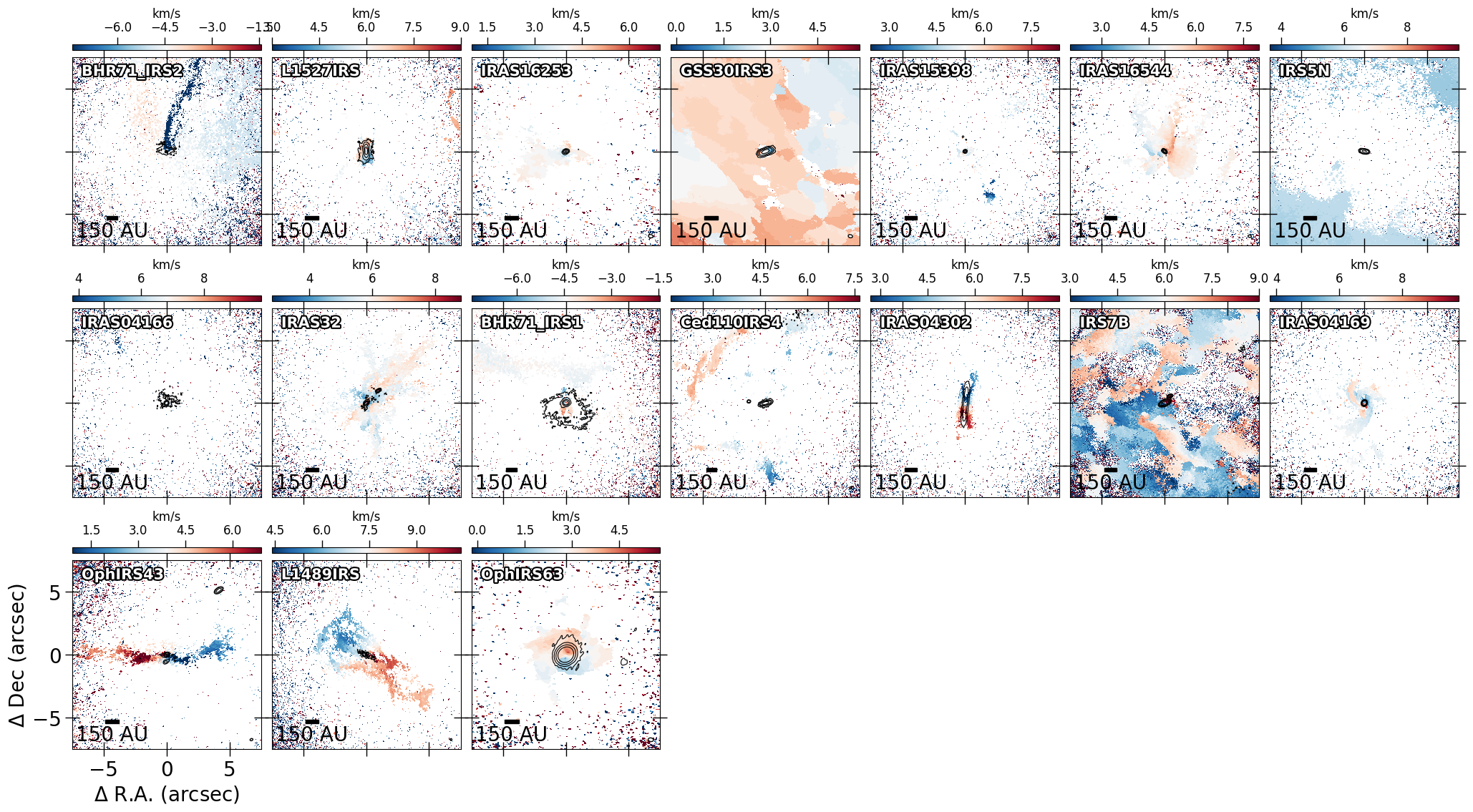}
  \caption{Same as Figure~\ref{appendix:12co_mom8_mom9} but for SO~($6_5$--$5_4$) instead. \label{appendix:so_mom8_mom9}}
\end{figure*}
\newpage

\begin{figure*}[htbp]
  \includegraphics[width=1.0\linewidth]{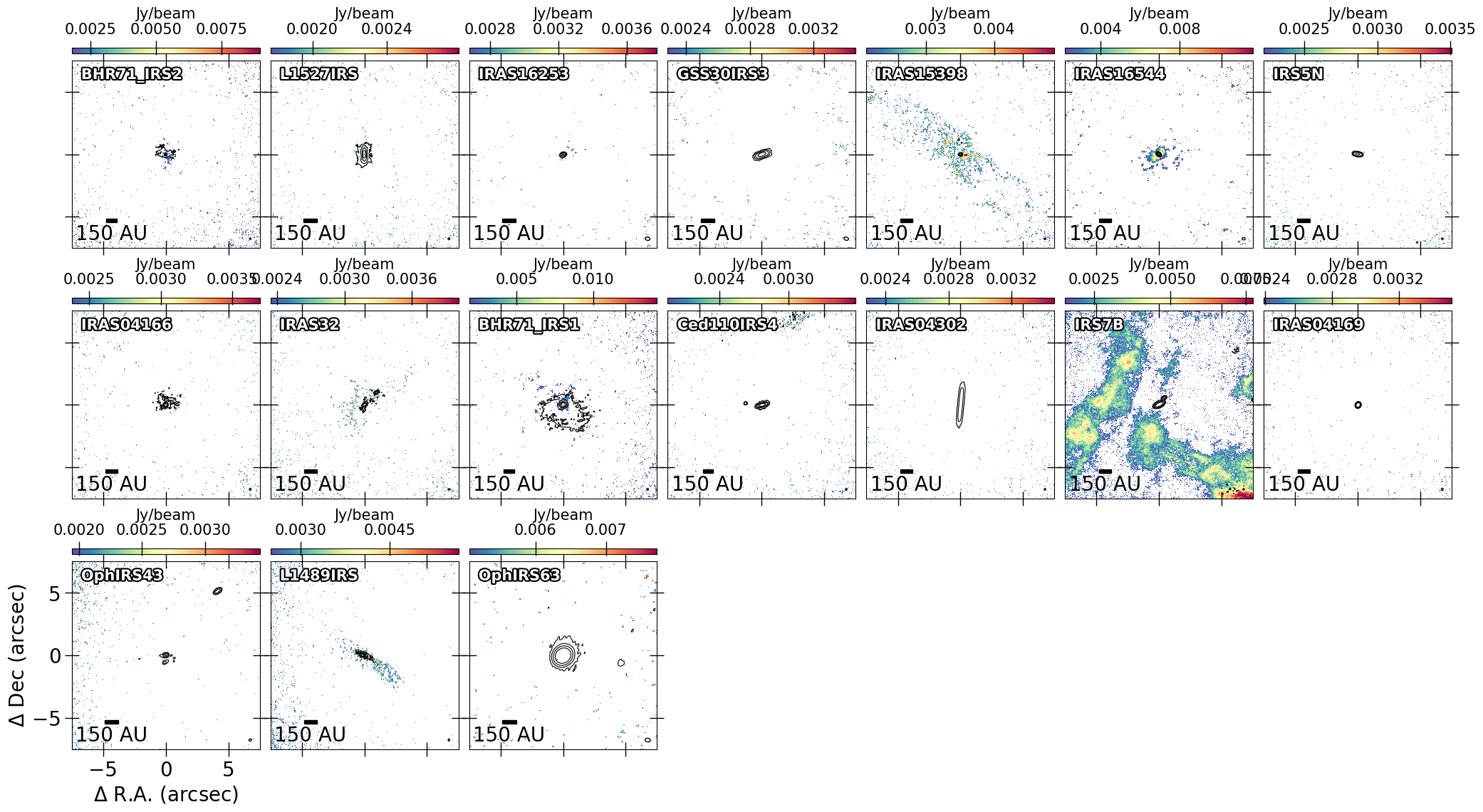}
  \includegraphics[width=1.0\linewidth]{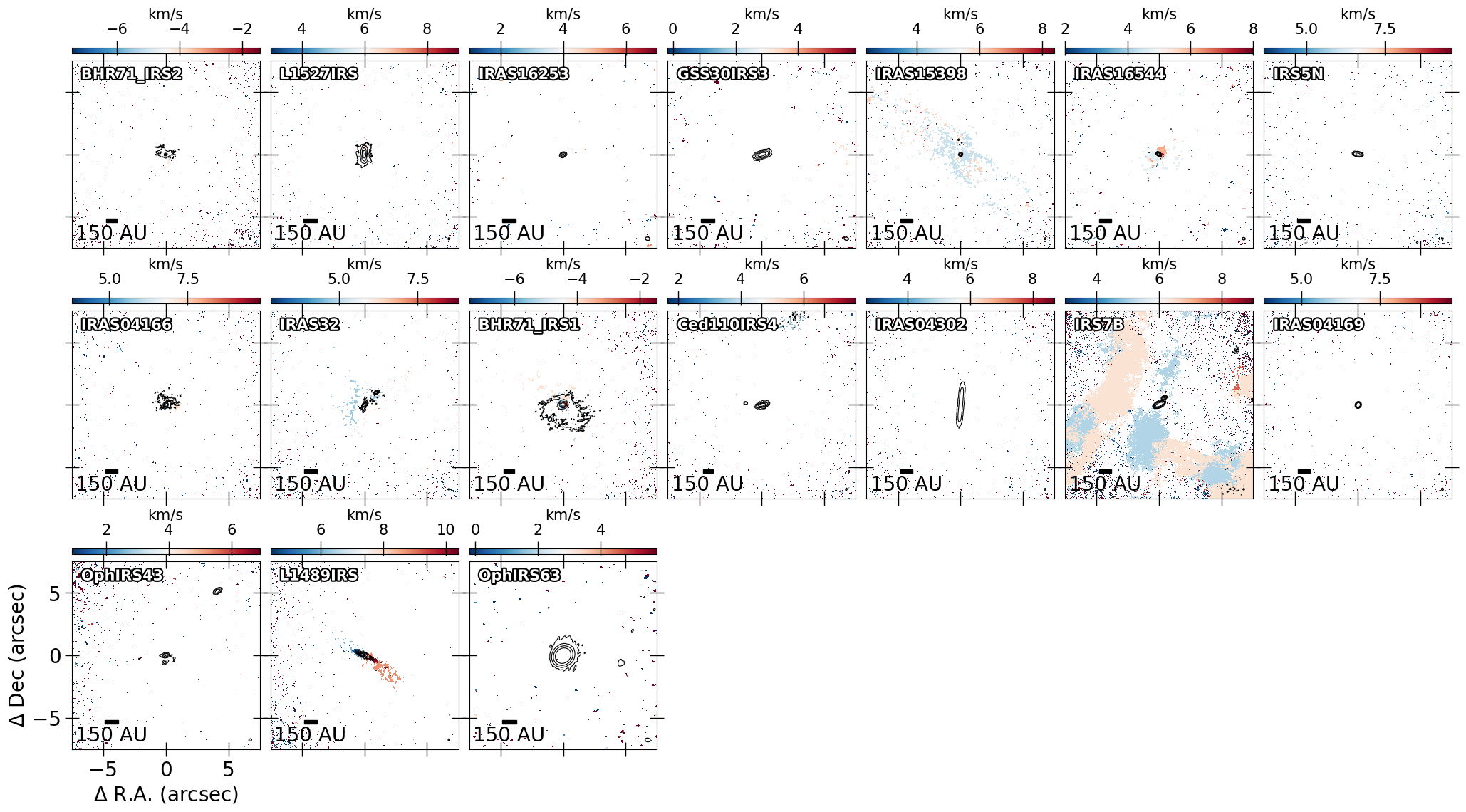}
  \caption{Same as Figure~\ref{appendix:12co_mom8_mom9} but for DCN~($3$--$2$) instead. \label{appendix:dcn_mom8_mom9}}
\end{figure*}
\newpage

\begin{figure*}[htbp]
  \includegraphics[width=1.0\linewidth]{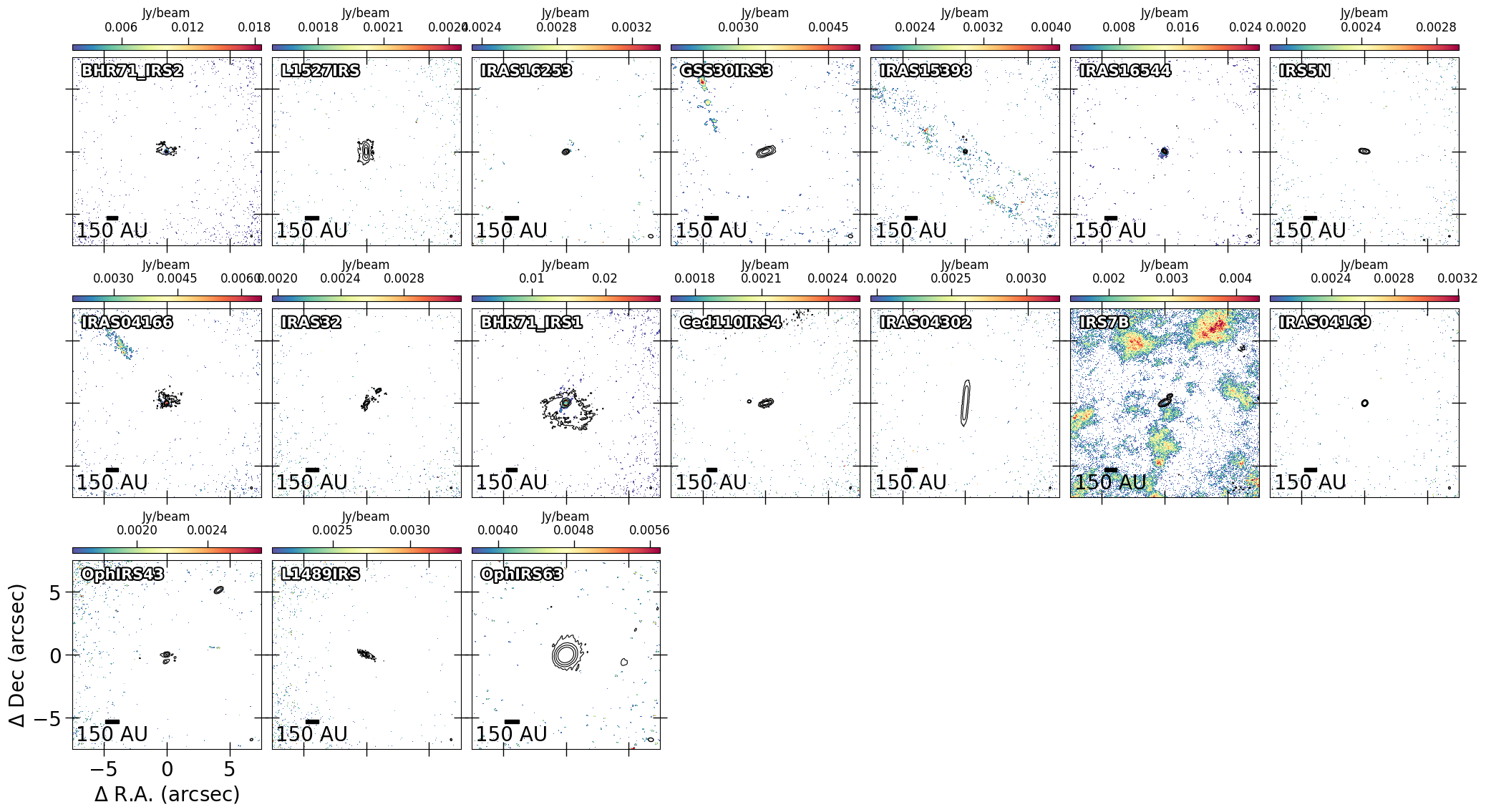}
  \includegraphics[width=1.0\linewidth]{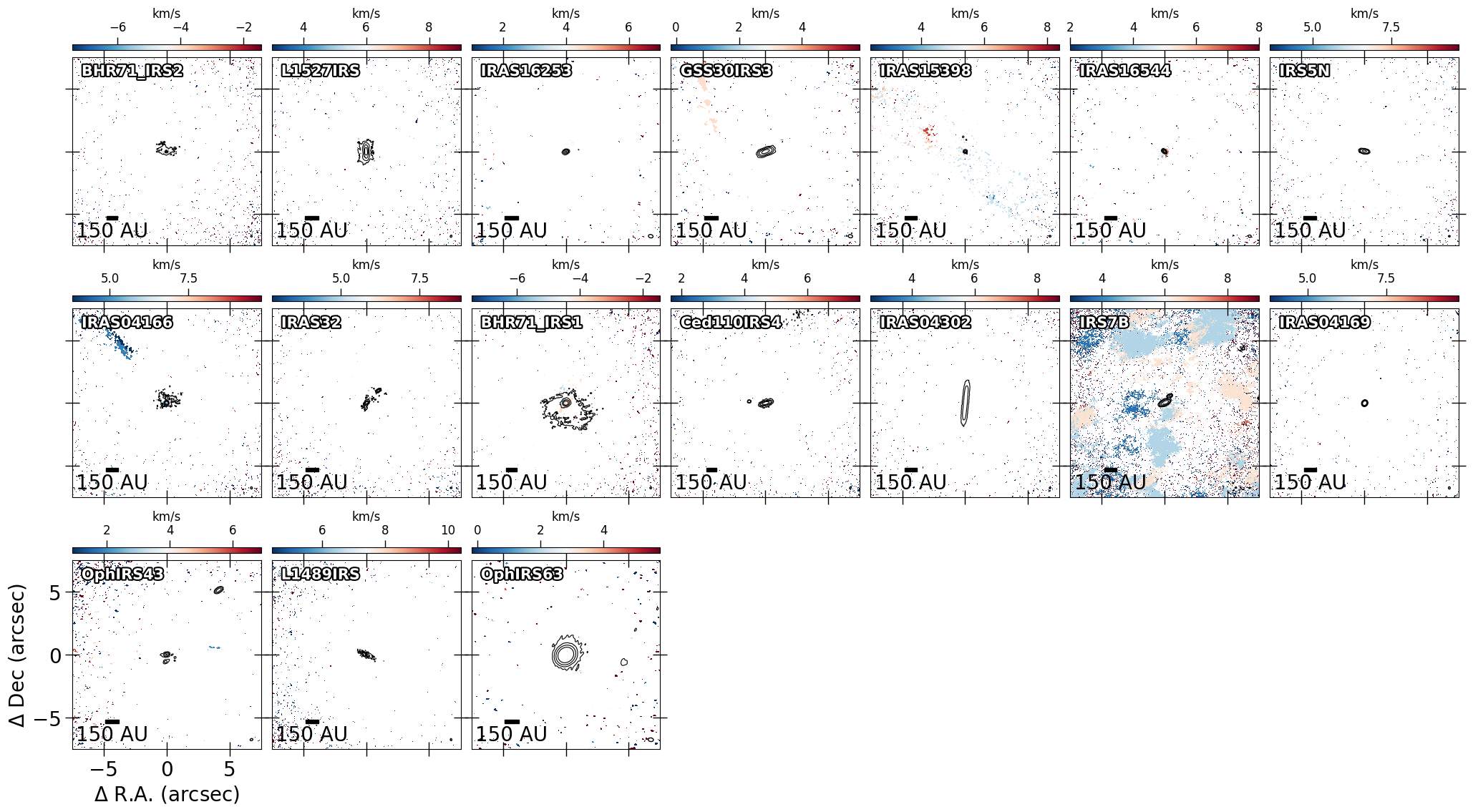}
  \caption{Same as Figure~\ref{appendix:12co_mom8_mom9} but for \ch3oh~($4_2$--$3_1$) instead. \label{appendix:ch3oh_mom8_mom9}}
\end{figure*}
\newpage

\begin{figure*}[htbp]
  \includegraphics[width=1.0\linewidth]{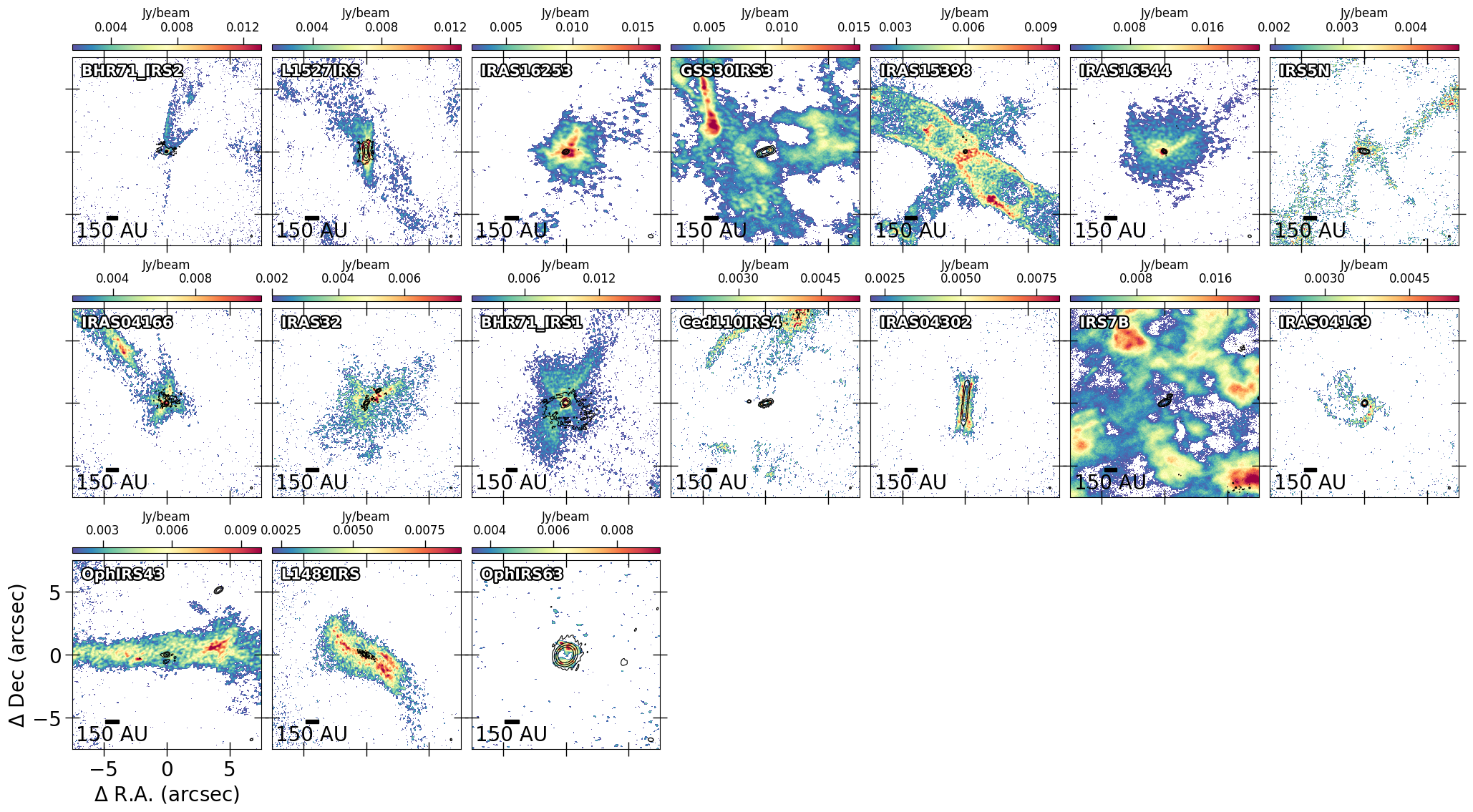}
  \includegraphics[width=1.0\linewidth]{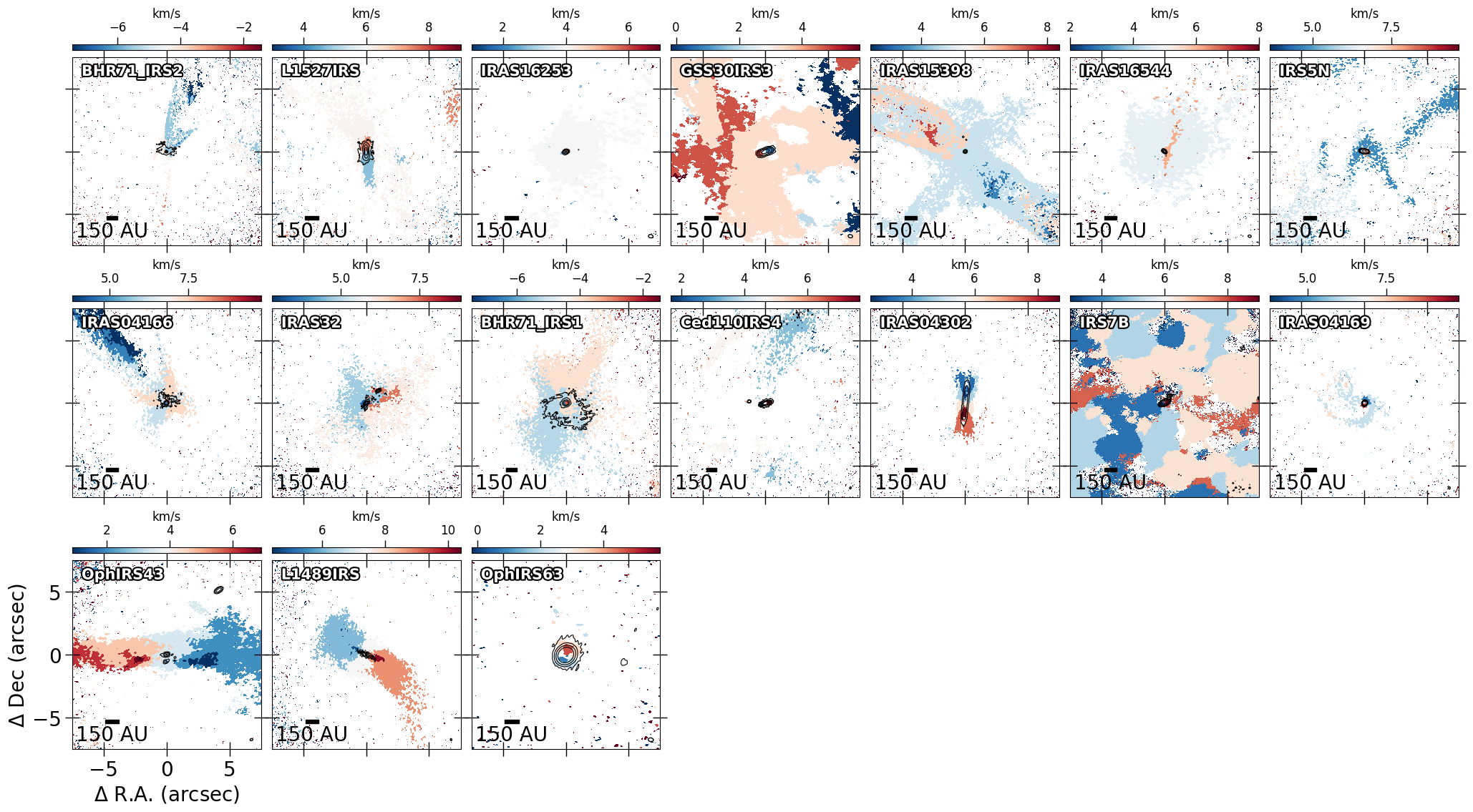}
  \caption{Same as Figure~\ref{appendix:12co_mom8_mom9} but for \h2co~($3_{0,3}$--$2_{0,2}$) instead. \label{appendix:h2co_3_03-2_02_mom8_mom9}}
\end{figure*}
\newpage

\begin{figure*}[htbp]
  \includegraphics[width=1.0\linewidth]{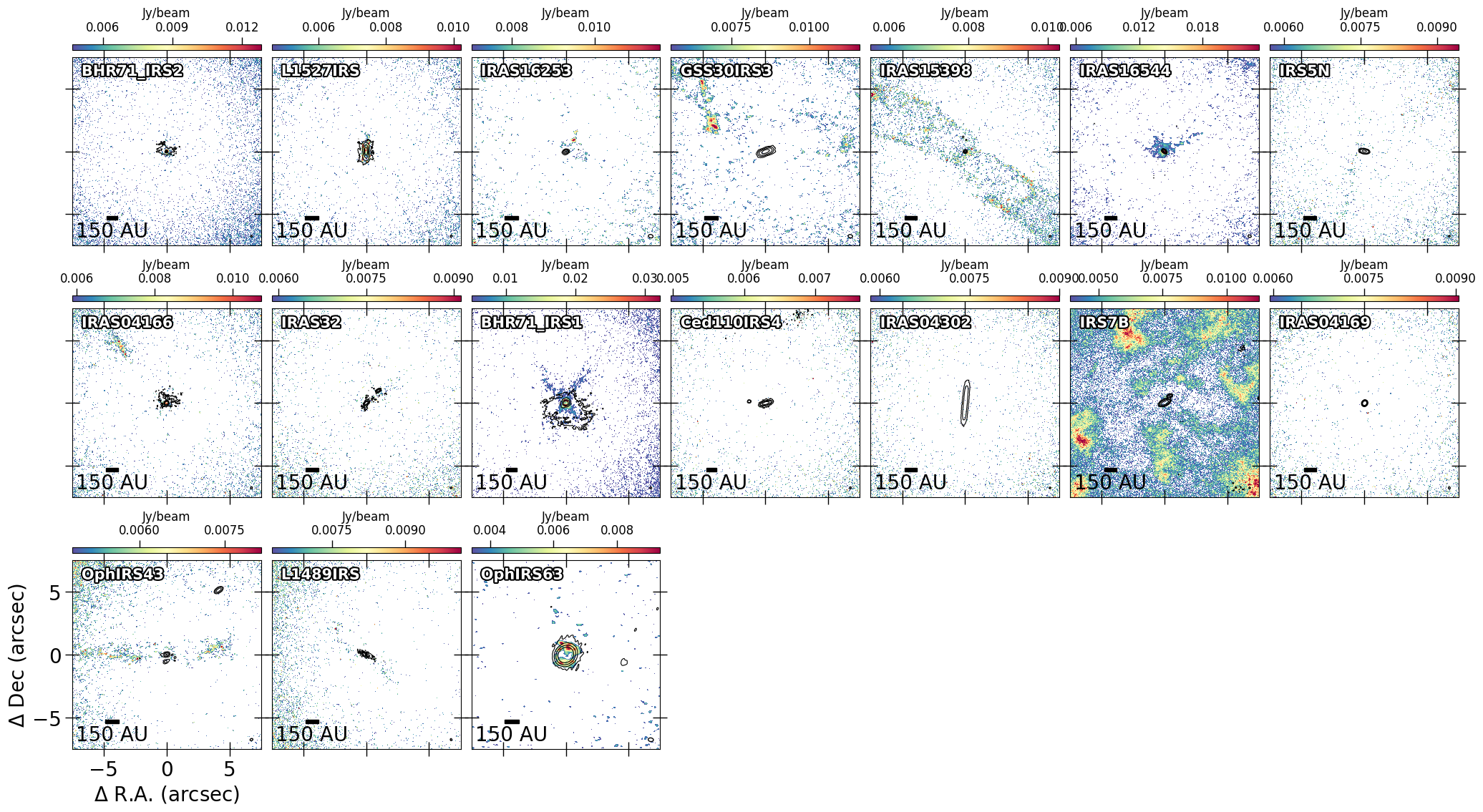}
  \includegraphics[width=1.0\linewidth]{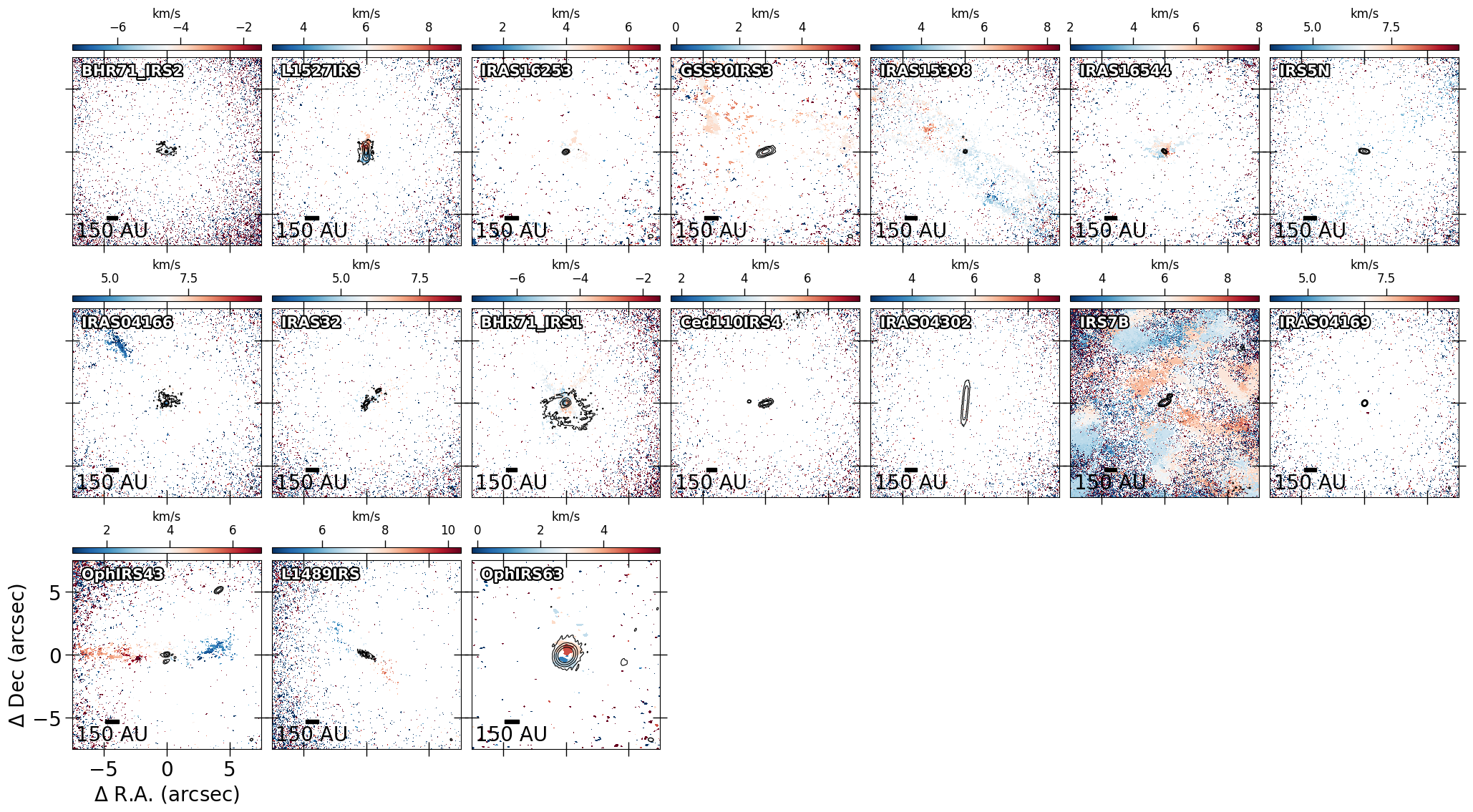}
  \caption{Same as Figure~\ref{appendix:12co_mom8_mom9} but for \h2co~($3_{2,1}$--$2_{2,0}$) instead. \label{appendix:h2co_3_21-2_20_mom8_mom9}}
\end{figure*}
\newpage

\begin{figure*}[htbp]
  \includegraphics[width=1.0\linewidth]{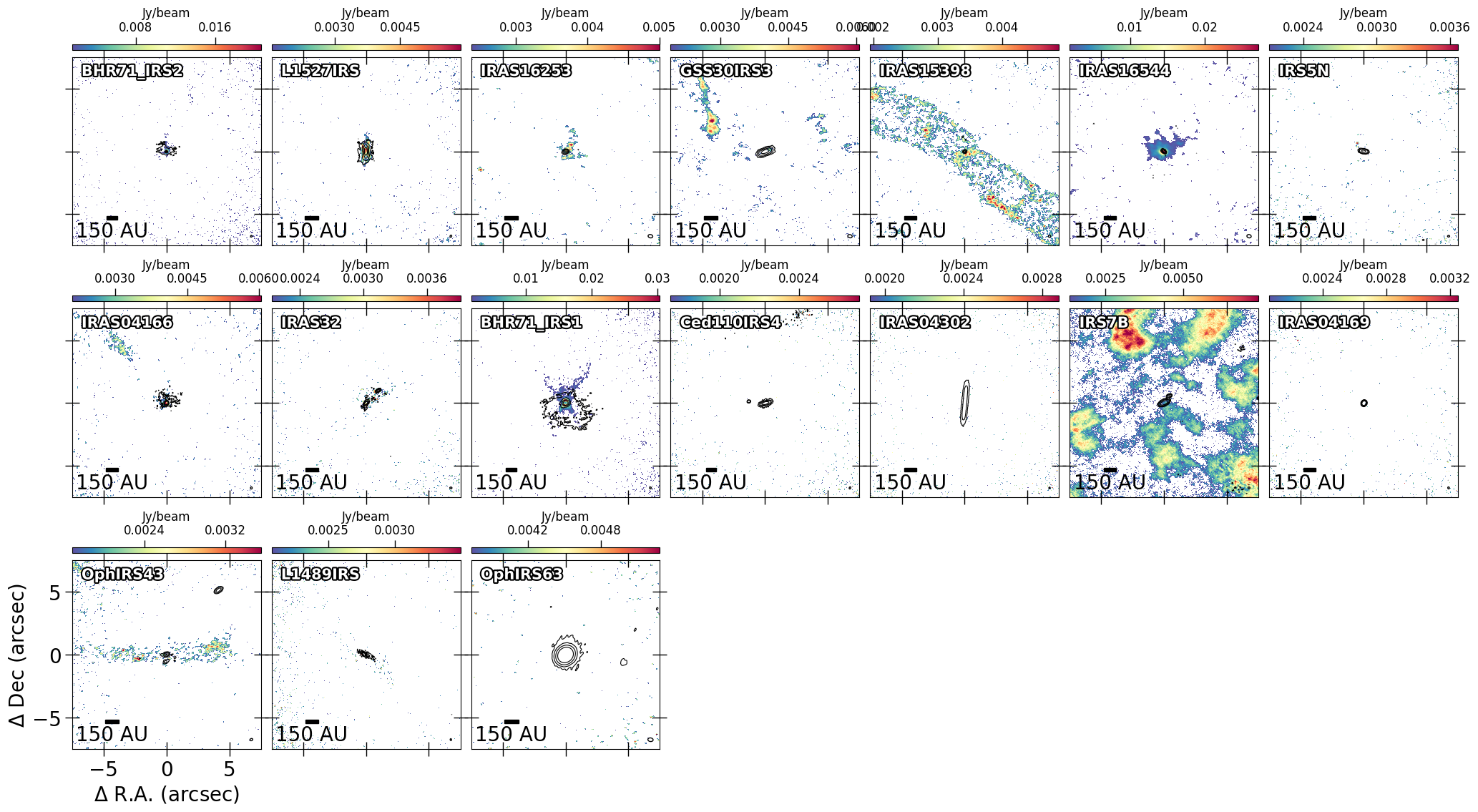}
  \includegraphics[width=1.0\linewidth]{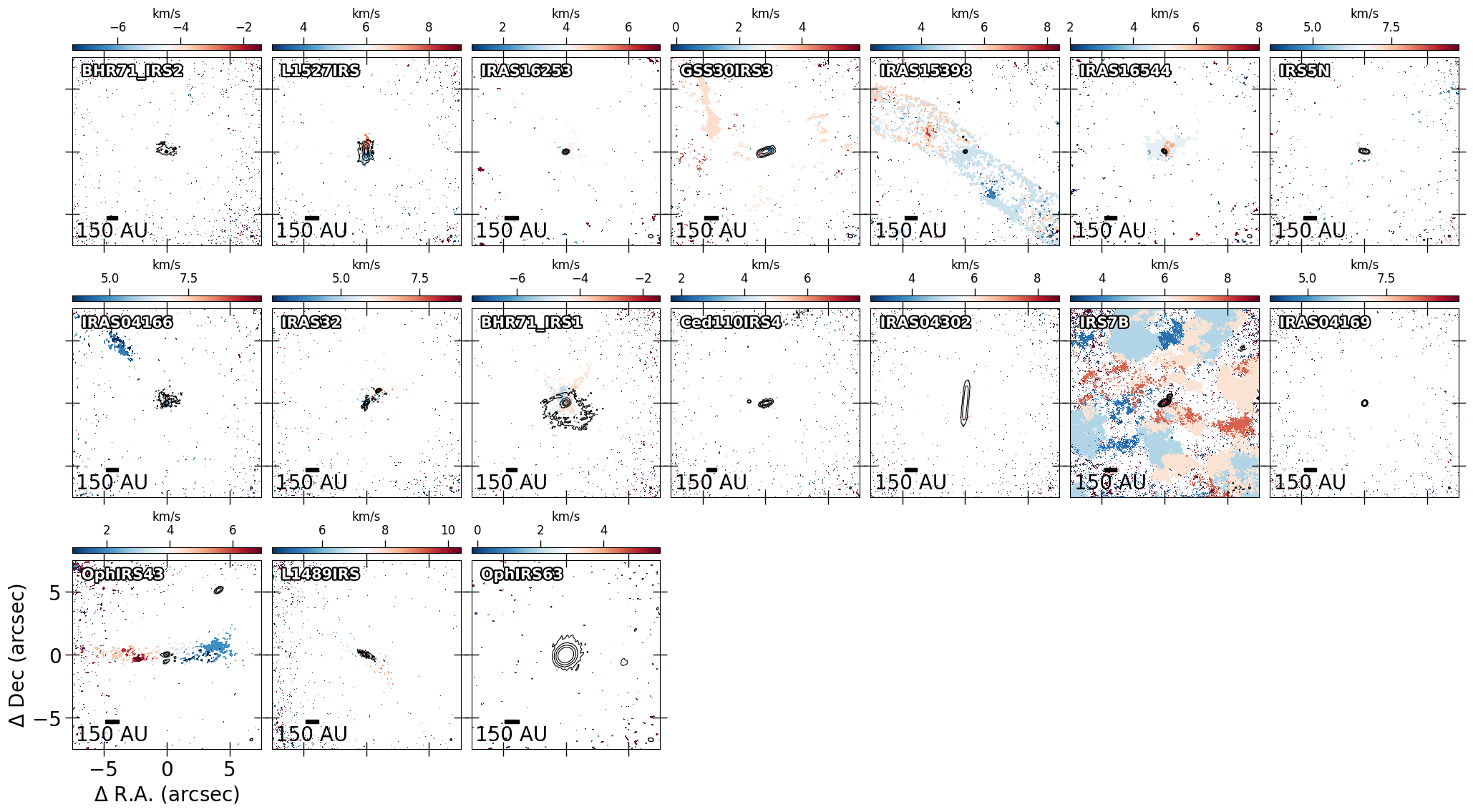}
  \caption{Same as Figure~\ref{appendix:12co_mom8_mom9} but for \h2co~($3_{2,2}$--$2_{2,1}$) instead. \label{appendix:h2co_3_22-2_21_mom8_mom9}}
\end{figure*}
\newpage

\begin{figure*}[htbp]
  \includegraphics[width=1.0\linewidth]{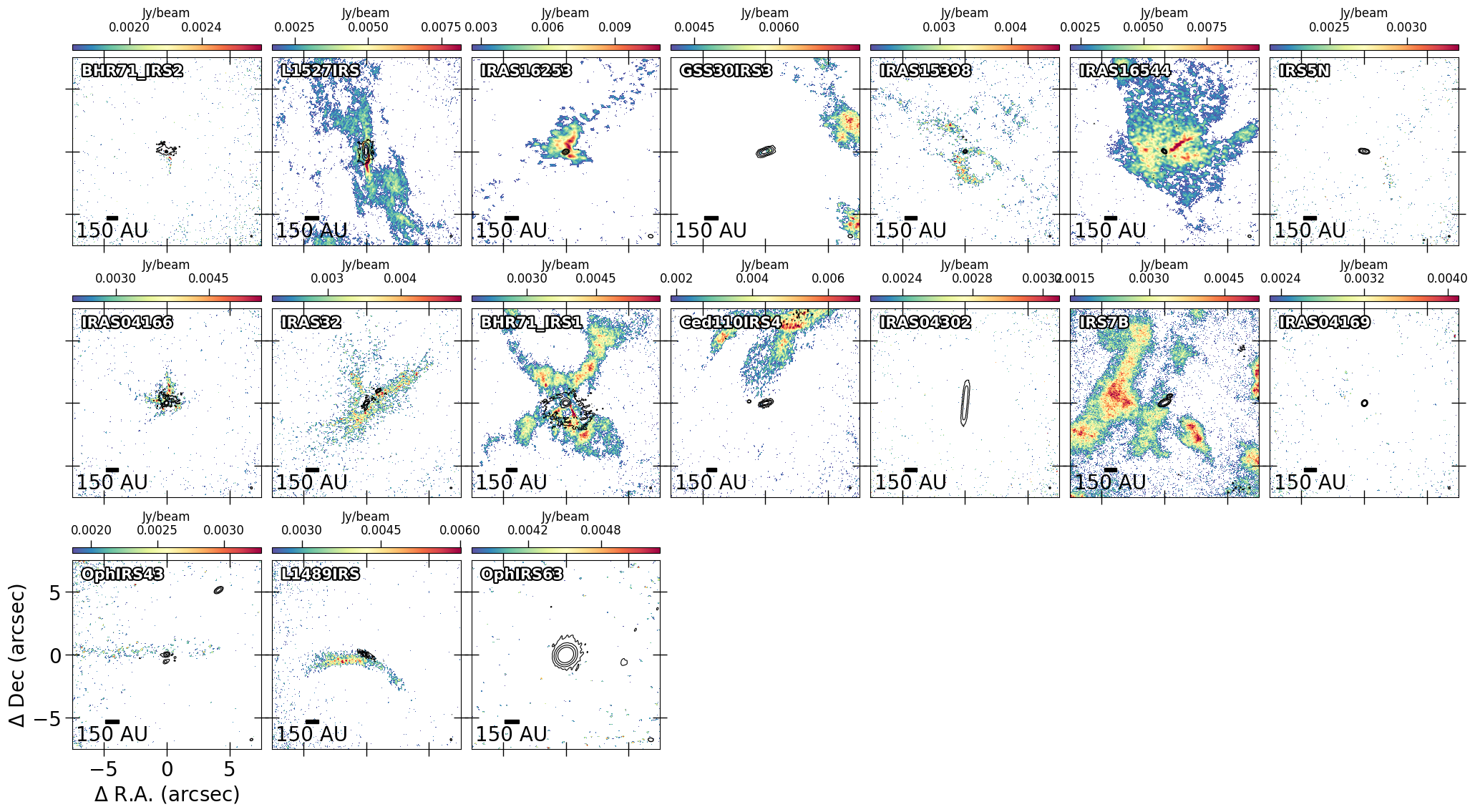}
  \includegraphics[width=1.0\linewidth]{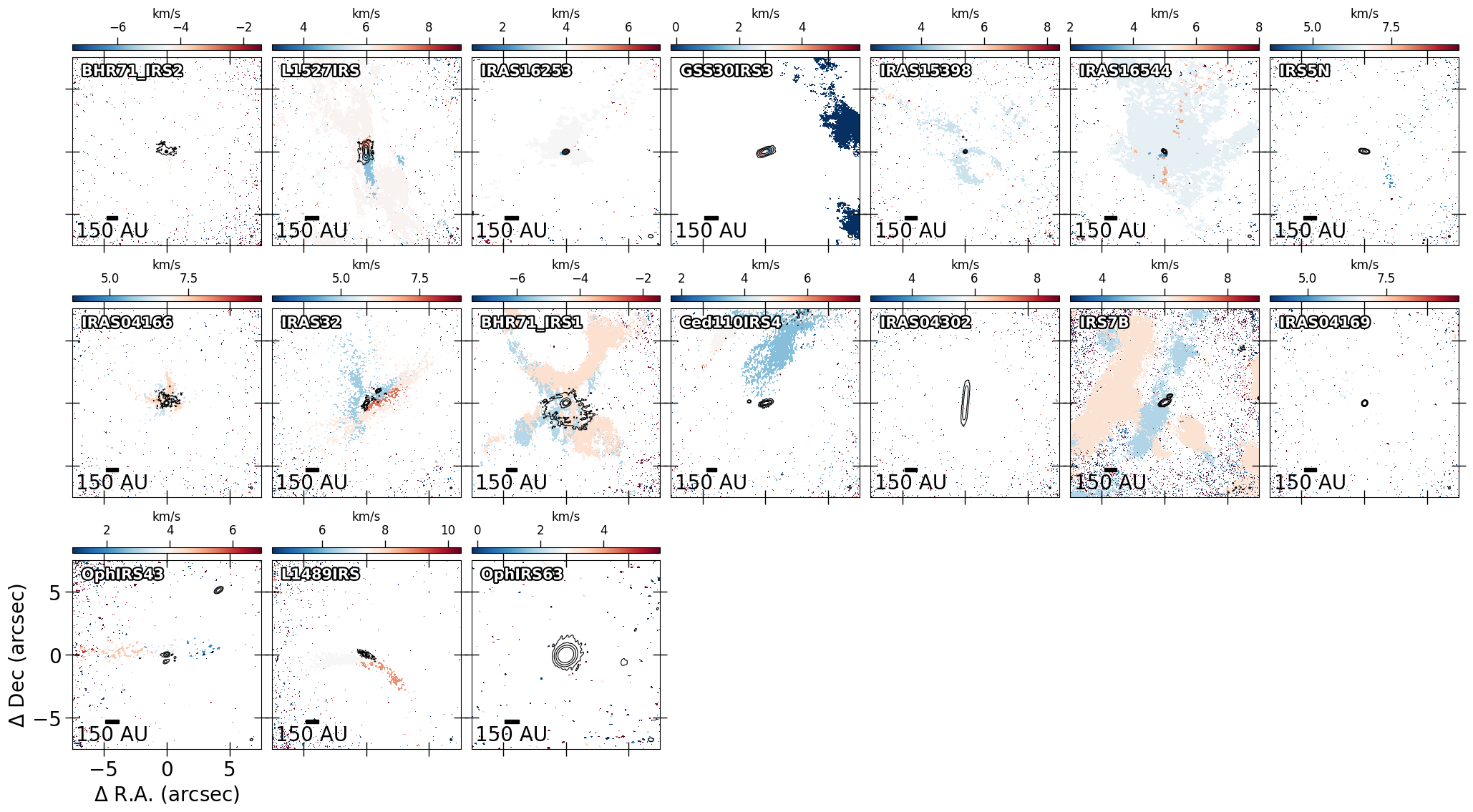}
  \caption{Same as Figure~\ref{appendix:12co_mom8_mom9} but for the blended \C3h2~($6_{0,6}$--$5_{1,5}$) and ($6_{1,6}$--$5_{0,5}$) instead. \label{appendix:c3h2_217.82_mom8_mom9}}
\end{figure*}
\newpage

\begin{figure*}[htbp]
  \includegraphics[width=1.0\linewidth]{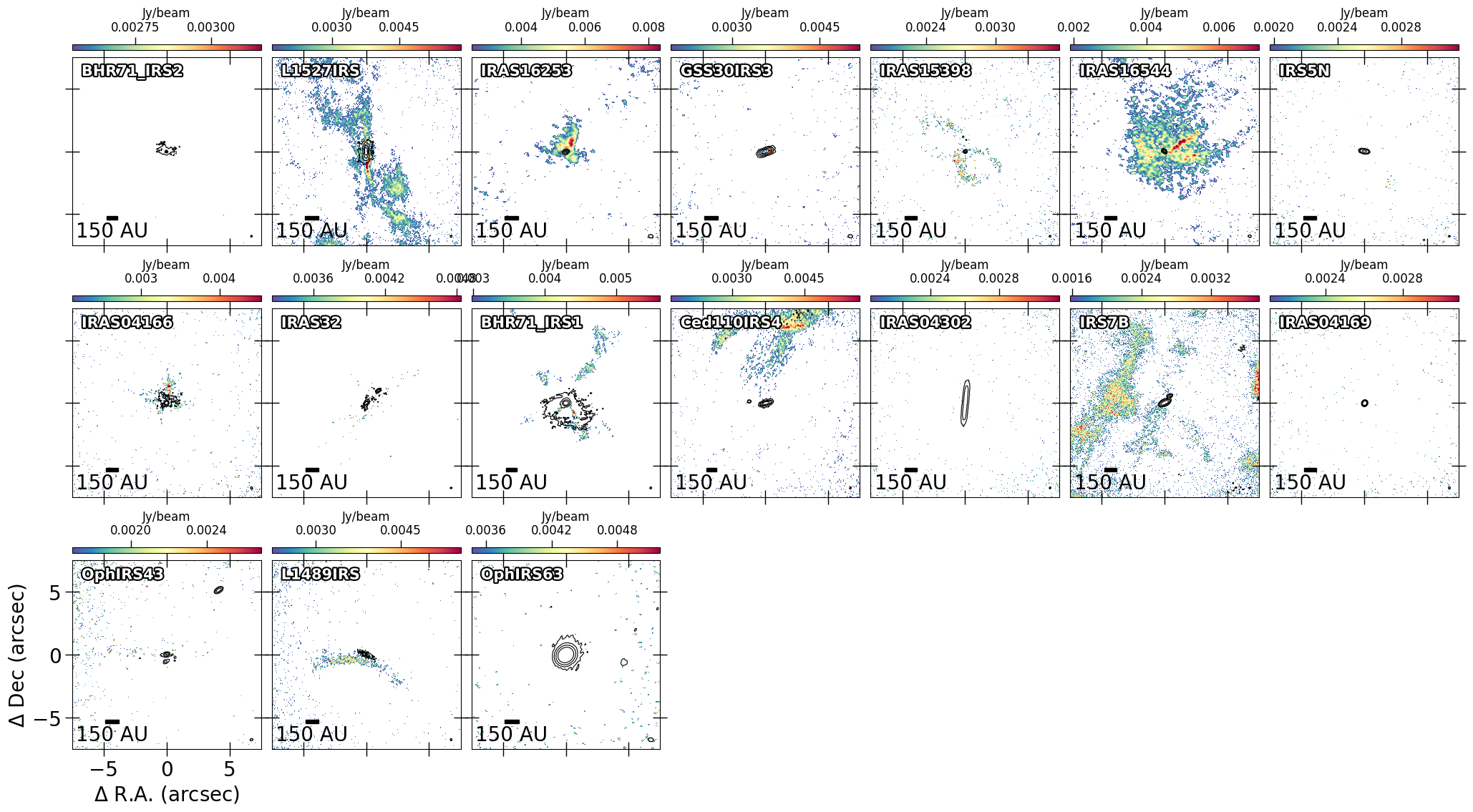}
  \includegraphics[width=1.0\linewidth]{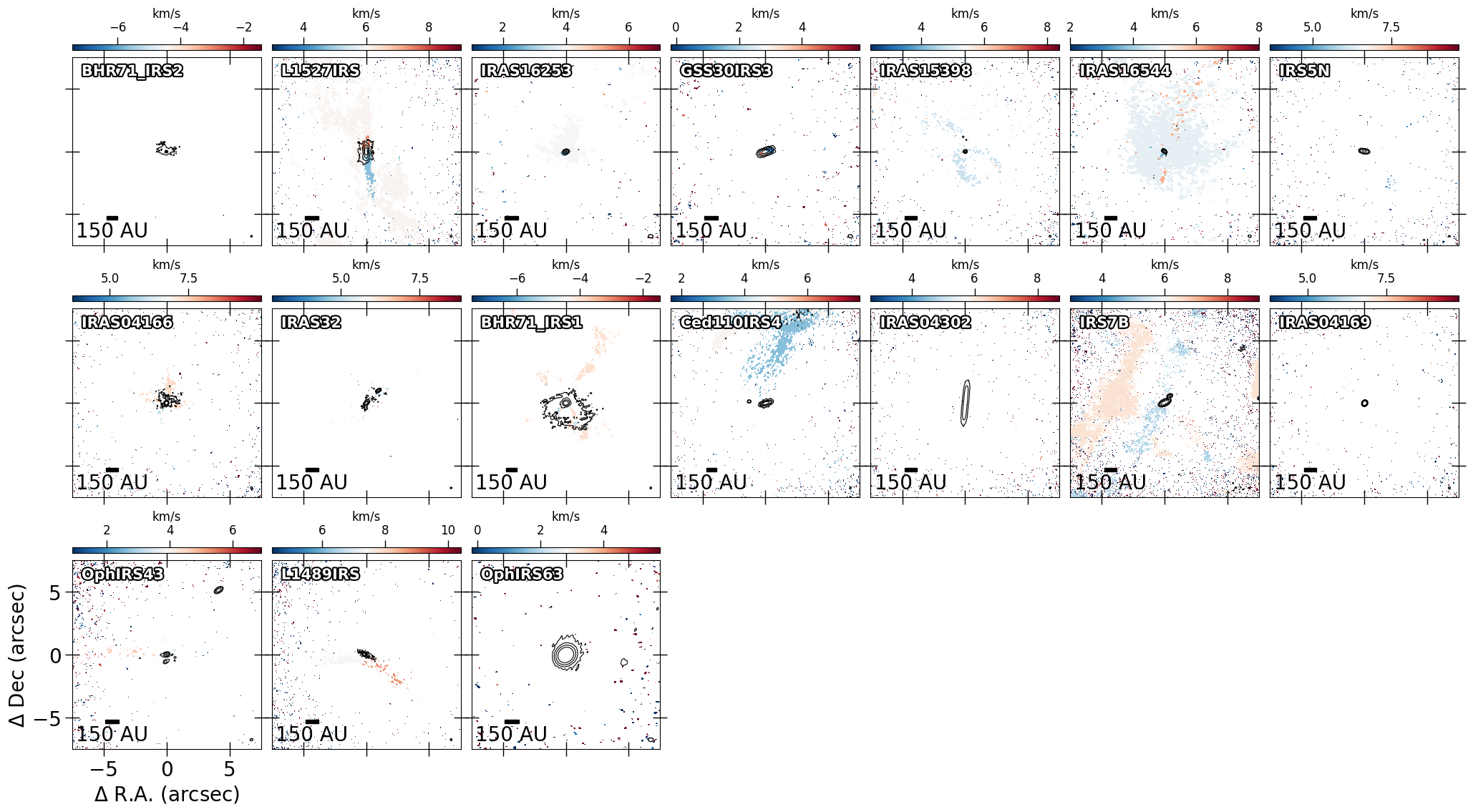}
  \caption{Same as Figure~\ref{appendix:12co_mom8_mom9} but for \C3h2~($5_{1,4}$--$4_{2,3}$) instead. \label{appendix:c3h2_217.94_mom8_mom9}}
\end{figure*}
\newpage

\begin{figure*}[htbp]
  \includegraphics[width=1.0\linewidth]{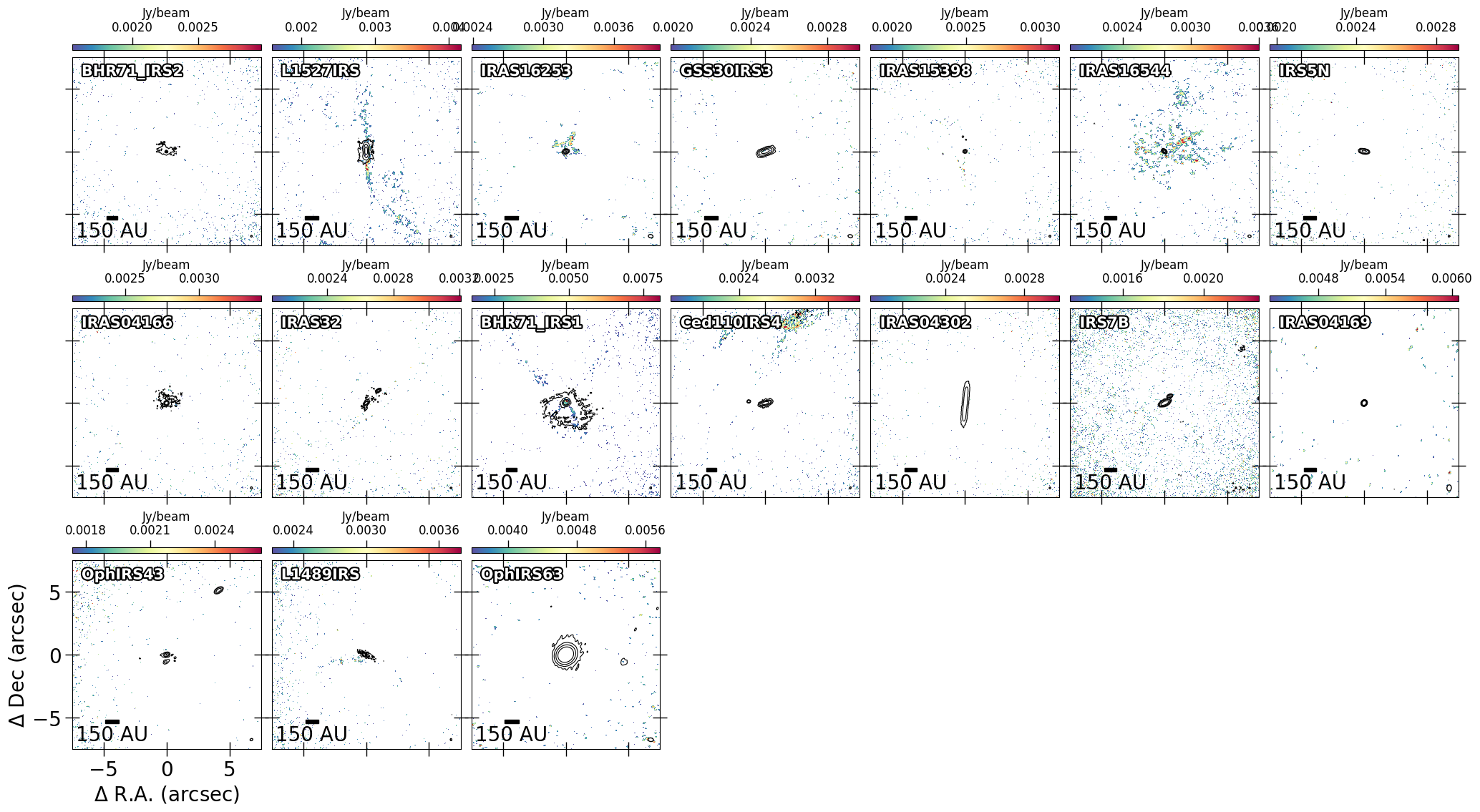}
  \includegraphics[width=1.0\linewidth]{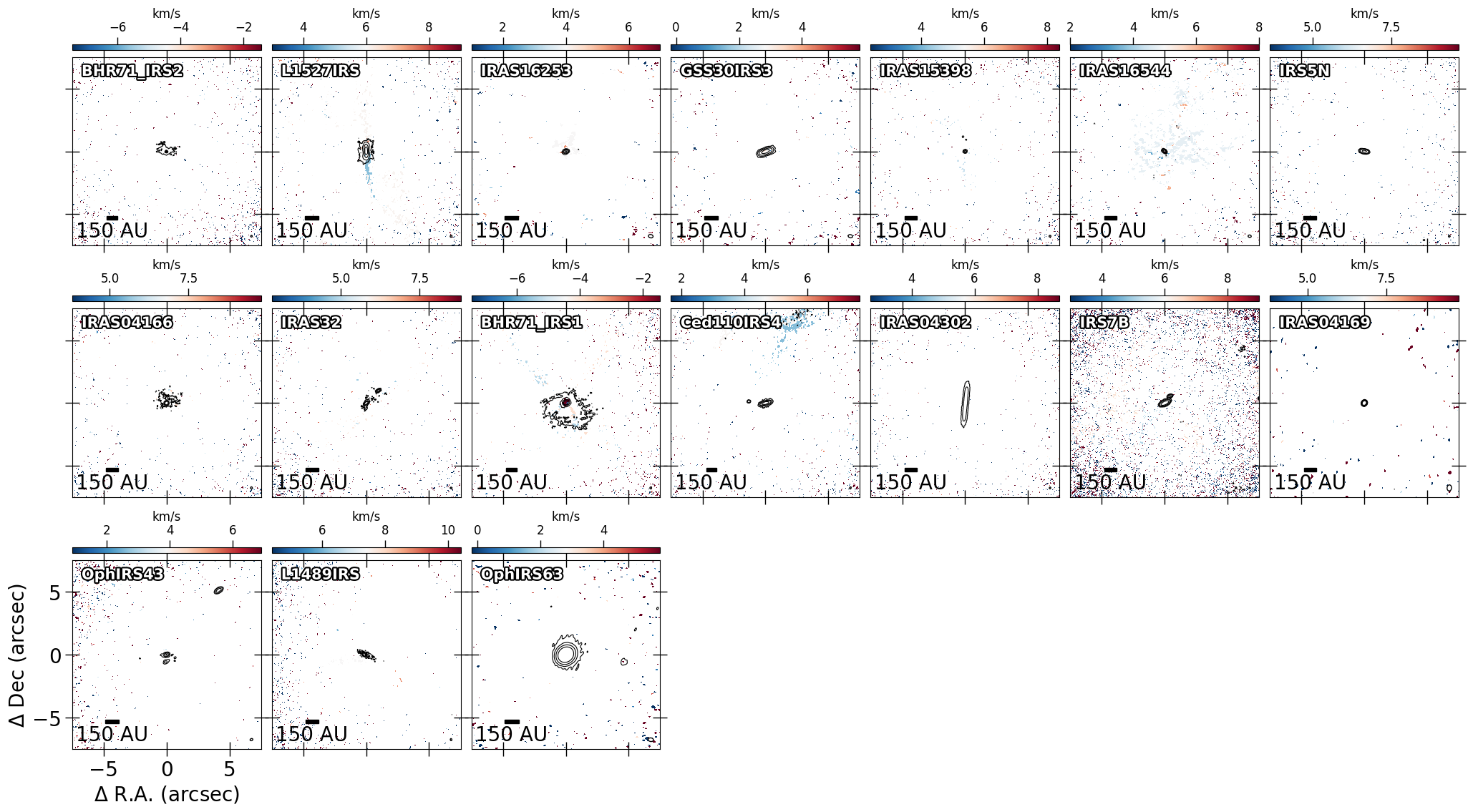}
  \caption{Same as Figure~\ref{appendix:12co_mom8_mom9} but for \C3h2~($5_{2,4}$--$4_{1,3}$) instead. \label{appendix:c3h2_218.16_mom8_mom9}}
\end{figure*}
\newpage

\section{Zoomed-in small-scale moment 8 and moment 9 maps}\label{appendix:small_scale_emissions}
\begin{figure*}[htbp]
  \includegraphics[width=1.0\linewidth]{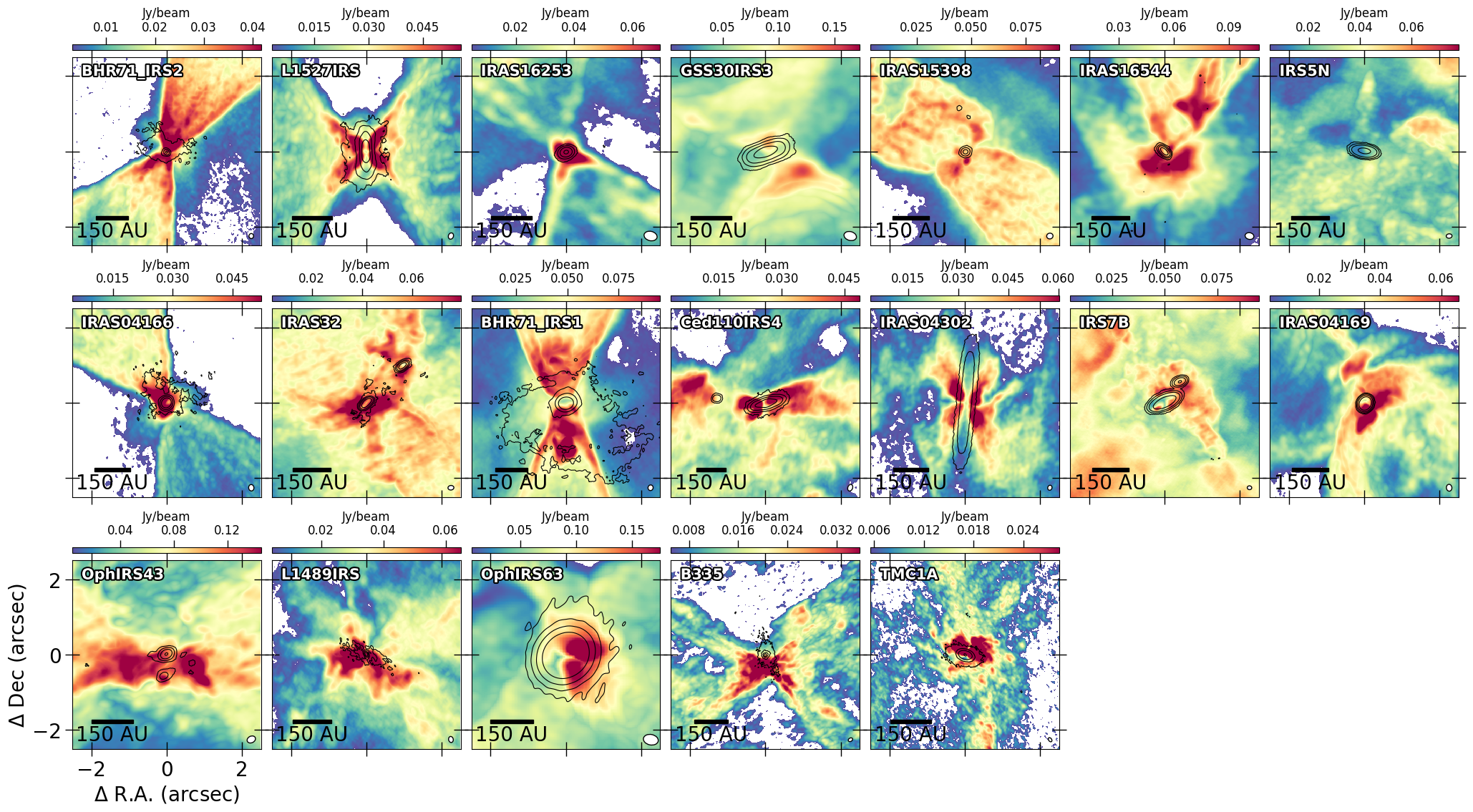}
  \includegraphics[width=1.0\linewidth]{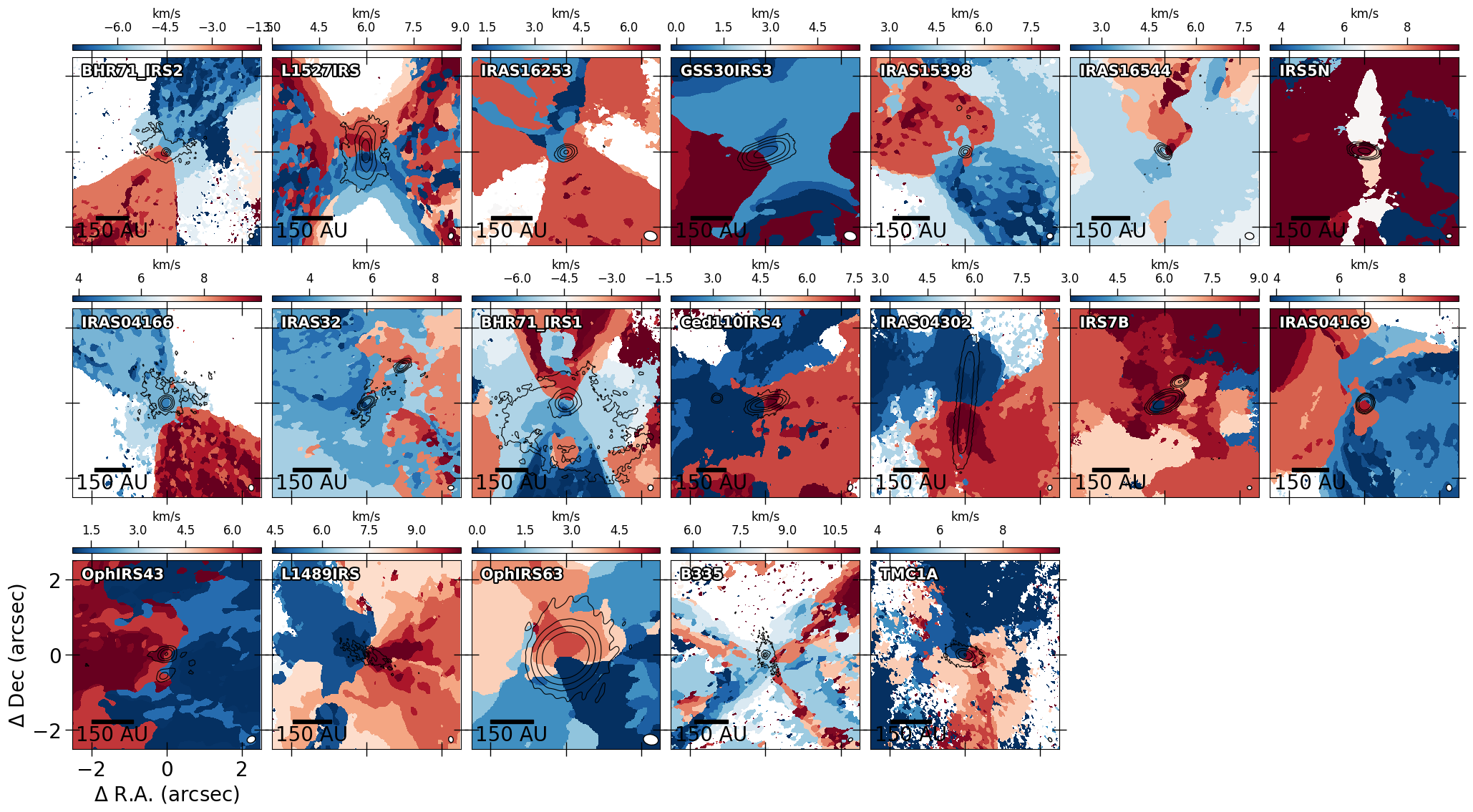} 
  \caption{Moment 8 (\textit{top}) and moment 9 (\textit{bottom}) maps depicting the \tlvco~(2--1) emission in the inner 5\asec region of the nineteen eDisk sources. The moment maps were generated by integrating the regions where $I_{\nu} > 3\sigma$, where $\sigma$ is the rms per channel. The contour lines display the continuum emission at thresholds of 5$\sigma$, 20$\sigma$, 80$\sigma$, and 320$\sigma$ for each source. The scale bar is located at the bottom left, and the synthesized beam is indicated in white at the bottom-right corner of each image. \label{appendix:12co_mom8_mom9_zoomed}}
\end{figure*}

\begin{figure*}[htbp]
  \includegraphics[width=1.0\linewidth]{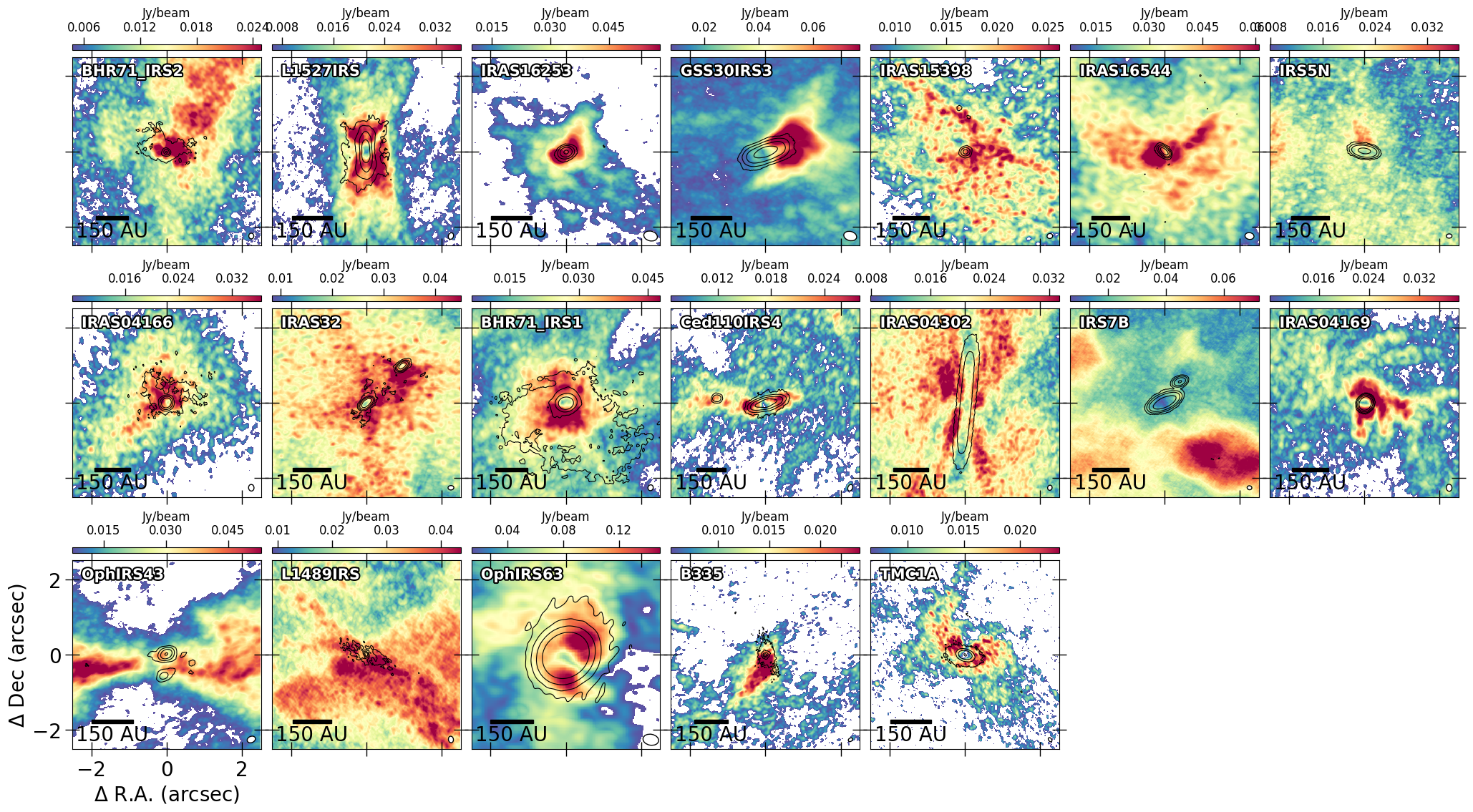}
  \includegraphics[width=1.0\linewidth]{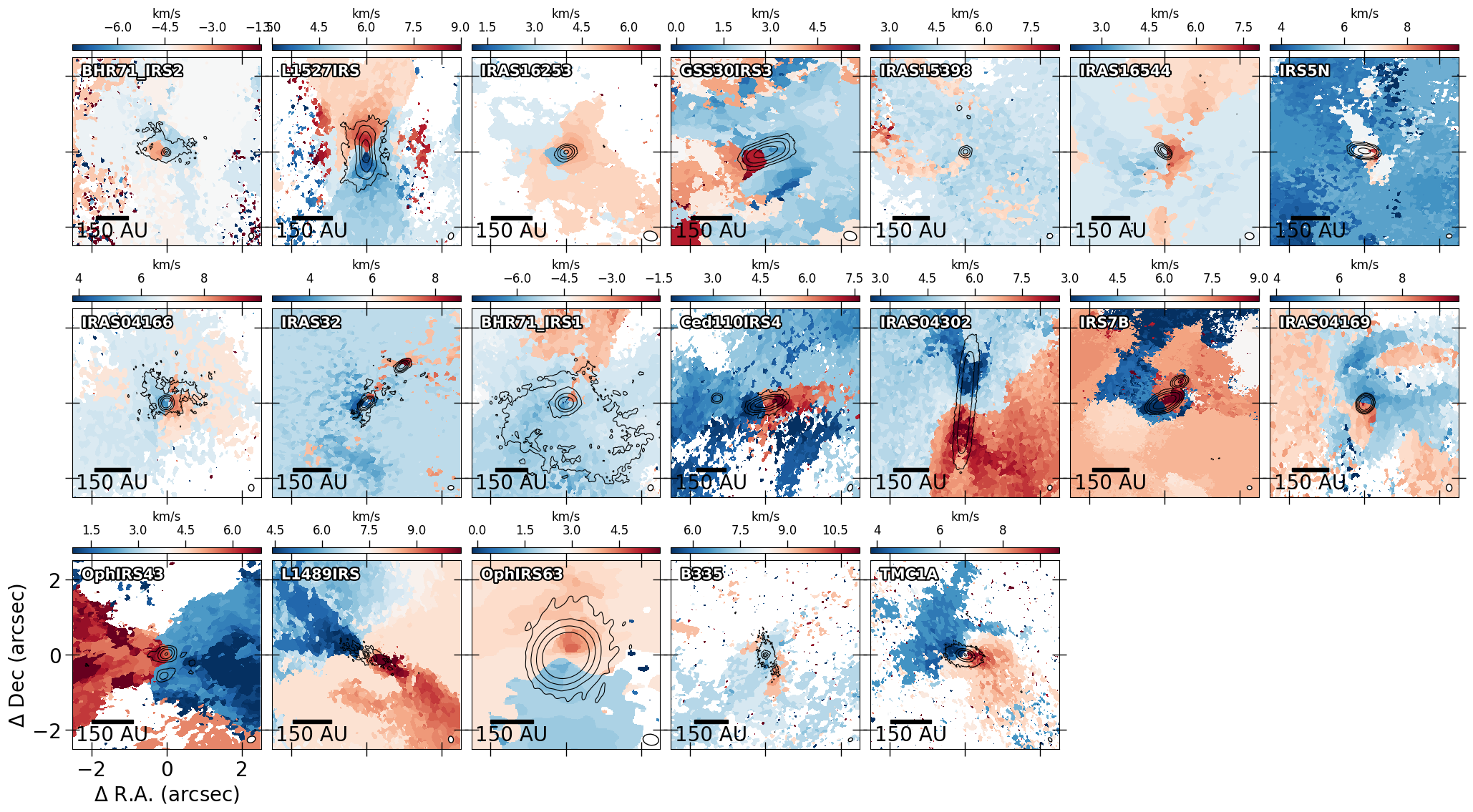}
  \caption{Same as Figure~\ref{appendix:12co_mom8_mom9_zoomed} but for \thrco~($2$--$1$) instead. \label{appendix:13co_mom8_mom9_zoomed}}
\end{figure*}

\begin{figure*}[htbp]
  \includegraphics[width=1.0\linewidth]{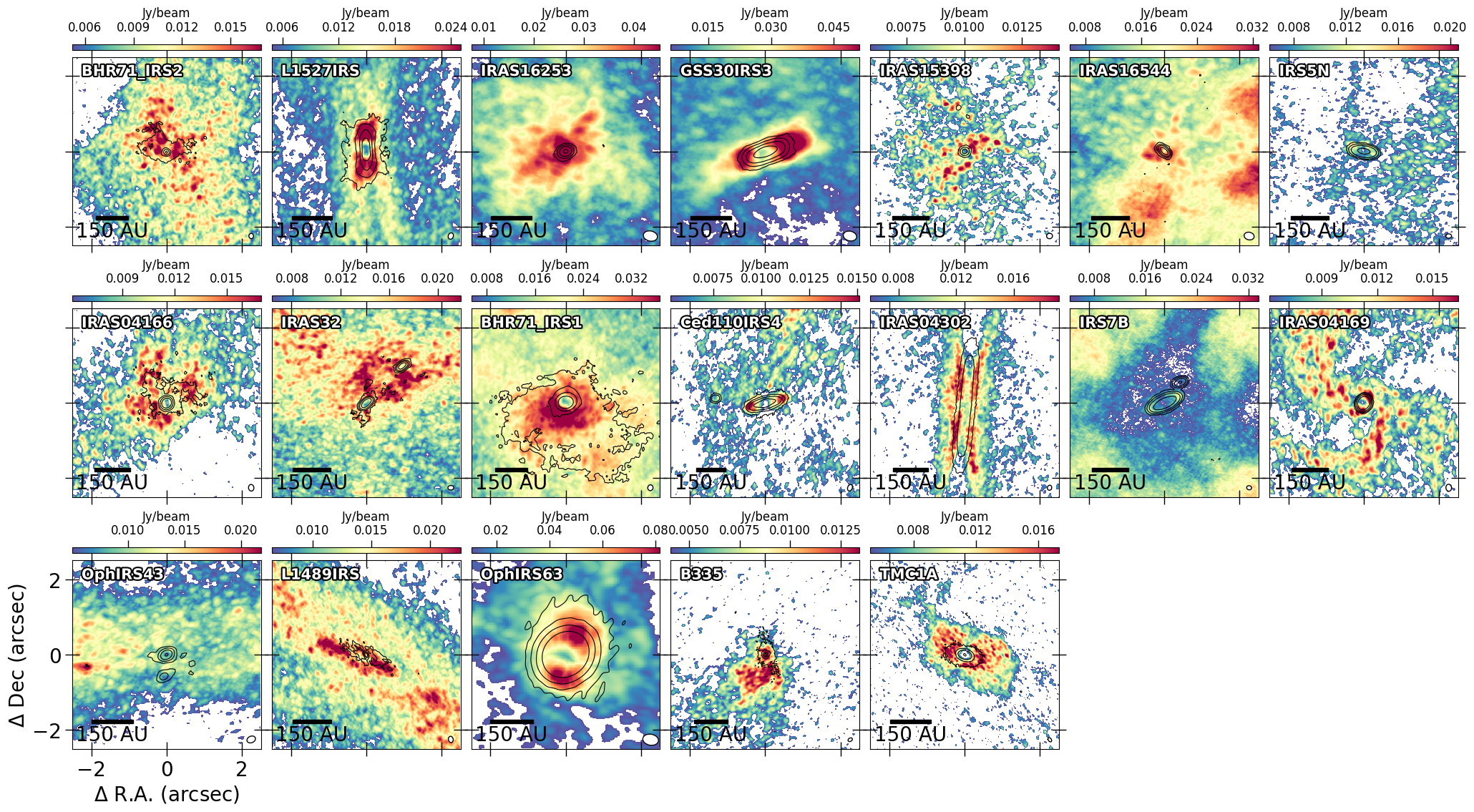}
  \includegraphics[width=1.0\linewidth]{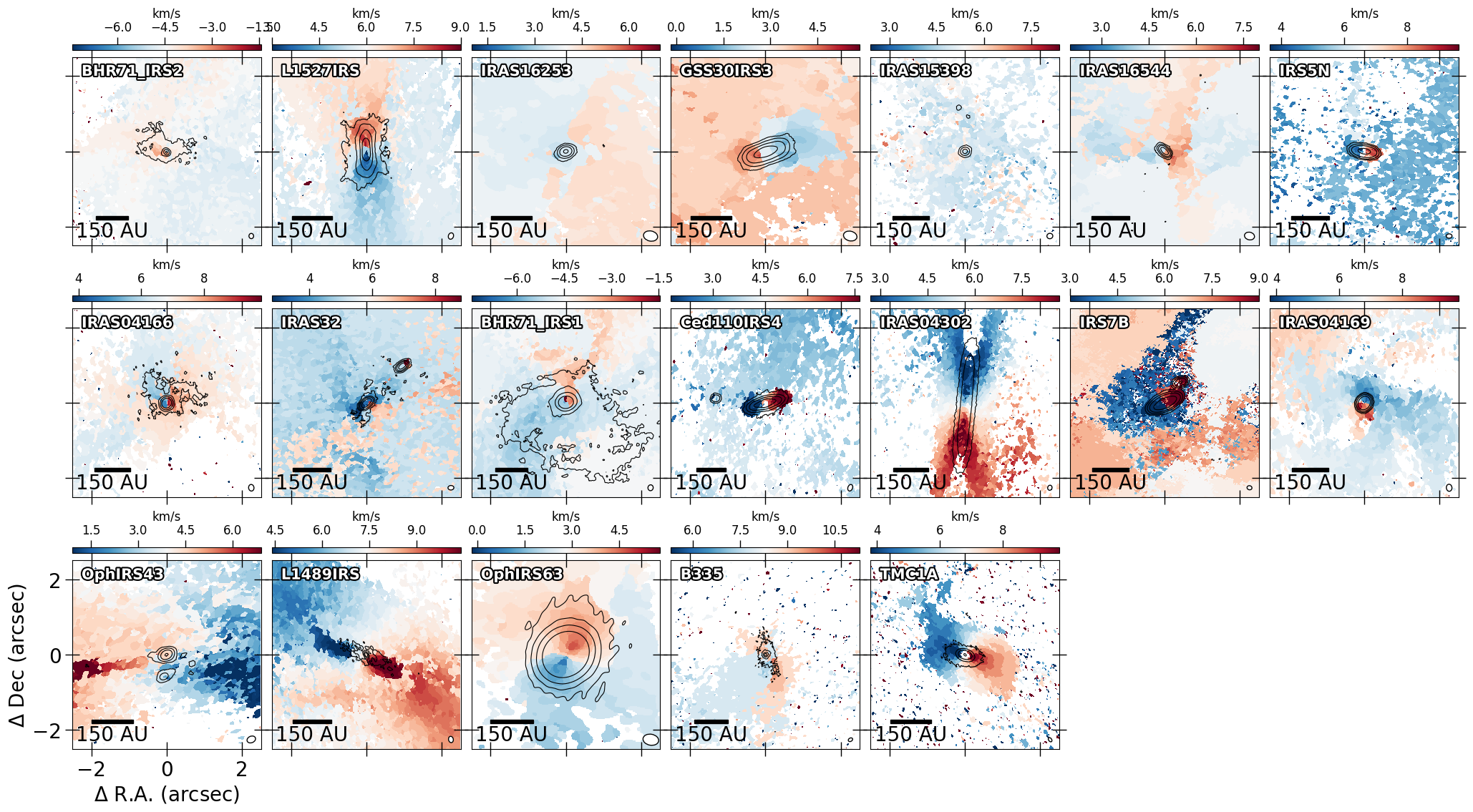}
  \caption{Same as Figure~\ref{appendix:12co_mom8_mom9_zoomed} but for \ceteno~($2$--$1$) instead. \label{appendix:c18o_mom8_mom9_zoomed}}
\end{figure*}

\begin{figure*}[htbp]
  \includegraphics[width=1.0\linewidth]{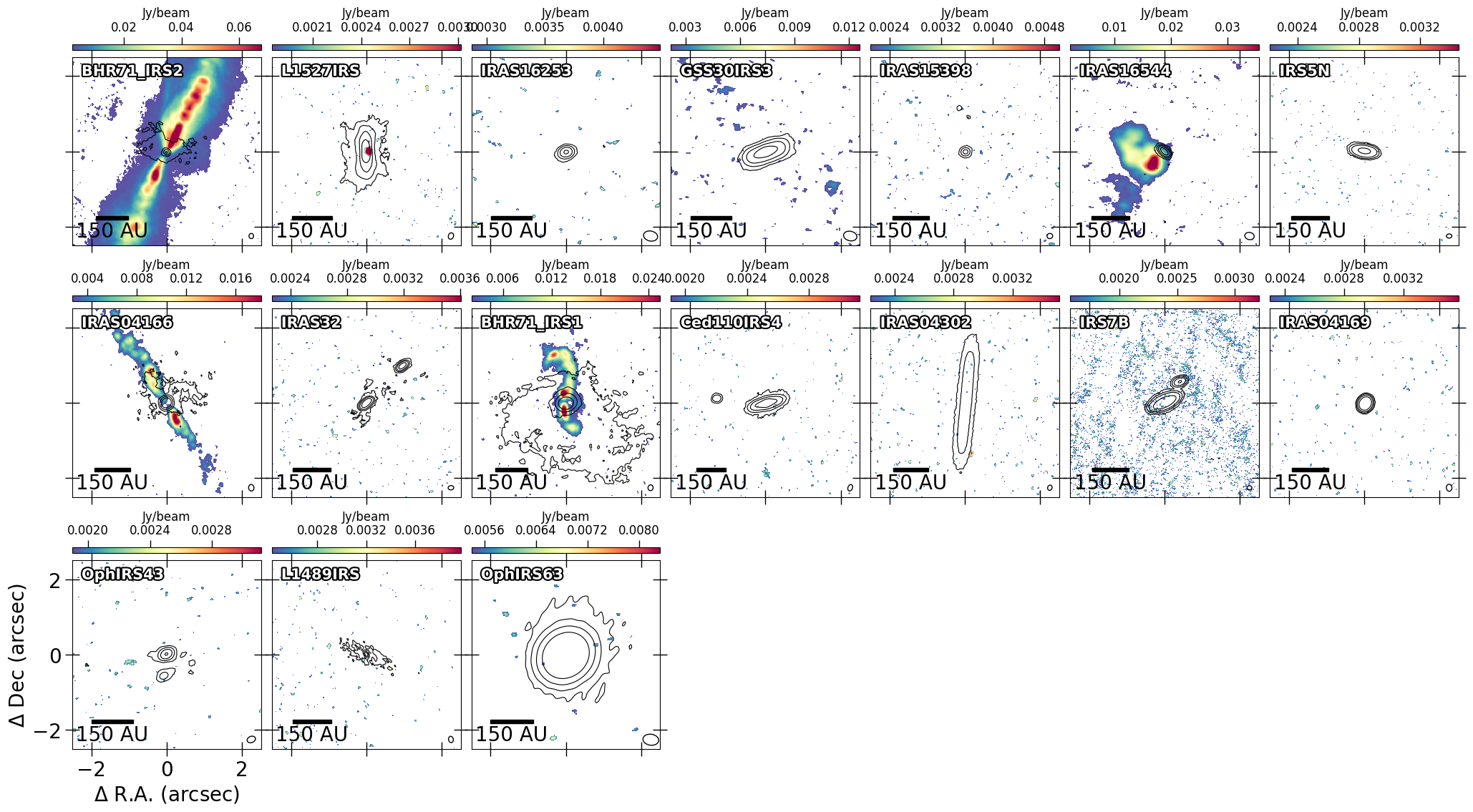} 
  \includegraphics[width=1.0\linewidth]{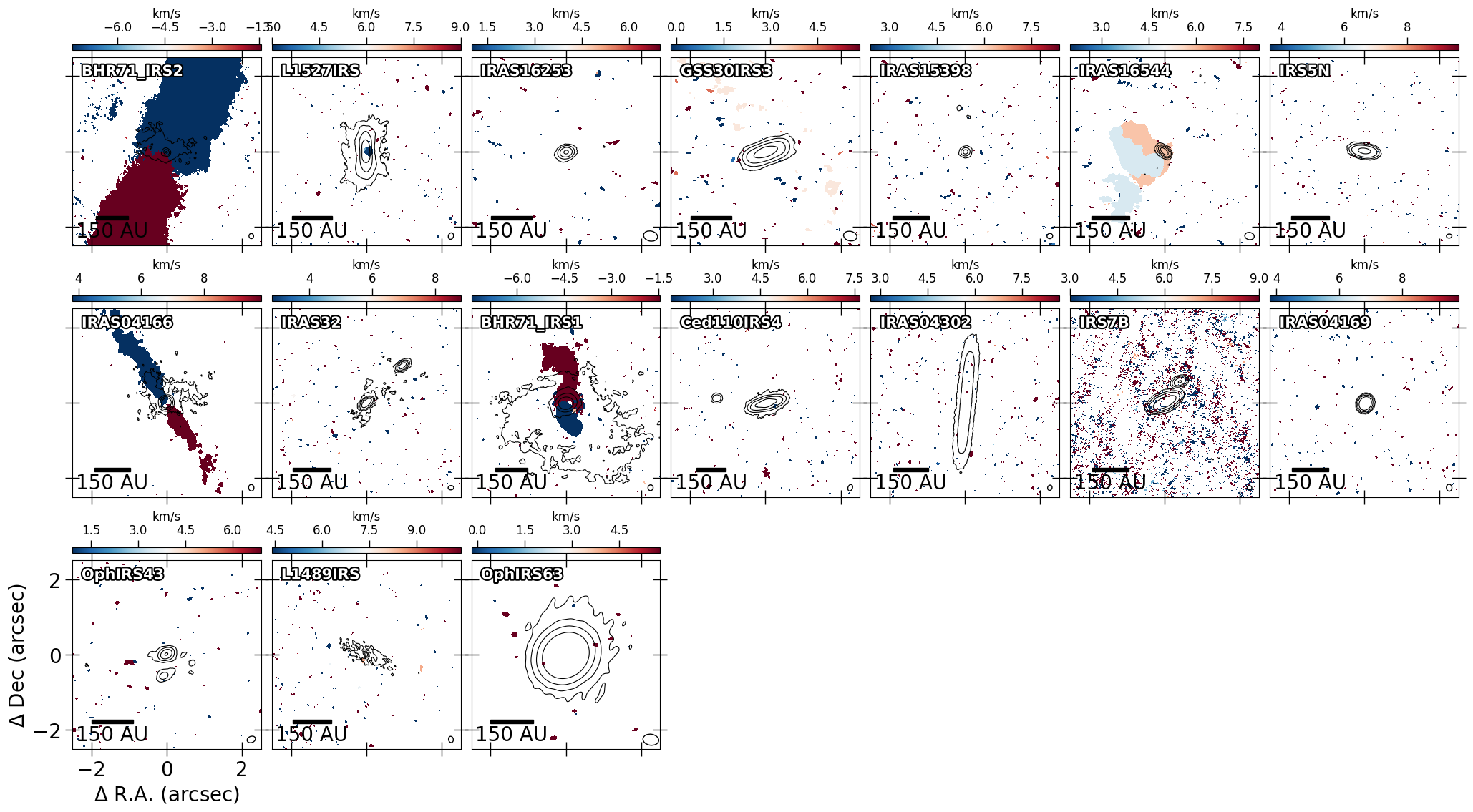}
  \caption{Same as Figure~\ref{appendix:12co_mom8_mom9_zoomed} but for SiO~(5--4) instead. \label{appendix:sio_mom8_mom9_zoomed}}
\end{figure*}

\begin{figure*}[htbp]
  \includegraphics[width=1.0\linewidth]{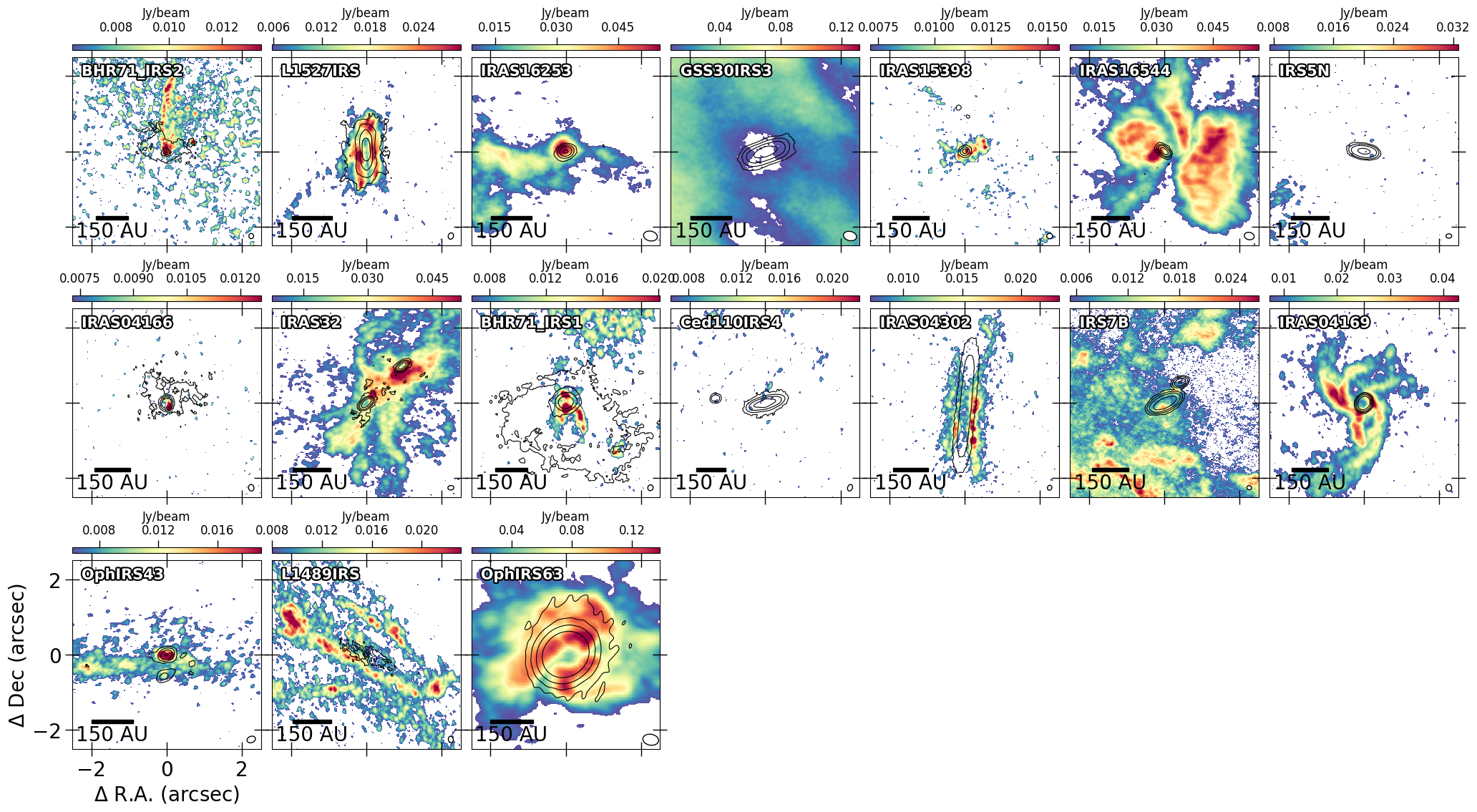}
  \includegraphics[width=1.0\linewidth]{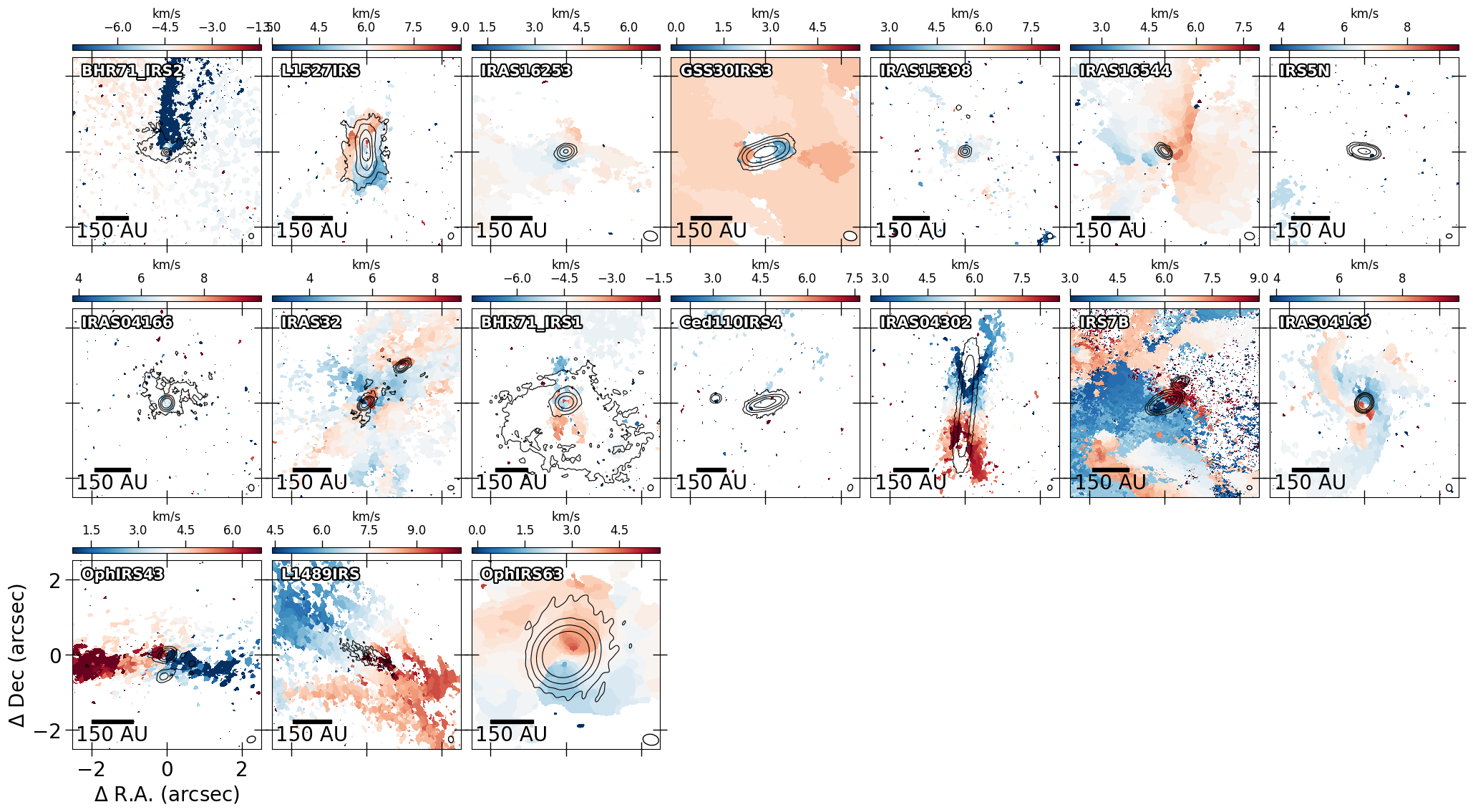}
  \caption{Same as Figure~\ref{appendix:12co_mom8_mom9_zoomed} but for SO~($6_5$--$5_4$) instead. \label{appendix:so_mom8_mom9_zoomed}}
\end{figure*}

\begin{figure*}[htbp]
  \includegraphics[width=1.0\linewidth]{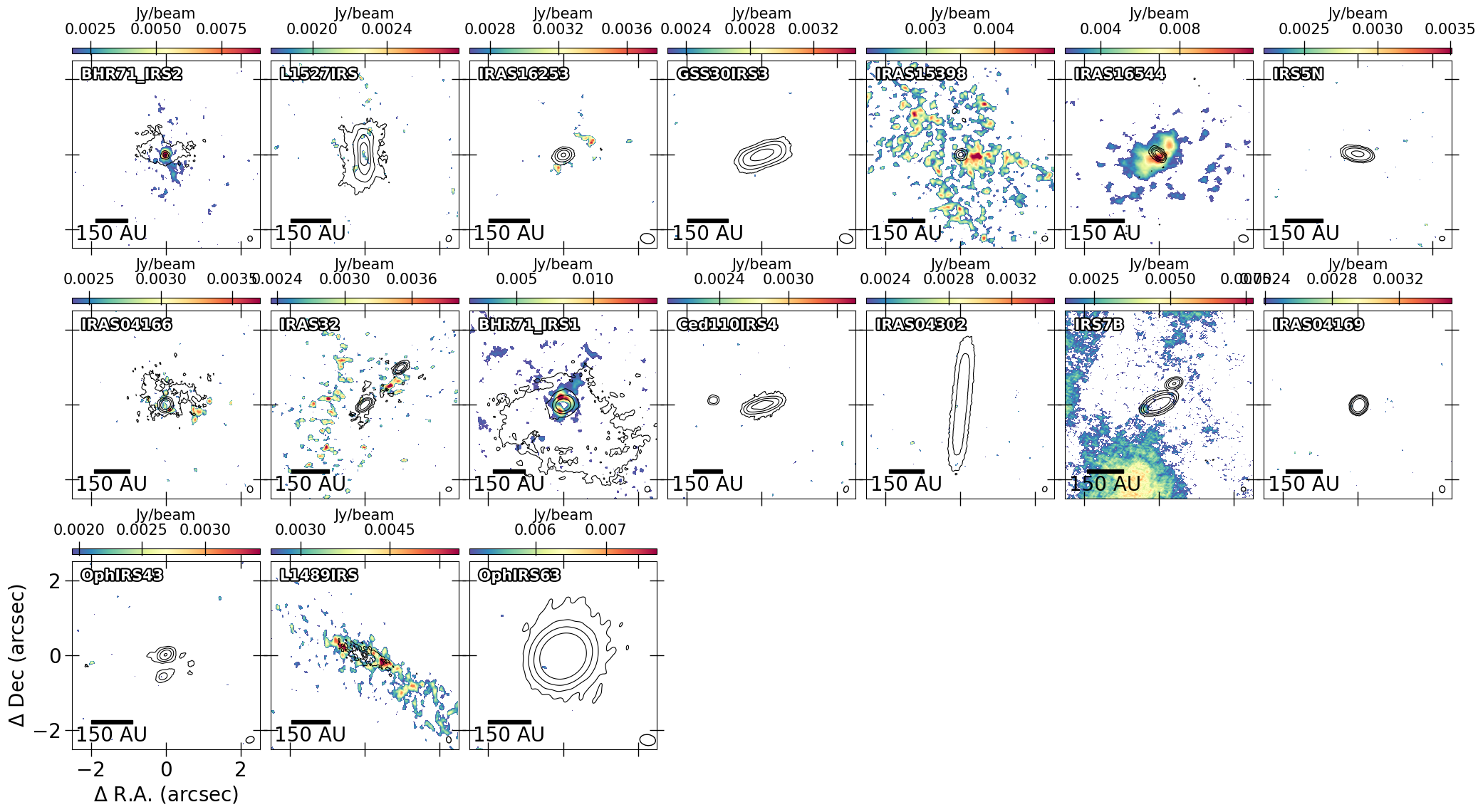}
  \includegraphics[width=1.0\linewidth]{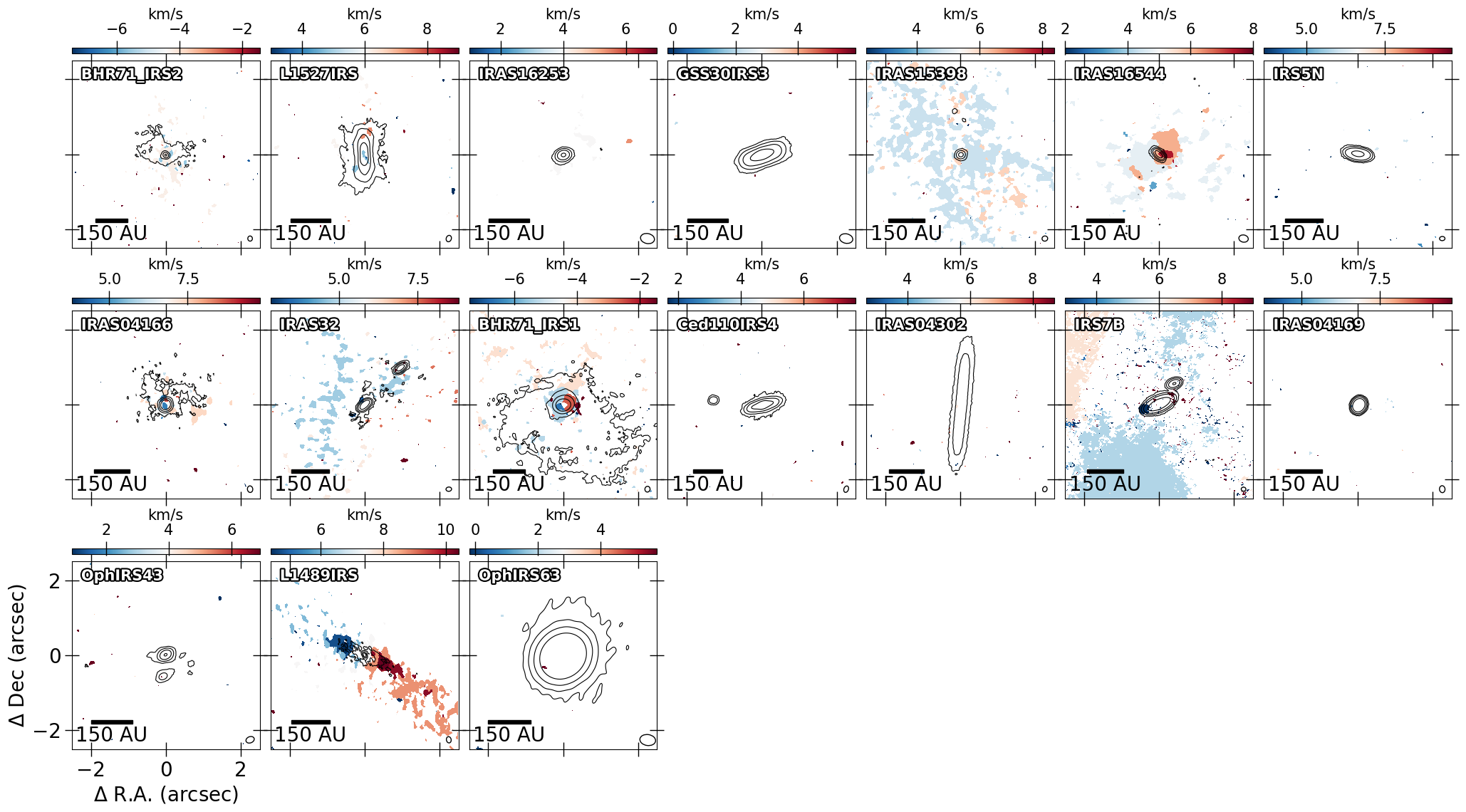}
  \caption{Same as Figure~\ref{appendix:12co_mom8_mom9_zoomed} but for DCN~($3$--$2$) instead. \label{appendix:dcn_mom8_mom9_zoomed}}
\end{figure*}

\begin{figure*}[htbp]
  \includegraphics[width=1.0\linewidth]{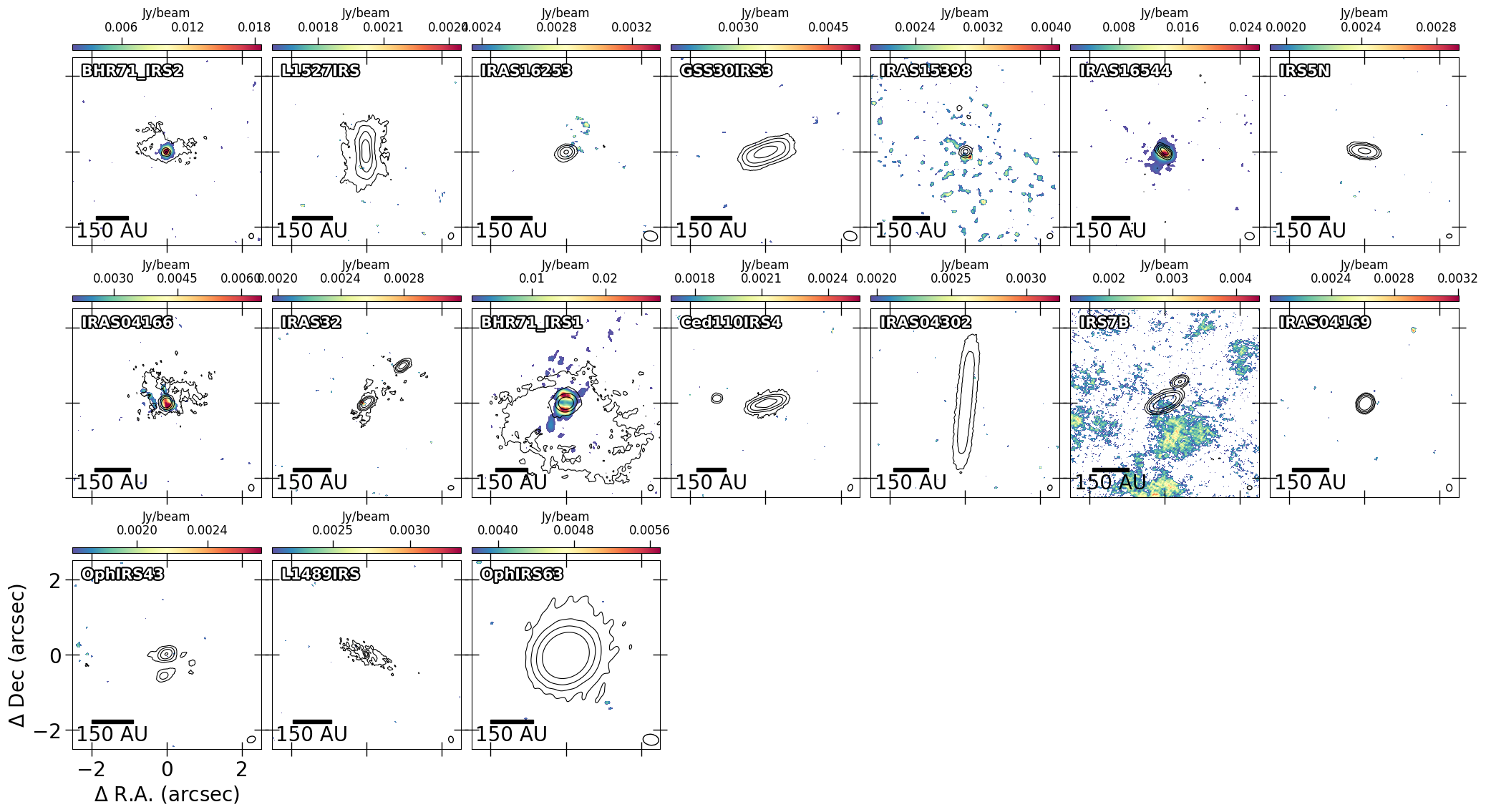}
  \includegraphics[width=1.0\linewidth]{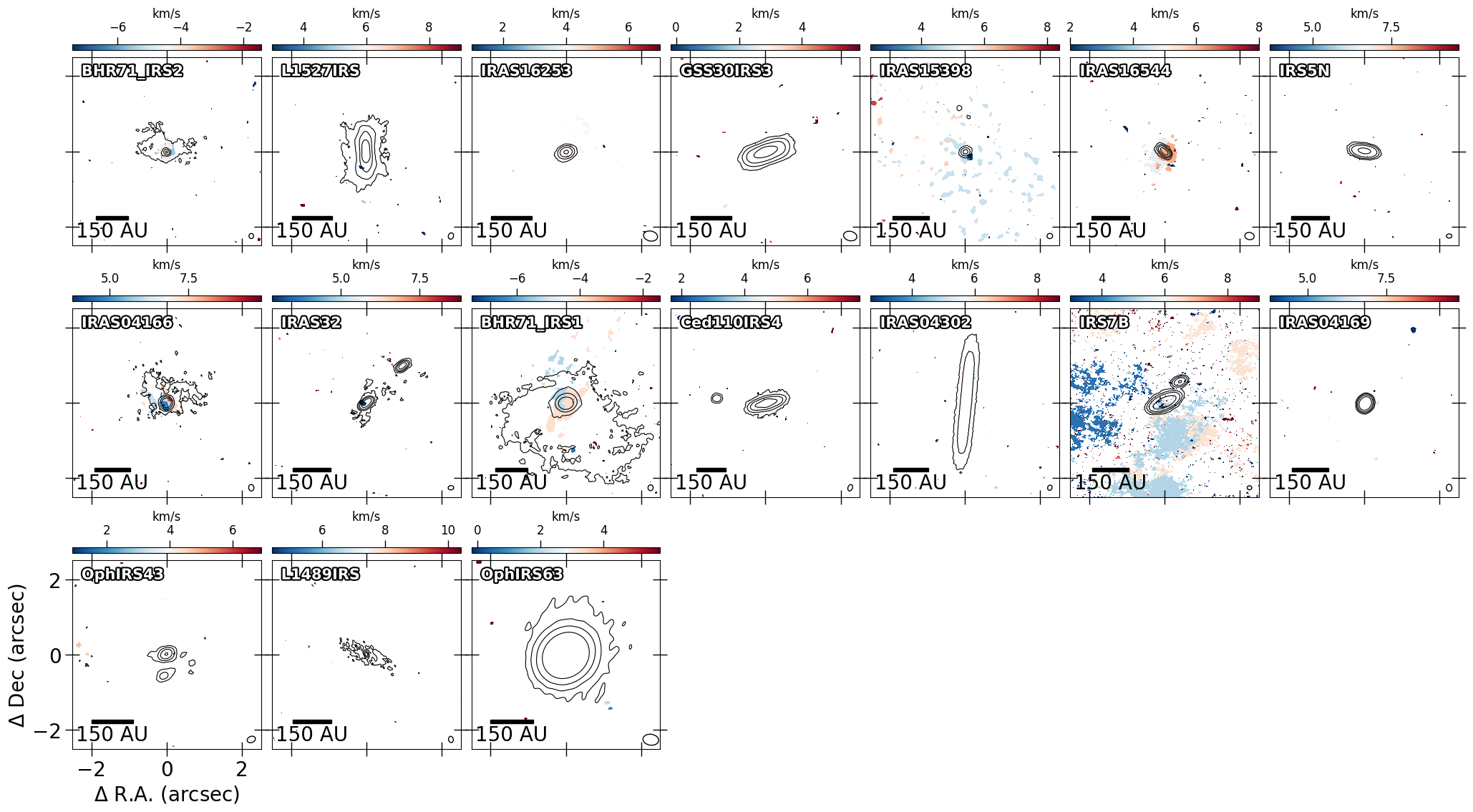}
  \caption{Same as Figure~\ref{appendix:12co_mom8_mom9_zoomed} but for \ch3oh~($4_2$--$3_1$) instead. \label{appendix:ch3oh_mom8_mom9_zoomed}}
\end{figure*}

\begin{figure*}[htbp]
  \includegraphics[width=1.0\linewidth]{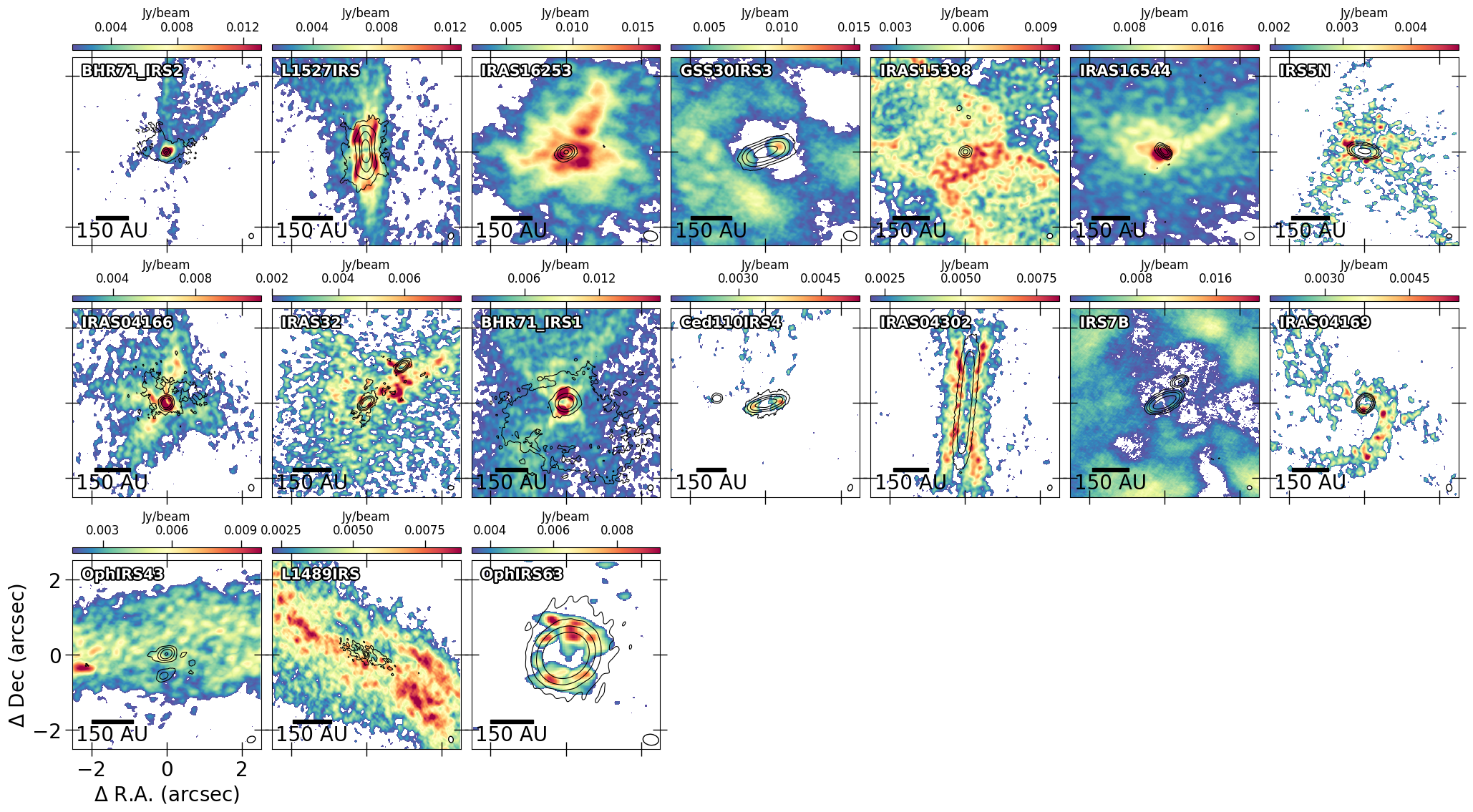}
  \includegraphics[width=1.0\linewidth]{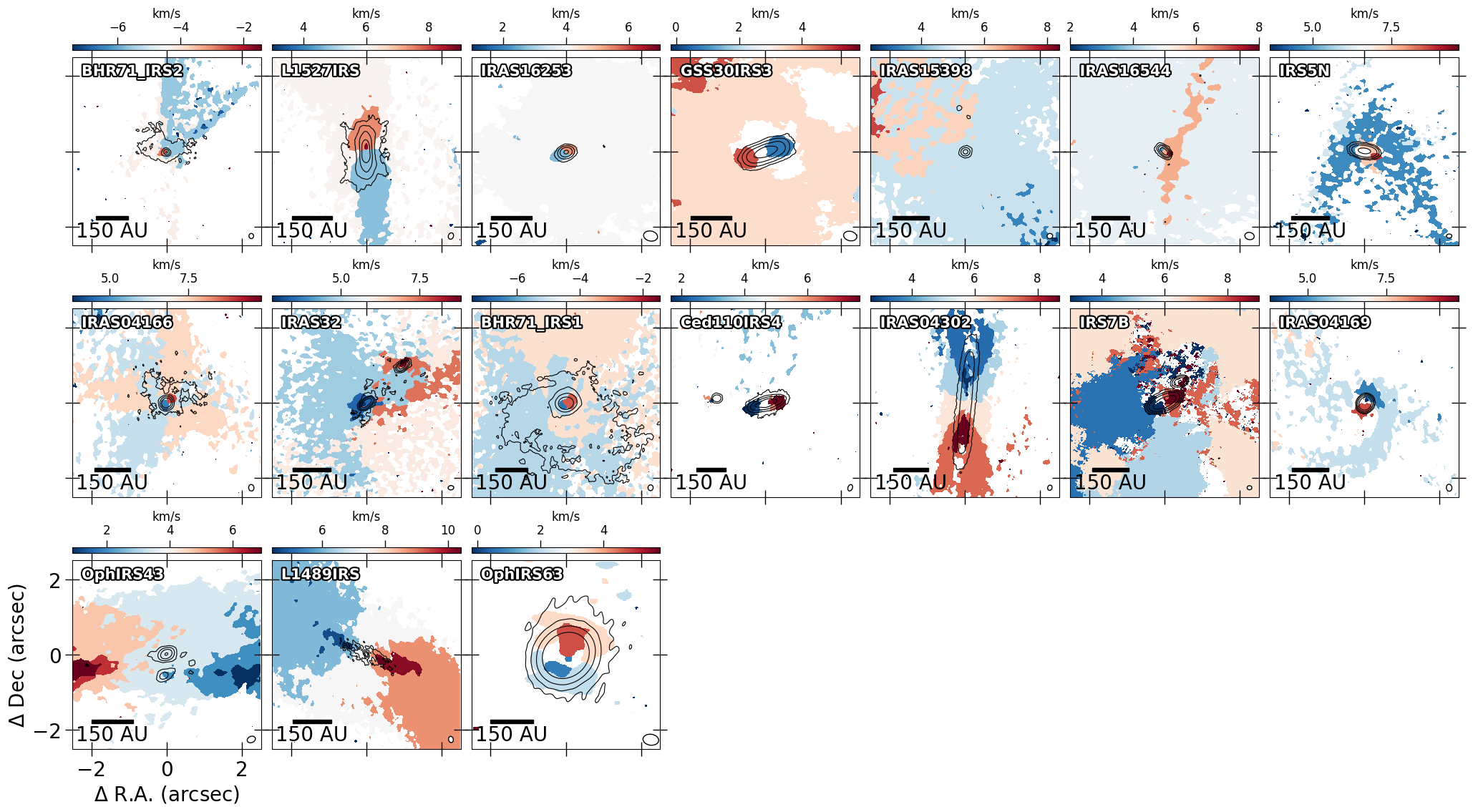}
  \caption{Same as Figure~\ref{appendix:12co_mom8_mom9_zoomed} but for \h2co~(\htwocolow) instead. \label{appendix:h2co_3_03-2_02_mom8_mom9_zoomed}}
\end{figure*}

\begin{figure*}[htbp]
  \includegraphics[width=1.0\linewidth]{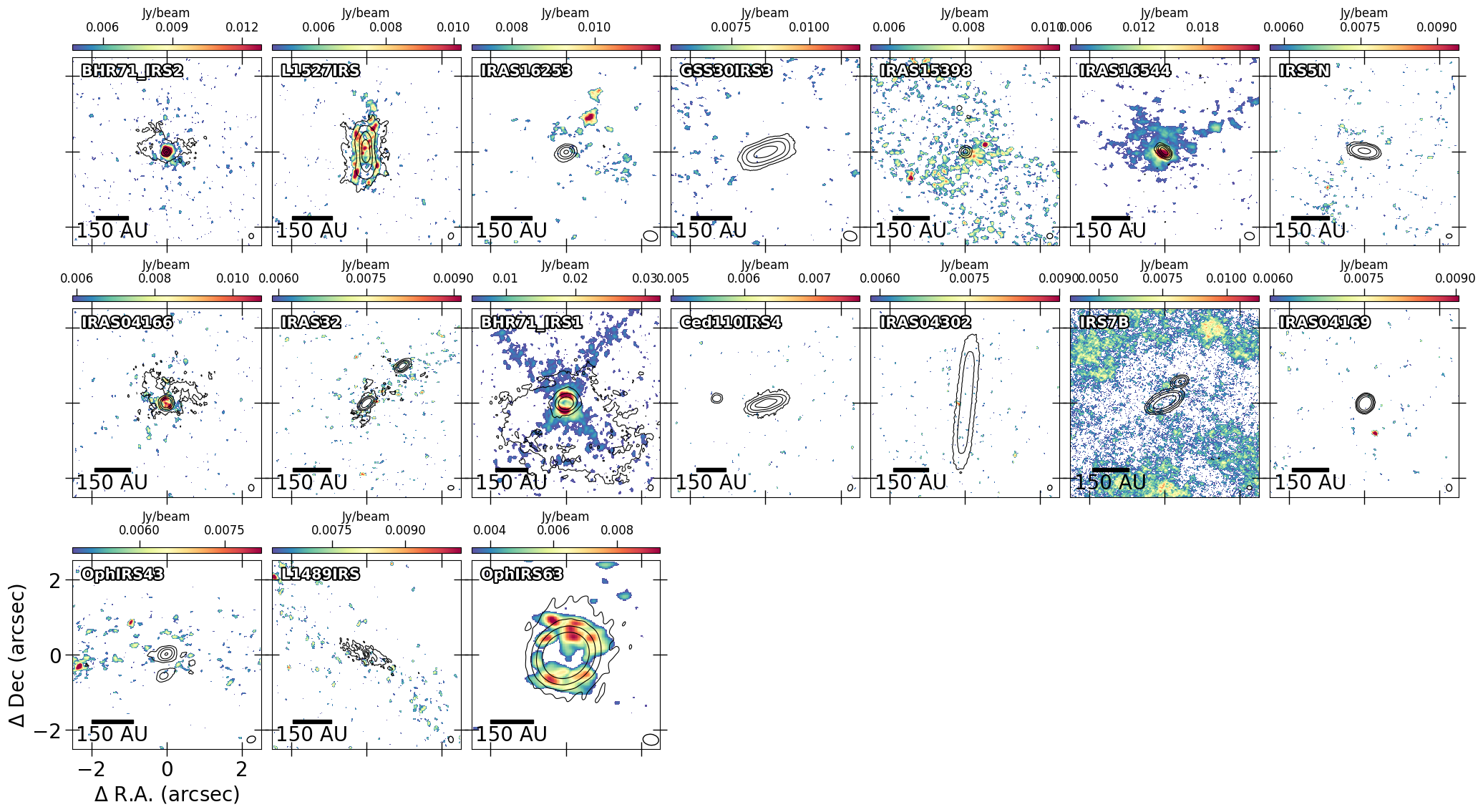}
  \includegraphics[width=1.0\linewidth]{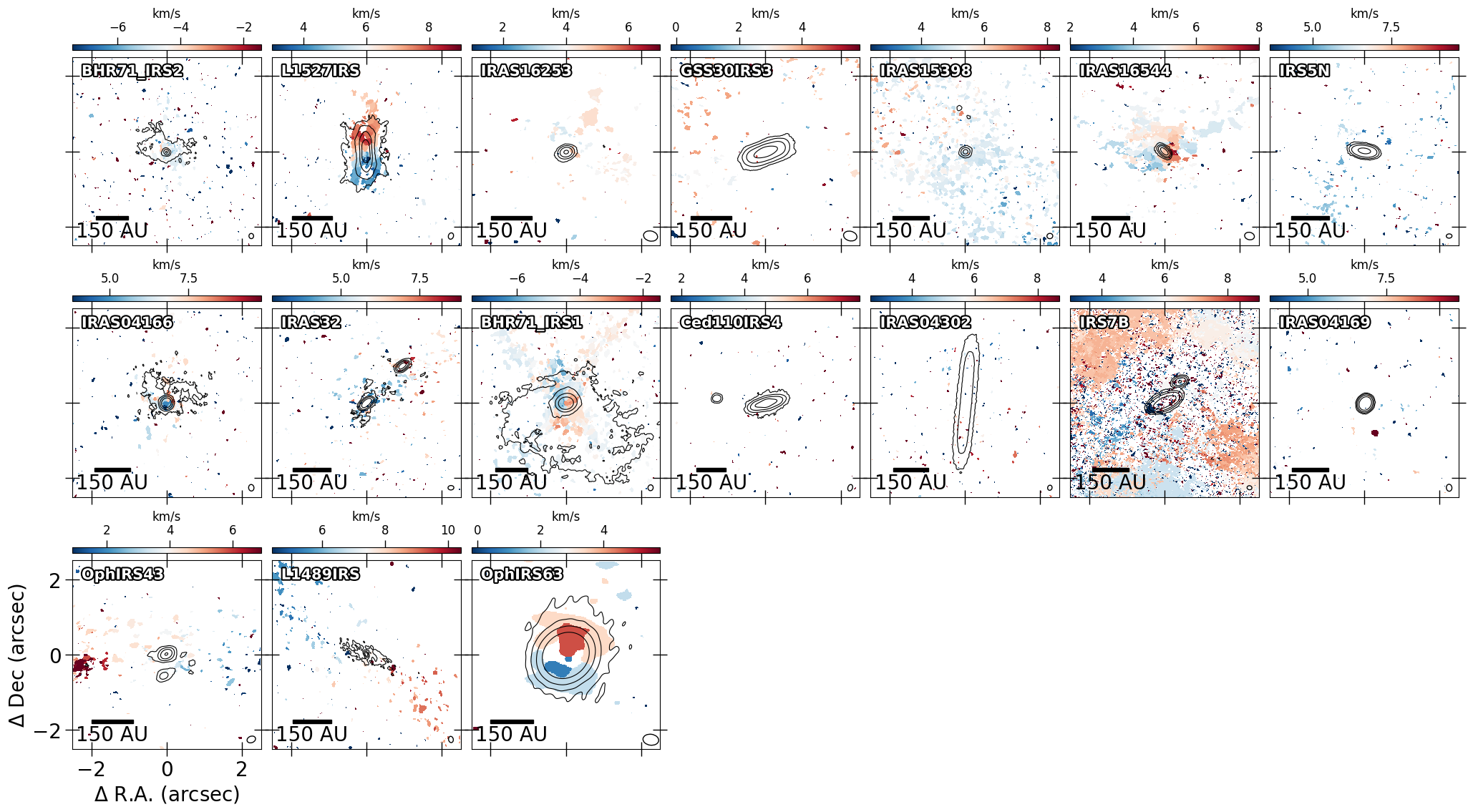}
  \caption{Same as Figure~\ref{appendix:12co_mom8_mom9_zoomed} but for \h2co~(\htwocomid) instead. \label{appendix:h2co_3_21-2_20_mom8_mom9_zoomed}}
\end{figure*}

\begin{figure*}[htbp]
  \includegraphics[width=1.0\linewidth]{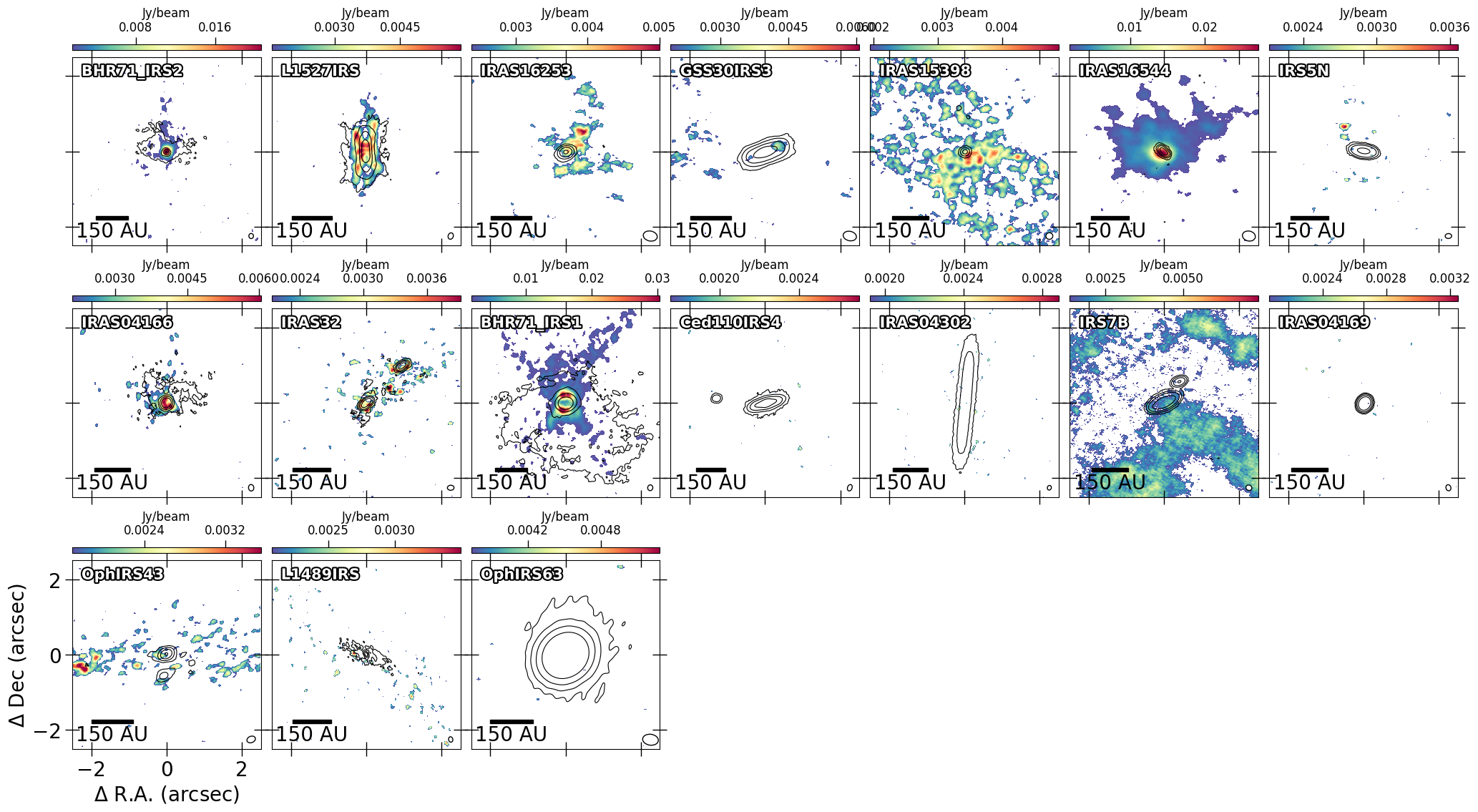}
  \includegraphics[width=1.0\linewidth]{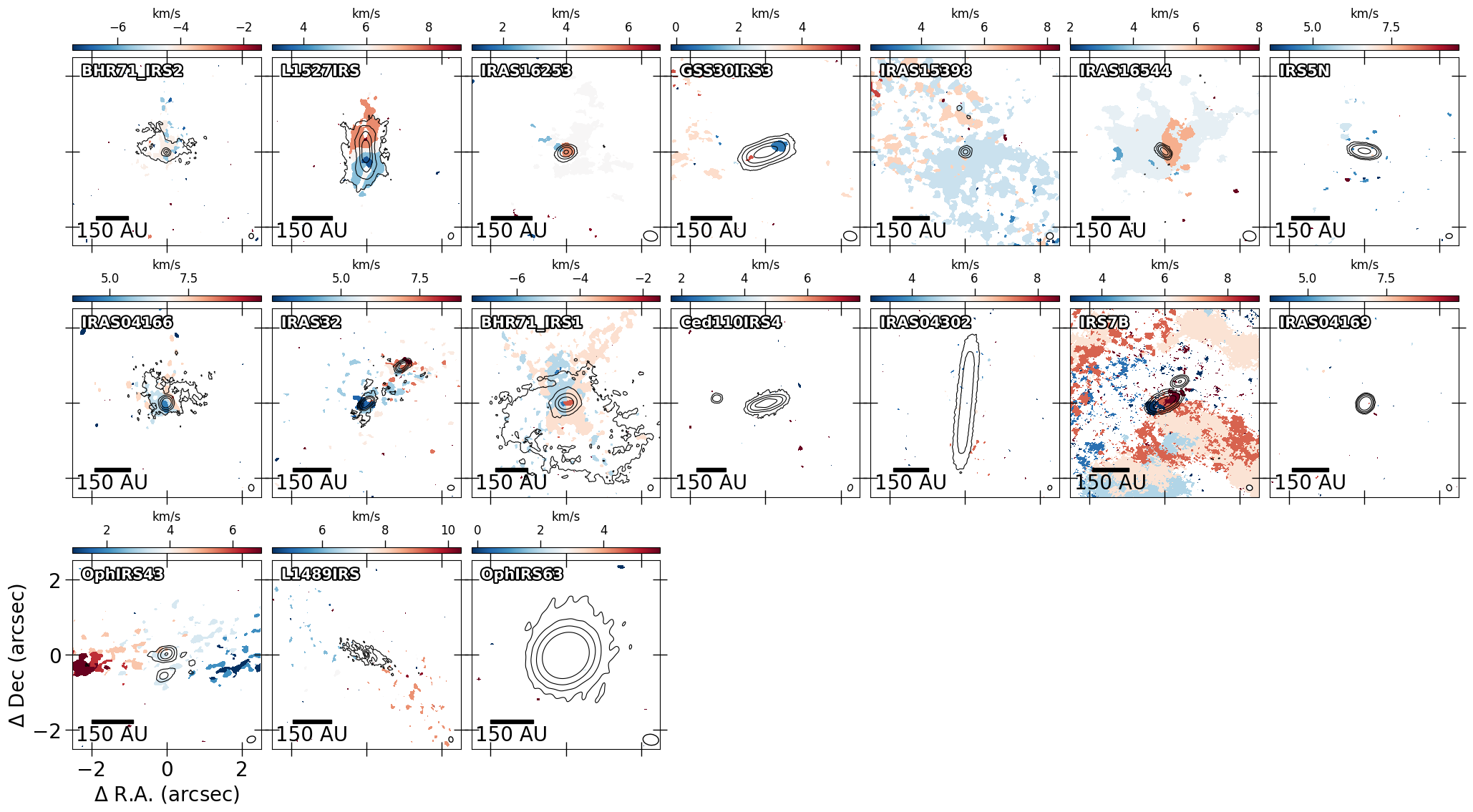}
  \caption{Same as Figure~\ref{appendix:12co_mom8_mom9_zoomed} but for \h2co~(\htwocohigh) instead. \label{appendix:h2co_3_22-2_21_mom8_mom9_zoomed}}
\end{figure*}

\begin{figure*}[htbp]
  \includegraphics[width=1.0\linewidth]{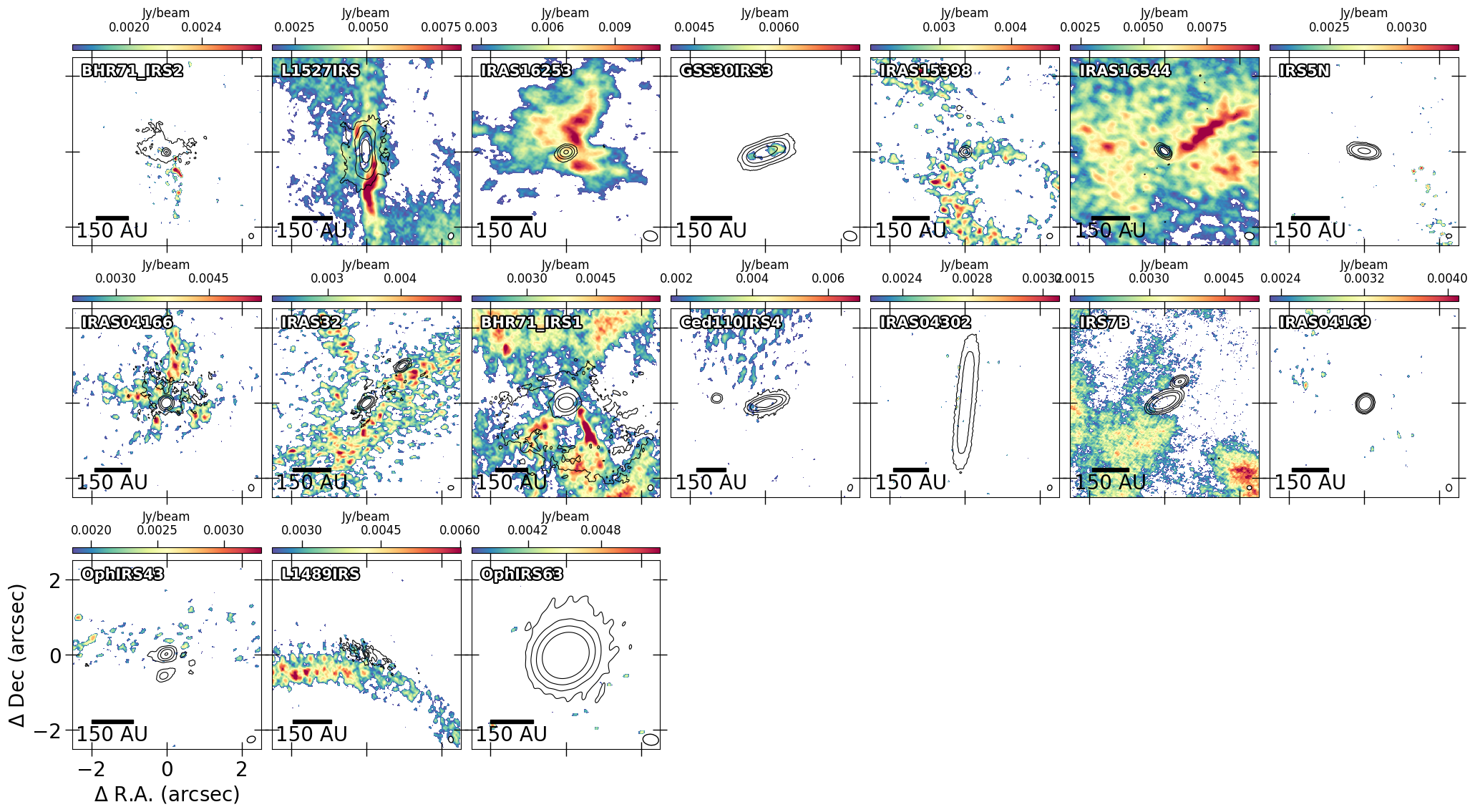}
  \includegraphics[width=1.0\linewidth]{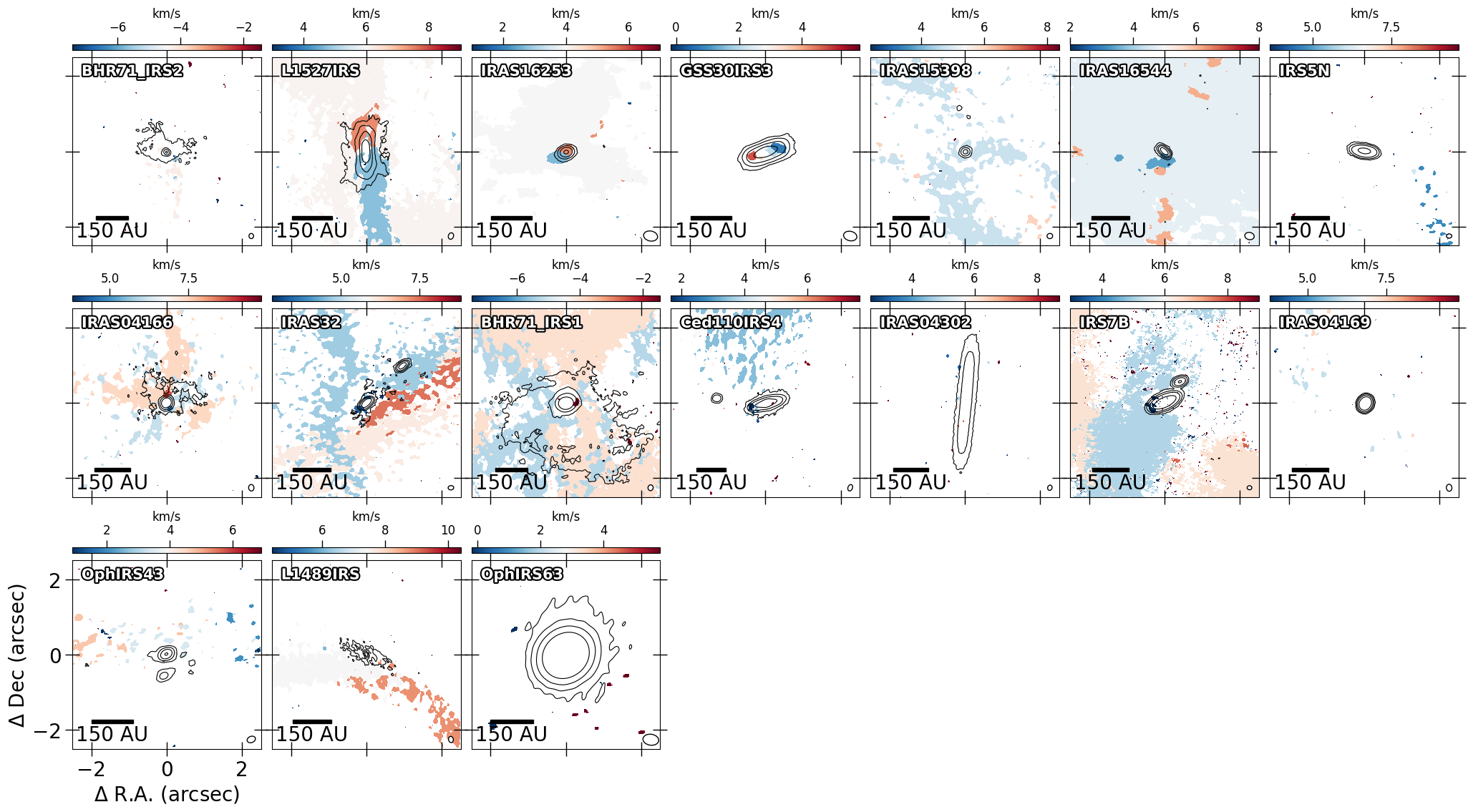}
  \caption{Same as Figure~\ref{appendix:12co_mom8_mom9_zoomed} but for the blended \C3h2~($6_{0,6}$--$5_{1,5}$) and ($6_{1,6}$--$5_{0,5}$) transitions instead. \label{appendix:c3h2_217.82_mom8_mom9_zoomed}}
\end{figure*}

\begin{figure*}[htbp]
  \includegraphics[width=1.0\linewidth]{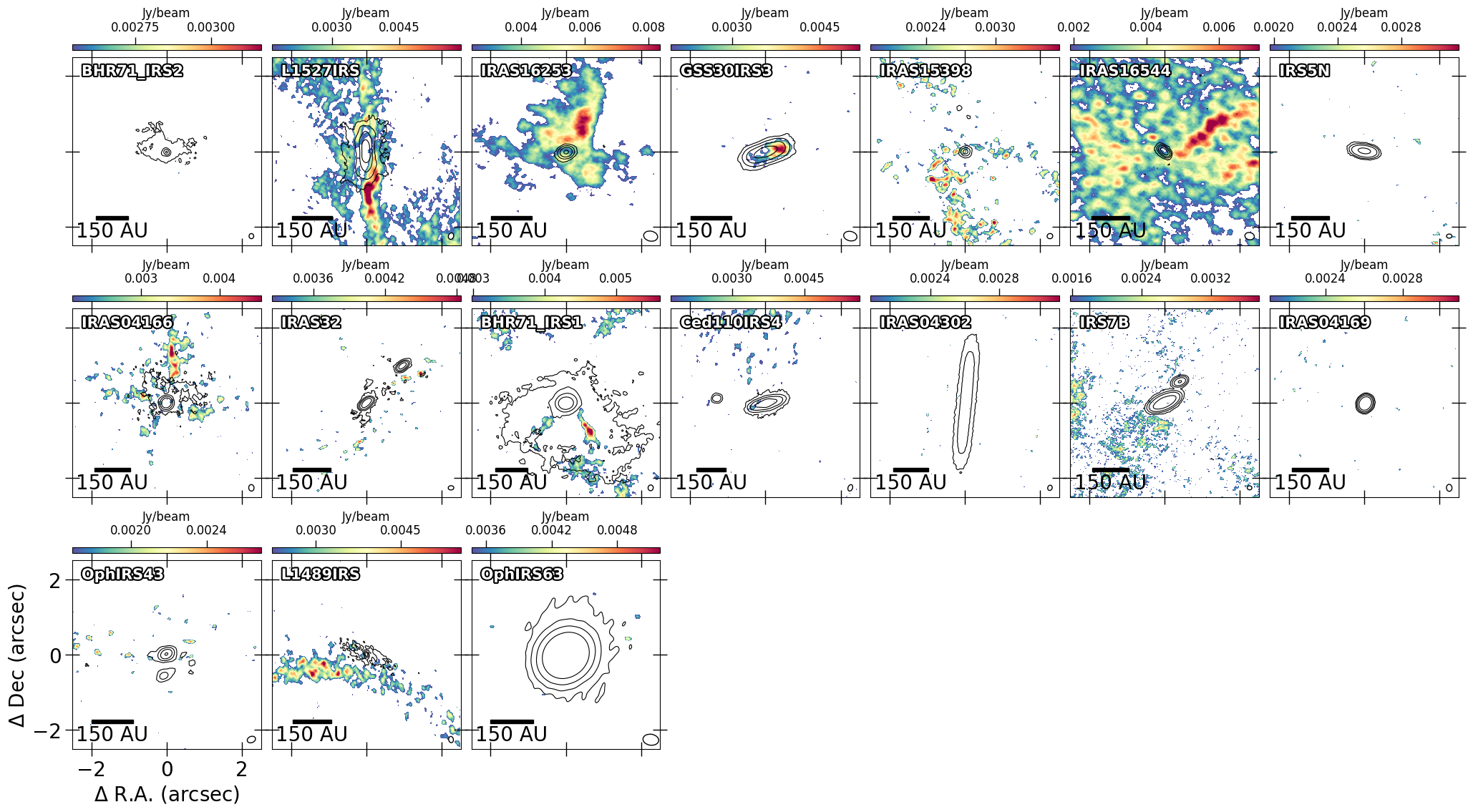}
  \includegraphics[width=1.0\linewidth]{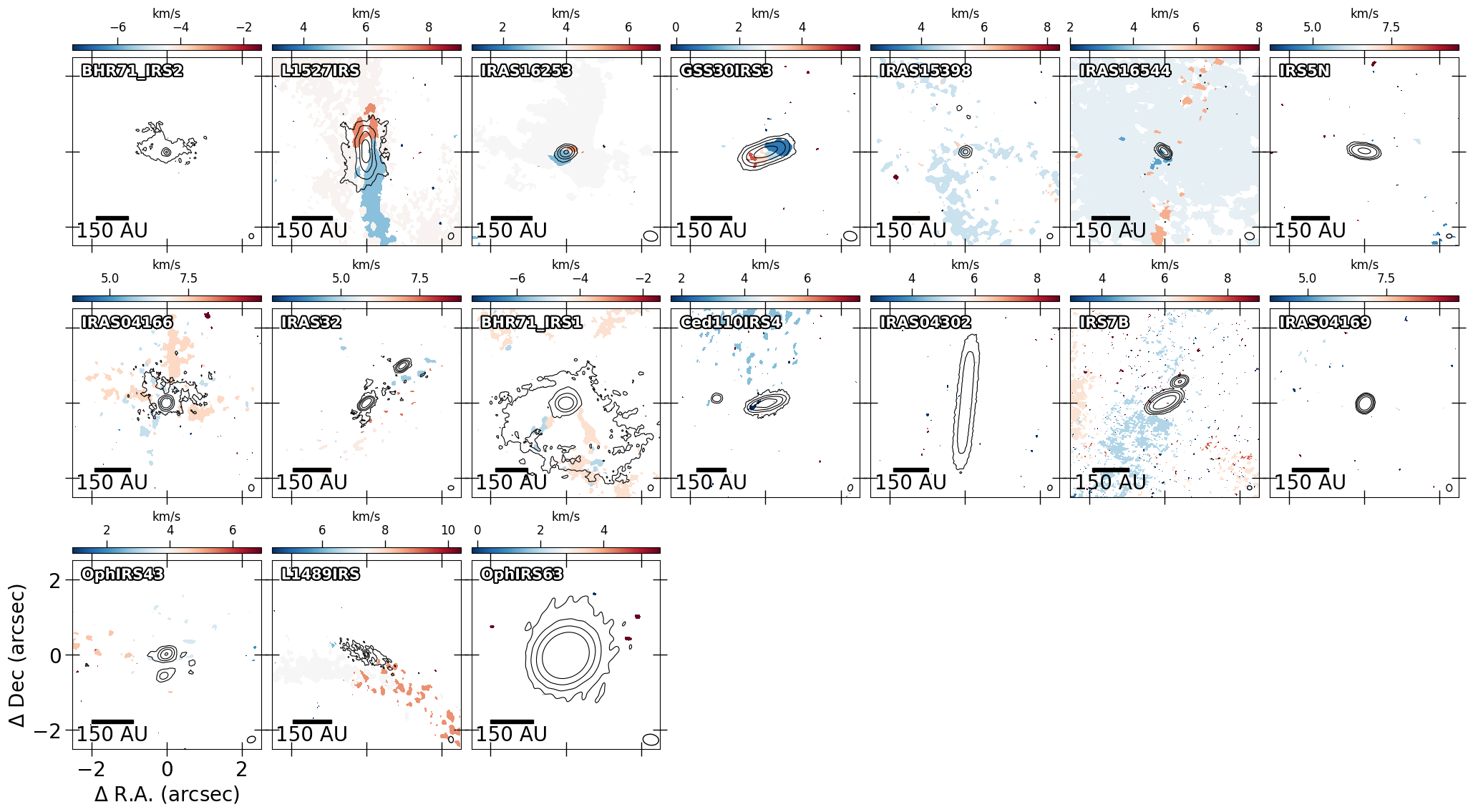}
  \caption{Same as Figure~\ref{appendix:12co_mom8_mom9_zoomed} but for \C3h2~($5_{1,4}$--$4_{2,3}$) instead. \label{appendix:c3h2_217.94_mom8_mom9_zoomed}}
\end{figure*}

\begin{figure*}[htbp]
  \includegraphics[width=1.0\linewidth]{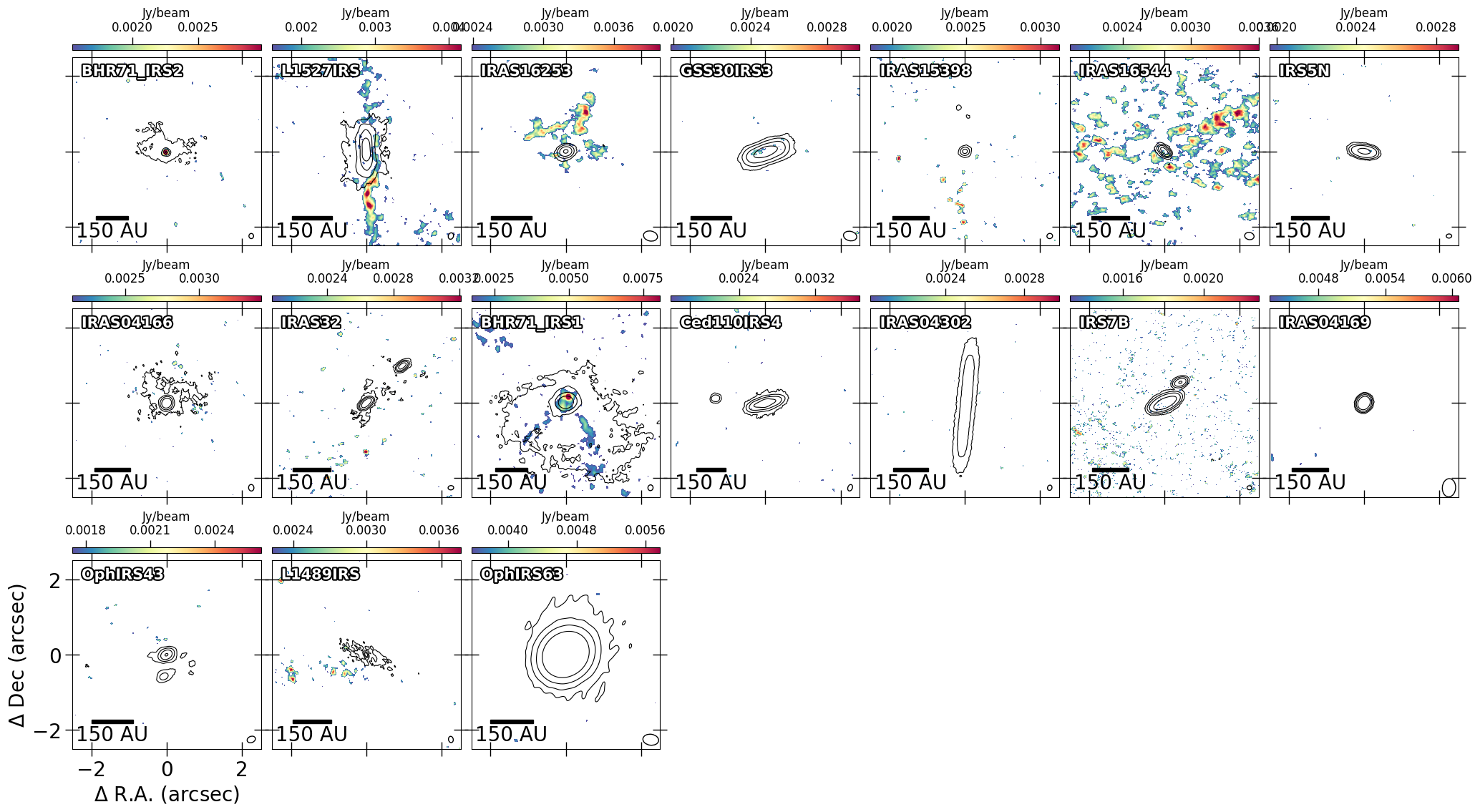}
  \includegraphics[width=1.0\linewidth]{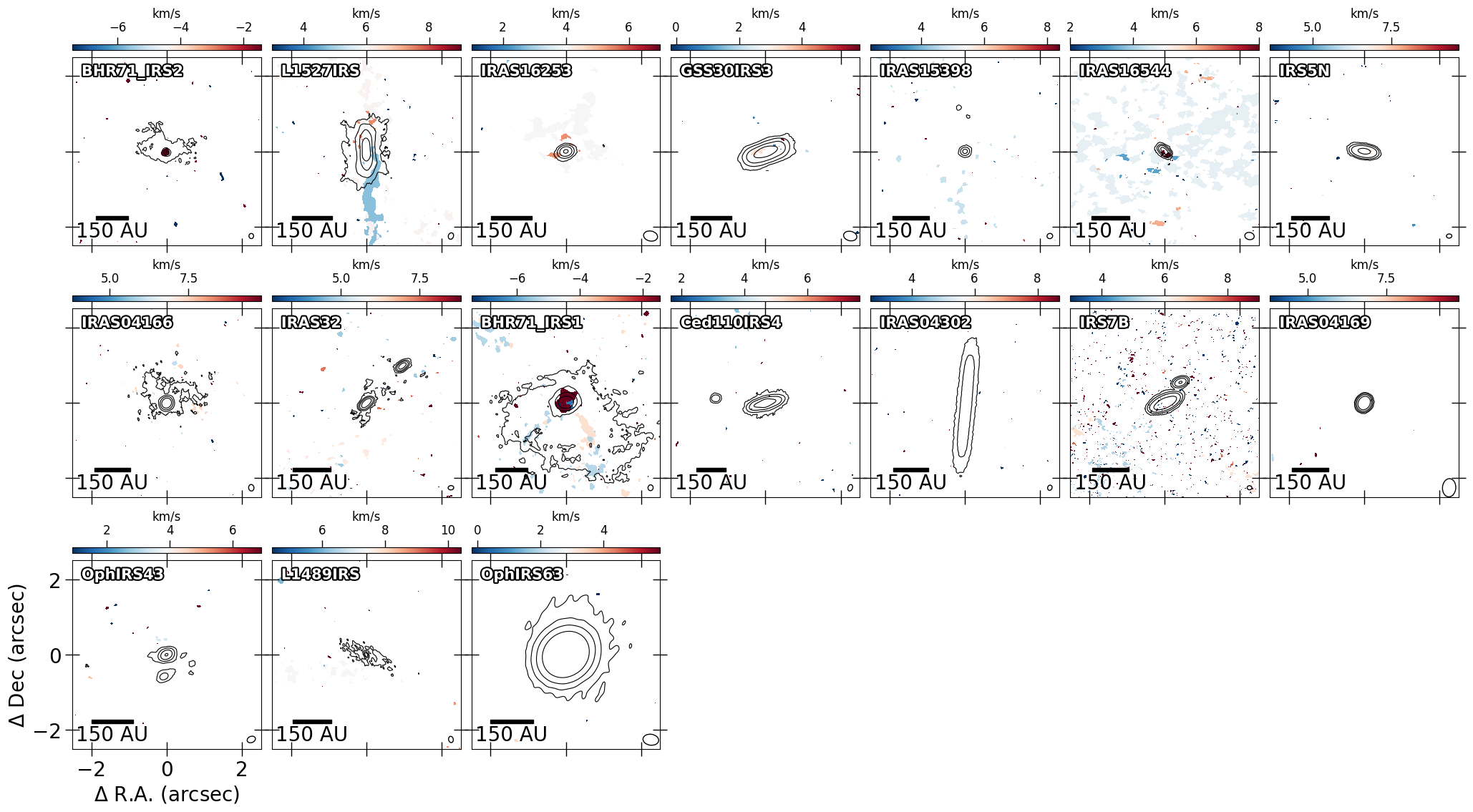}
  \caption{Same as Figure~\ref{appendix:12co_mom8_mom9_zoomed} but for \C3h2~($5_{2,4}$--$4_{1,3}$) instead. \label{appendix:c3h2_218.16_mom8_mom9_zoomed}}
\end{figure*}

\end{document}